\documentclass[namedreferences]{SolarPhysics}
\usepackage[optionalrh]{spr-sola-addons} % For Solar Physics
\usepackage{epsfig}          % For eps figures, old commands
\usepackage{graphicx}        % For eps figures, newer & more powerfull
\usepackage{amssymb}        % useful mathematical symbols
\usepackage{color}           % For color text: \color command
\usepackage{url}             % For breaking URLs easily trough lines
\usepackage{rotating}

\begin{document}

\begin{article}

\begin{opening}

\title{Solar Energetic Particle Events in the 23rd Solar Cycle: Interplanetary Magnetic Field Configuration and Statistical Relationship with Flares and CMEs}

\author{R.~\surname{Miteva}$^{1}$\sep
        K.-L.~\surname{Klein}$^{1}$\sep
        O.~\surname{Malandraki}$^{2}$\sep
        G.~\surname{Dorrian}$^{2}$
       }
\runningauthor{Miteva et al.}
\runningtitle{SEP Events in the 23rd Solar Cycle}

   \institute{$^{1}$ LESIA-Observatoire de Paris, 5 place Jules Janssen, 92195 Meudon, CNRS, UPMC Univ. Paris 06, Univ. Paris-Diderot, France
                     email: \url{rositsa.miteva@obspm.fr} email: \url{ludwig.klein@obspm.fr}\\
              $^{2}$ Institute of Astronomy, Astrophysics, Space Applications and Remote Sensing, National Observatory of Athens, Greece
                     email: \url{omaland@astro.noa.gr} email: \url{gdorrian@astro.noa.gr} \\
             }

\begin{abstract}
We study the influence of the large-scale interplanetary magnetic field configuration on the solar energetic particles (SEPs) as detected at different satellites near Earth and on the correlation of their peak intensities with the pa\-rent solar activity. We selected SEP events associated with X and M-class flares at western longitudes, in order to ensure good magnetic connection to Earth. These events were classified into two categories according to the global interplanetary magnetic field (IMF) configuration present during the SEP pro\-pagation to 1~AU: standard solar wind or interplanetary coronal mass ejections (ICMEs). Our analy\-sis shows that around 20\% of all particle events are detected when the spacecraft is immersed in an ICME. The correlation of the peak particle intensity with the projected speed of the SEP-associated coronal mass ejection is similar in the two IMF categories of proton and electron events, $\approx 0.6$. The SEP events within ICMEs show stronger correlation between the peak proton intensity and the soft X-ray flux of the associated solar flare, with correlation coefficient $r=\,$0.67$\pm$0.13, compared to the SEP events propagating in the standard solar wind, $r=\,$0.36$\pm$0.13. The difference is more pronounced for near-relativistic electrons. The main reason for the different correlation behavior seems to be the larger spread of the flare longitude in the SEP sample detected in the solar wind as compared to SEP events within ICMEs. We discuss to which extent observational bias, different physical processes (particle injection, transport, etc.), and the IMF configuration can influence the relationship between SEPs and coronal activity.
\end{abstract}
\end{opening}
%-------------------------------------------------

\section{Introduction} %%%%%%%%%%%%%%%%%%%%%%%%%%%%%%%%%%%%%%%%%%%%%
     \label{S-Introduction}

Solar energetic particles (SEPs) are transient enhancements of the intensities of energetic protons, ions, and electrons observed in the interplanetary (IP) space. They are known to follow in time eruptive phenomena in the solar corona, such as flares and coronal mass ejections (CMEs). Both small scale processes during flares and CME-driven shock waves are used to explain the particle acceleration (see, {\it e.g.}, the review of \opencite{2006SSRv..123..217K}). The question how flares and CMEs affect SEPs is, however, largely unresolved, because particle measurements near 1~AU are related to the coronal accelerator through a poorly understood chain of processes of acceleration, access to, and propagation in the dynamic interplanetary medium.

One approach to identify physical relationships between SEPs and the parent solar activity is statistical. Numerous studies have shown that SEP events are associated both with flares, as manifested, {\it e.g.}, by their soft X-ray \cite{2004SpWea...2.6003G} or radio (\opencite{1982ApJ...261..710K}, \citeyear{1982JGR....87.3439K}) emission, and with fast and broad coronal mass ejections \cite{1992ARA&A..30..113K,1999SSRv...90..413R}. The recent global study of SEP events in the 23rd solar cycle by \inlinecite{2010JGRA..11508101C} confirmed this. Pure cases of SEP events in association with fast CMEs lacking evidence of other, flare-like, acceleration processes in the corona are rare \cite{1986ApJ...302..504K} or non existent \cite{2006ApJ...642.1222M}. On the other hand, strong flares that are not accompanied by CMEs are not associated with SEP events (detectable by GOES) either, mostly because the flare-accelerated particles remain confined in coronal magnetic fields \cite{2010SoPh..263..185K,2011SoPh..269..309K}.

More detailed statistical studies tried to relate the peak intensity of SEP events to parameters of the flare or the CME. \inlinecite{1990AN....311..379C} showed a close correlation between peak proton intensities measured in space and gamma-ray line fluxes, but others concluded that the ratio between the numbers of deka-MeV protons emitting gamma-ray lines and detected in situ varied considerably from event to event \cite{1989ApJ...343..953C,1993AdSpR..13..275R}. There are also indications for some statistical correlations between SEP intensities and microwave burst parameters \cite{1982JGR....87.3439K}. \inlinecite{2004SpWea...2.6003G} developed a very detailed empirical analysis relating the proton intensity at energies above 10~MeV to a combination of parameters of the soft X-ray (SXR) burst (peak flux, duration, and emission measure). Correlations between SEP peak intensities and CME parameters were also found, especially with the plane-of-the-sky speed of CMEs \cite{2001JGR...10620947K}. The reported correlation coefficients range between 0.6 and 0.7. Since the significance of  correlation coefficients was either not assessed, or only confidence levels were given, it is hard to see if any difference in the correlation coefficients is statistically significant. The large scatter in most correlations was considered as an argument that other factors contribute to the efficiency of SEP acceleration. \inlinecite{2001JGR...10620947K} showed that correlations also exist between the SEP peak intensity and the pre-event intensity, and interpreted this as evidence that a pre-accelerated seed population increases the acceleration efficiency of a CME shock. \inlinecite{2004JGRA..10912105G} argued that CME interaction enhances SEP intensities.

The main and inevitable limitation of such statistical studies is that SEP intensities are generally measured at only one point. Its magnetic connection to the accelerator in the corona is not well known. It is often approximated by a Parker spiral. But this may not be true, as has been shown by event studies where SEPs reach the detector in transient interplanetary magnetic field (IMF) structures, {\it i.e.} interplanetary coronal mass ejections or ICMEs \cite{1987JGR....92....6T,2004ApJ...600L..83T,2005JGRA..11009S06M,2011JGRA..11601104K}. \inlinecite{2012A&A...538A..32M} showed recently that the majority of relativistic SEP events of solar cycle 23 were detected within or in the vicinity of ICMEs. These ICMEs stem from solar activity that occurred one or several days before the SEP event, so that the magnetic configuration had the time to expand and reach the Earth.

This paper presents a re-assessment of statistical relationships between the peak intensities (and fluences) of particle events ({\it i.e.}, near-relati\-vistic electrons of tens to hundreds of keV and deka-MeV protons), and the parameters of the associated coronal activity ({\it i.e.}, the peak SXR flux of the flare and the speed and width of the CME). Two categories of IMF configuration guiding the particles through the IP space are distinguished: standard solar wind and ICMEs. All SEP events of solar cycle 23 (1997$-$2006) that occurred with flares of classes M and X\footnote{GOES X-ray classification in the 1$-$8 \AA $\,$channel: M class flares have peak flux that exceeds $10^{-5}\,$W$\,$m$^{-2}$, whereas the X class flares are 10 times more intense.} in the western solar hemisphere are considered. The data sets and analysis technique are described in Section~\ref{S-Data}. Section~\ref{S-Observations} presents the observational findings: the identification of SEP events within ICMEs and within the standard solar wind, together with the distributions of peak particle intensities (Section~\ref{S-IP_field}), rise times (Section~\ref{S-Rise_time}), and connection distances to the parent flare (Section~\ref{S-Conn_dist}). In se\-parate subsections we address the following statistical relationships: between the flares and CMEs (Section~\ref{S-correlfc}), between the SEP intensity (Section~\ref{S-correlations}), and rise-to-peak fluence (Section~\ref{S-Fluence}), on one hand, and the parameters of the associated solar activity, on the other. The ICME category is discussed in more details in Section~\ref{S-ICME-category} and the effect of the connection distance on the correlations in Section~\ref{S-Corr_conn_dist}. Section~\ref{S-disc} addresses the influence of observational bias and physical effects on the correlation between the intensity of deka-MeV protons and near-relativistic electrons and the parameters of the parent coronal activity.

\section{Data Analysis} %%%%%%%%%%%%%%%%%%%%%%%%%%%%%%%%%%%%%%%%
      \label{S-Data}

The data set for this statistical study is composed of SEP events between 1997 and 2006 (23rd solar cycle) associated with well-identified activity in the western solar hemisphere (flares at longitude $<90^{\circ}$), using the list of SEP events at energies above 25~MeV of \inlinecite{2010JGRA..11508101C}  based on IMP-8 and SOHO/ERNE data. Several events were excluded because of high background from a preceding particle event (typically when the two subsequent SEP events are less than 8$-$10~h apart). We excluded from the statistics also SEP events that occurred during a SOHO data gap. This leads to a data sample that contains 38 SEP events associated with X-class flares and 66 with M-class flares.

For the quantitative analysis of the proton data we use three complementary sets of observations: GOES 15$-$40 MeV data from the Ionising Particle ONERA DatabasE (IPODE) developed at the Office National d'Etudes et Recherches A\'erospatiales (ONERA) in Toulouse, provided by D.~Boscher. This database hosts GOES data that were carefully compared between simultaneously observing GOES spacecraft and corrected for evident outliers. Additionally we use the {\it Wind}/EPACT 19$-$28 MeV particle data available via the CDAWeb service\footnote{\url{http://cdaweb.gsfc.nasa.gov}}, reported there as preliminary browse data, and the SEP events in \inlinecite{2010JGRA..11508101C}. This yields a total of 81 SEP events observed by GOES, 96 by {\it Wind}/EPACT, and 104 from \inlinecite{2010JGRA..11508101C}. The three data sets from different instruments with different calibrations allow for consistency checks of statistical results. Energetic electron data are used as measured by the 38$-$53 keV and 175$-$315 keV energy channels of the EPAM experiment\footnote{\url{http://www.srl.caltech.edu/ACE/ASC/level2/index.html}} aboard the {\it Advanced Composition Explorer} (ACE) spacecraft \cite{1998SSRv...86..541G}.

A constant pre-event background was subtracted from the particle intensities. While determining the peak particle intensity, we sought for the maximum of the SEP events, avoiding late peaks that could be associated with energetic storm particle (ESP) events. An additional difficulty in estimating the intensity peak was the presence of several maxima in the time evolution. In such cases (denoted with $m$ in the tables in the Appendix) we took the largest intensity maximum.

Data on flare size, times, and heliographic location were taken from the {\it Solar Geophysical Data} reports\footnote{\url{ftp://ftp.ngdc.noaa.gov/STP/SOLAR_DATA/SGD_PDFversion/}} compiled by NOAA. The CME speed is the projected one reported (as linear speed) in the CDAW catalog\footnote{\url{http://cdaw.gsfc.nasa.gov/CME_list/}} \cite{2004JGRA..10907105Y}. The CME angular widths are taken from \inlinecite{2010JGRA..11508101C}, who estimated the angular width of each CME in the LASCO-C2 instrument field of view (heliocentric distance 2.5$-$6 solar radii). The complete event list and the information on the related parameters is given in the Appendix (Tables~\ref{T-ICME_events}$-$\ref{T-Other_events}).

\section{Observational Findings} %%%%%%%%%%%%%%%%%%%%%%%%%%%%%%%%%%%%%%%%
      \label{S-Observations}

\subsection{Interplanetary Magnetic Field Configuration}
  \label{S-IP_field}

In order to classify the SEP events into categories according to the magnetic field configuration guiding the particles from the Sun to the Earth, we compared the times of onset and rise of the SEP profiles with the start and end times of ICMEs reported by \inlinecite{2010SoPh..264..189R}. The time boundaries of the ICME were inferred there primarily from plasma and magnetic field data measured by the ACE spacecraft and are reported to the nearest hour. This catalog gives, together with different ICME characteristics, the time of the disturbance at Earth, the start and end times of the ICME at ACE (and occasionally at the {\it Wind} spacecraft). We shifted the ICME times measured at ACE (positioned at the Lagrangian L1 point) to the GOES orbit taking the reported ICME speed as constant over the distance between the two spacecraft. When the onset and rise of the SEP event was detected while the Earth was within an ICME, we consider that the particles propagated in a transient flux tube of an ICME. These SEP events are called `ICME events' in the following. When the SEP onset occurred at least one day after the end of a previous ICME at the Earth and at least one day before the start of the next ICME, we consider that the particles propagated within the standard solar wind. Those events are called `SoWi events'. In the remaining events the IMF configuration is uncertain. In those cases the SEP may propagate in the solar wind, in the vicinity of the shock, in the sheath, or in disturbed interplanetary field lines behind the ICME, which may result from reconnection between the ICME and the ambient solar wind. Finally, the labels `All'/`All SEPs' denote the complete set of SEP events, comprising the categories of ICME, SoWi, and the SEP events in the vicinity on an ICME (Tables~\ref{T-ICME_events}$-$\ref{T-Other_events} in the Appendix).

\begin{table}[t!]
\caption{Number of SEP events in the different IMF configurations with respect to the associated flare size}
\label{T-IMF}
\begin{tabular}{lccc}
\hline
Flare              &  \multicolumn{3}{c}{IMF categories of SEP events}\\
size               & ICME  & SoWi & All SEPs \\
%(1)                & (2)   & (3)  & (4)\\
\hline
Protons & \multicolumn{3}{c}{GOES/{\it Wind}-EPACT/\inlinecite{2010JGRA..11508101C}} \\
M-class              & 7  / 11 / 9  & 24 / 29 / 36 & 46 / 60 / 66 \\
X-class              & 10 / 11 / 11 & 14 / 14 / 16 & 35 / 36 / 38 \\
(X+M)-class          & 17 / 22 / 20 & 38 / 43 / 52 & 81 / 96 / 104 \\
\hline
Electrons & \multicolumn{3}{c}{ACE/EPAM 38$-$53 and 175$-$315 keV} \\
M-class              &  10          & 34       & 65 \\
X-class              &  8           & 12       & 31 \\
(X+M)-class          &  18          & 46       & 96 \\
\hline
\end{tabular}
\end{table}

The distributions of the SEP events among the IMF categories and also with respect to the size of the associated flare (in GOES X-ray classification of the 1$-$8 \AA $\,$channel) are listed in Table~\ref{T-IMF}. On separate rows we give the number of SEP events associated with M, X, and (X+M)-class flares, respectively. The three proton instruments detected different numbers of SEP events related to M and X-class flares (probably due to their different sensitivities). The three proton entries separated by slashes refer to the SEP events observed by GOES, {\it Wind}/EPACT, and from the catalog of \inlinecite{2010JGRA..11508101C}, respectively. The table shows that a significant number of SEP events are observed when the Earth is within an ICME, and that the ratio of ICME events to the total number increases with the importance of the parent activity: 29\% of the proton events associated with X-class flares belongs to the ICME events ({\it e.g.}, 10/35 cases observed by GOES), but only between 14\% and 18\%, depending on the data set considered, of the proton events accompanied by M-class flares. For electron events this ratio is 26\% for X-class flares and 15\% for M-class.

In the following figures we present statistical properties and correlations of the SEP events that are in general given as a comparative set of plots for the protons (on the left) and electrons (right) in the different IMF categories, namely ICME, SoWi, and All events. Unless stated otherwise in the text, the proton data in the figures is the 15$-$40 MeV peak intensity measured by the GOES instrument, whereas the electron data is from the low energy channel of ACE/EPAM. Since the numbers of proton and electron events are slightly different (for a few SEP events there are proton but not electron data and vice versa), we carry on the analysis for both particle species separately.

\begin{figure}[!t]
   \centerline{\hspace*{-0.02\textwidth}
               \includegraphics[width=0.6\textwidth,clip=]{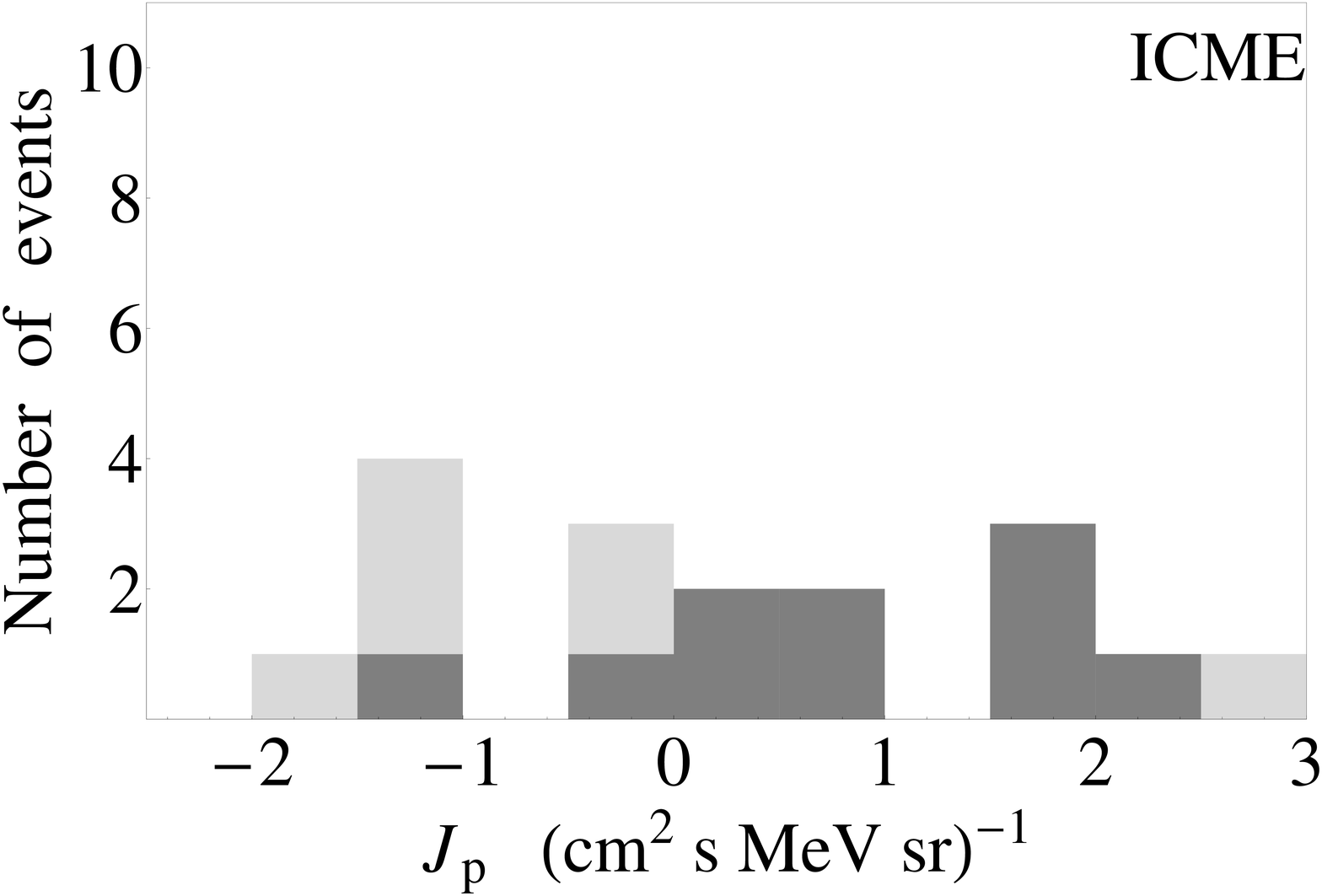}
               \hspace*{-0.1\textwidth}
               \includegraphics[width=0.6\textwidth,clip=]{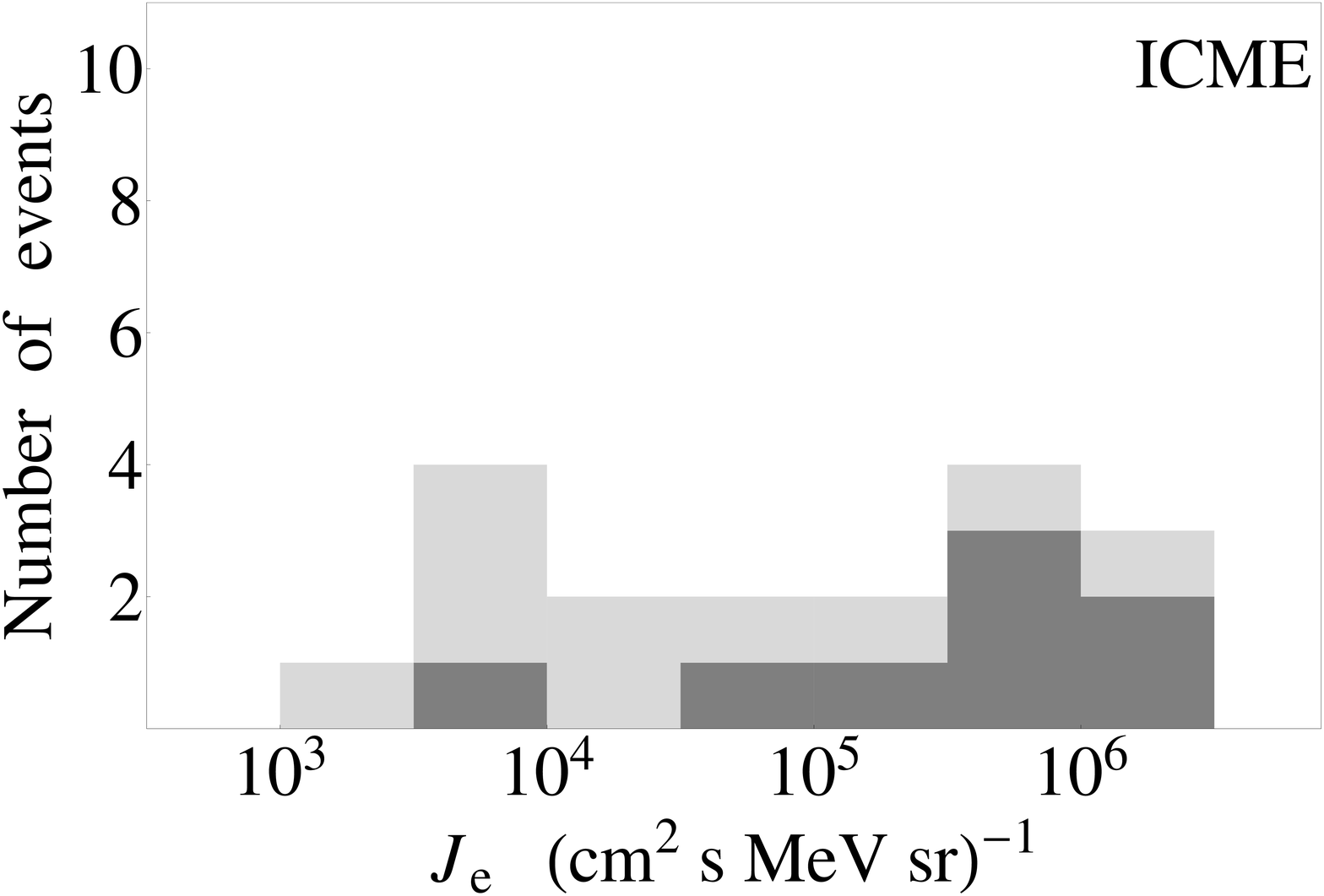}
               }
              \vspace*{-0.05\textwidth}
   \centerline{\hspace*{-0.02\textwidth}
               \includegraphics[width=0.6\textwidth,clip=]{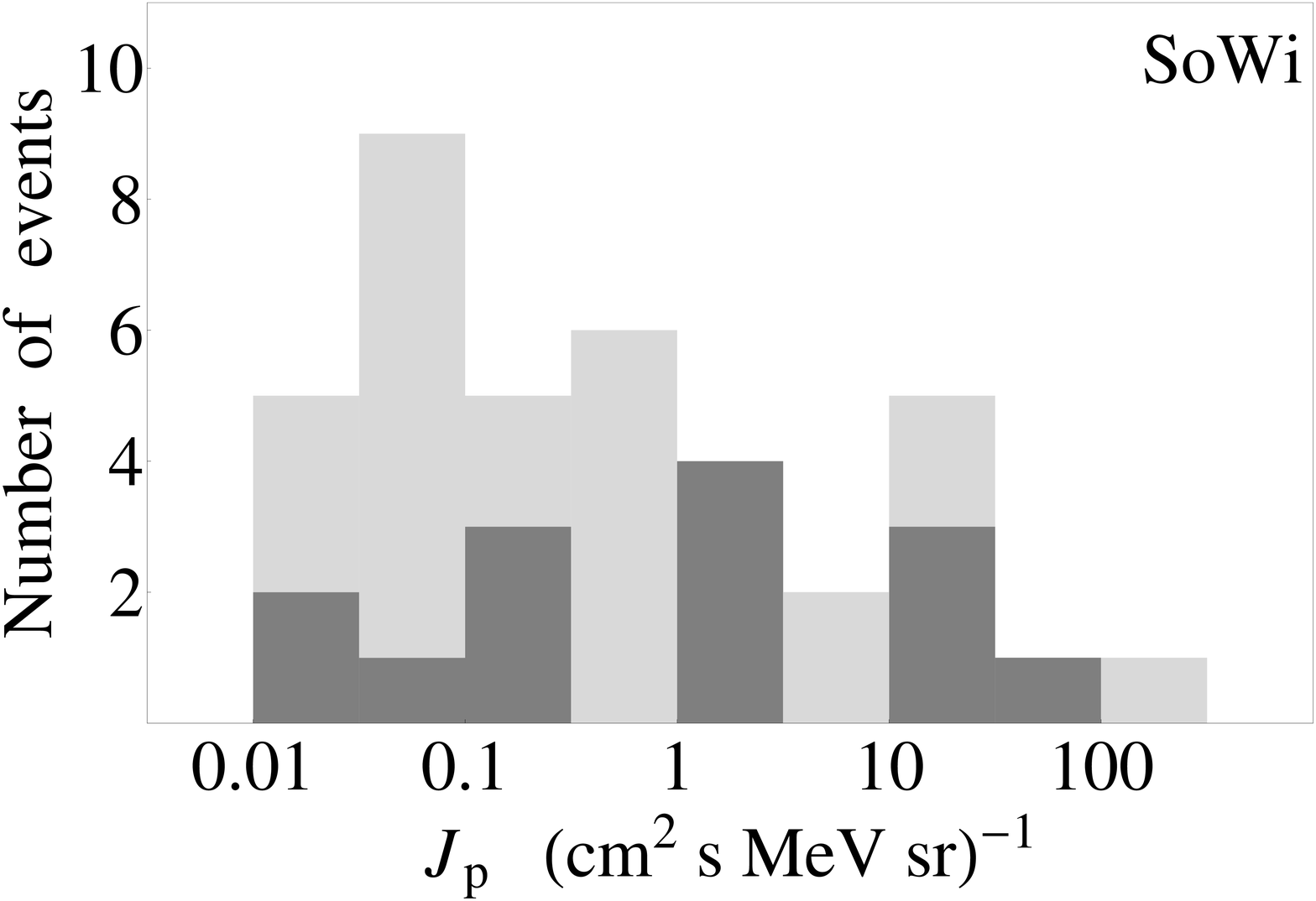}
               \hspace*{-0.1\textwidth}
               \includegraphics[width=0.6\textwidth,clip=]{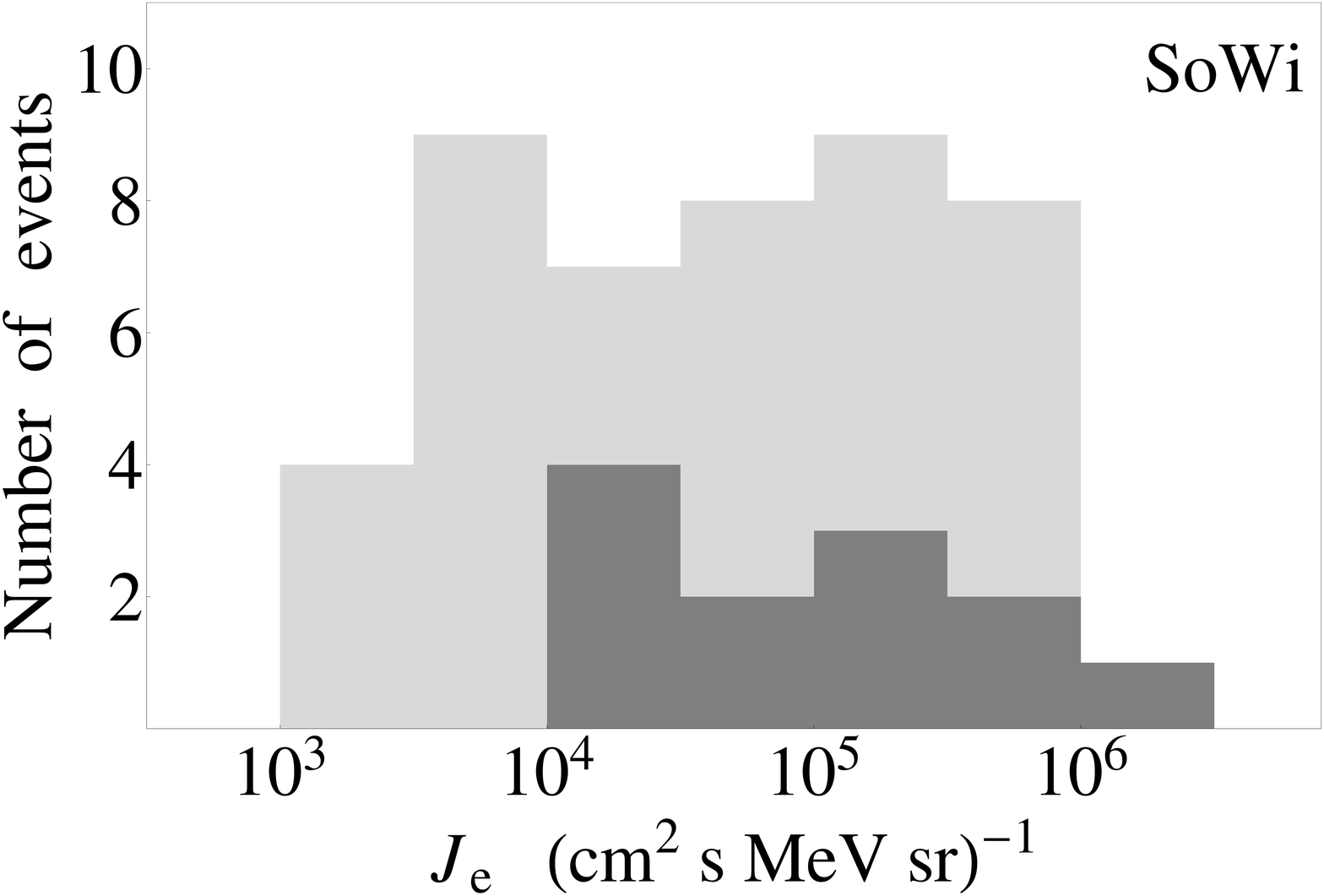}
              }
              \vspace*{-0.05\textwidth}
   \centerline{\hspace*{-0.02\textwidth}
               \includegraphics[width=0.6\textwidth,clip=]{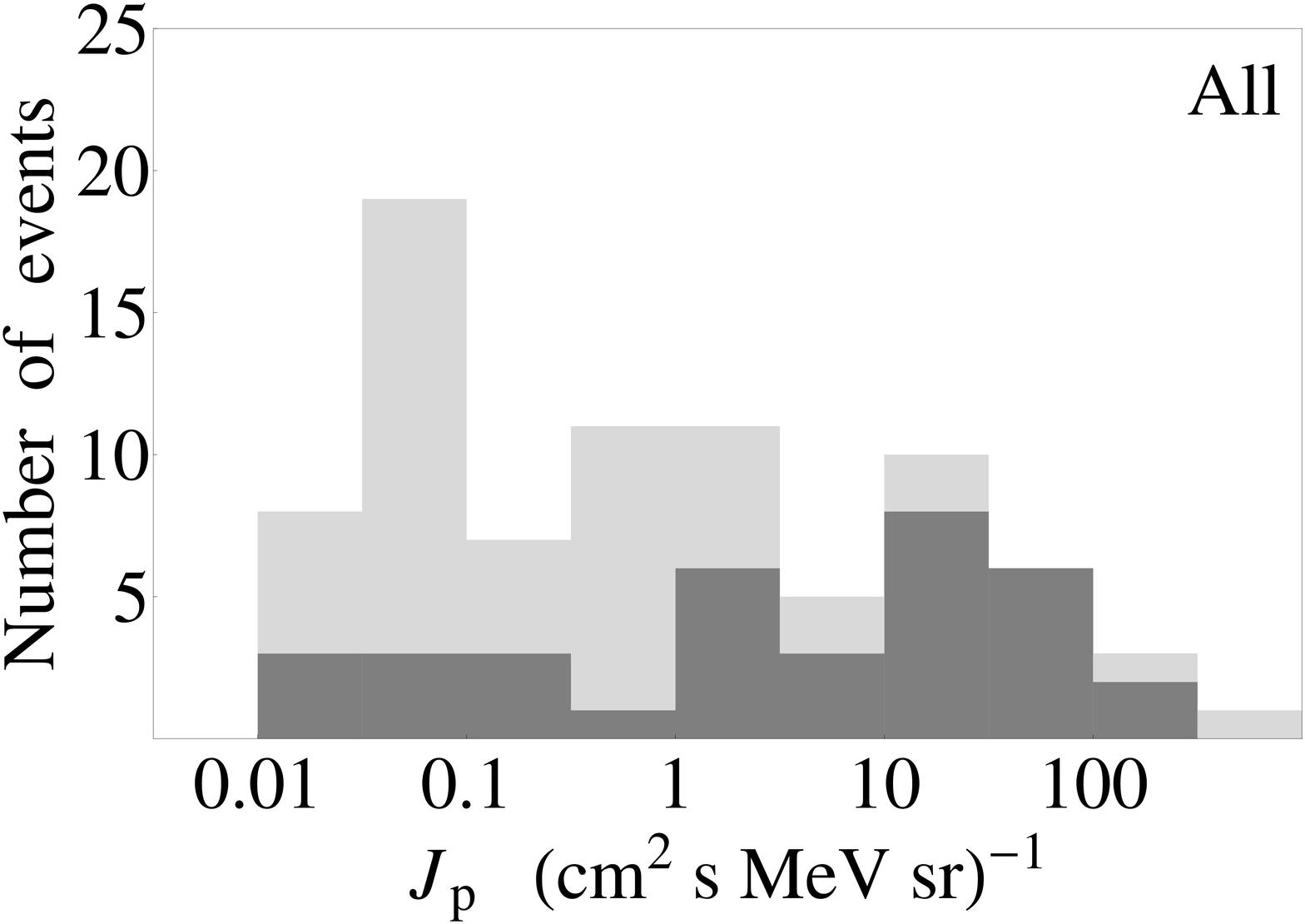}
               \hspace*{-0.1\textwidth}
               \includegraphics[width=0.6\textwidth,clip=]{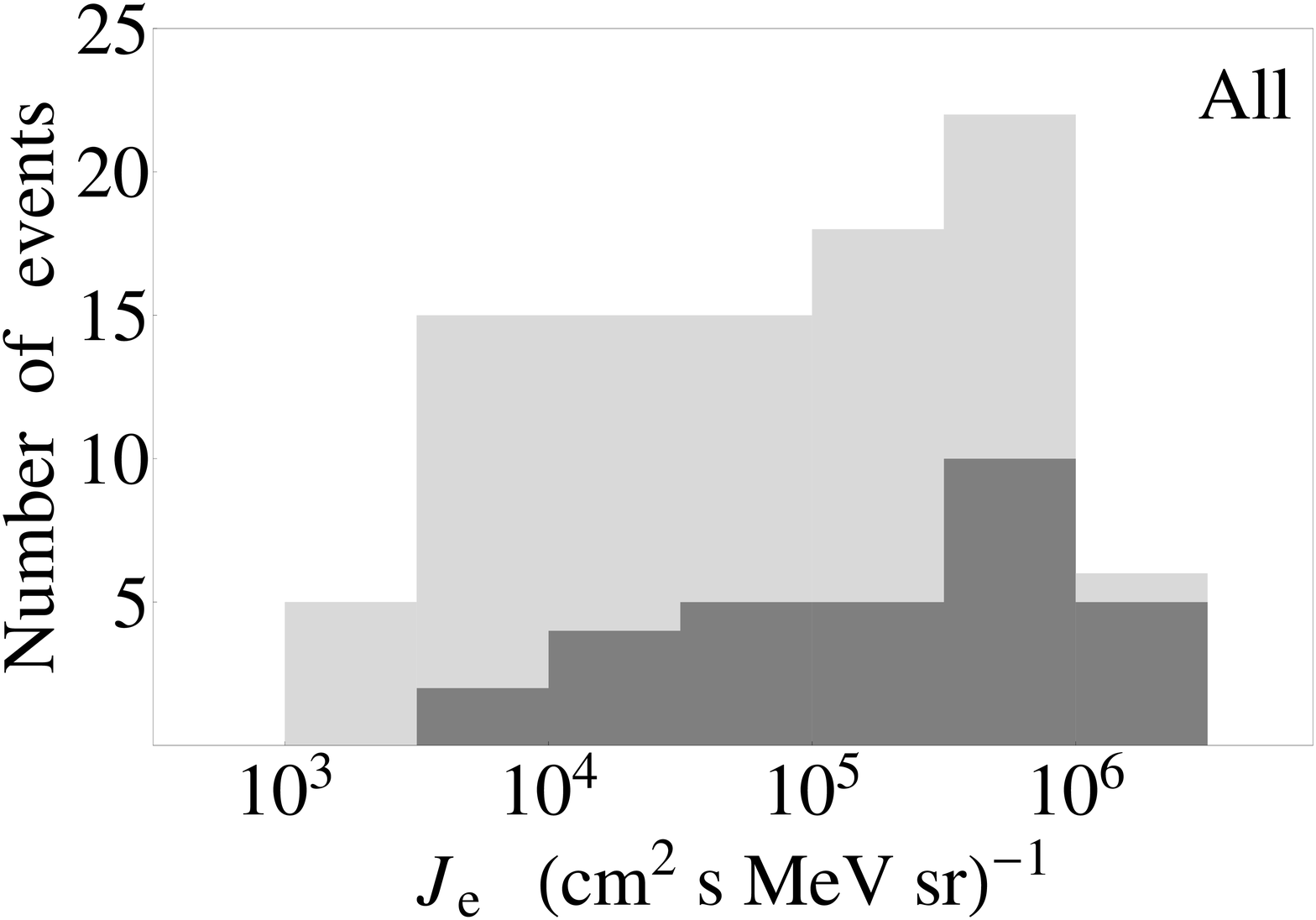}
              }
\caption{Distributions of the proton (left) and electron (right) peak intensity for the different IMF categories of SEP events (proton data from GOES and electron data from low energy channel of ACE/EPAM). The number of the SEP events in each category is given by the length of the corresponding color bar in each bin. SEP events associated with X-class flares are highlighted in dark gray color, whereas M-class associated SEP events are given in light gray.}
\label{F-peak_int}
   \end{figure}

The distributions of the peak differential intensities of protons, $J_{\rm p}$, and electrons, $J_{\rm e}$, are given in Figure~\ref{F-peak_int} in `stacked form'. Dark shading shows SEP associated with X-class flares and light shading with M-class flares. The total distribution is the envelope of the two colored sections. The proton and electron plots in Figure~\ref{F-peak_int} show essentially the same overall ranges of peak intensities in ICME events and SoWi events, both for electrons and protons.  The details of the histograms show some differences: Proton distributions of the entire data set and of the SoWi events peak at low differential peak intensities, whereas the protons of the ICME category and the electrons of both categories are more evenly distributed. There seems to be a tendency of a larger fraction of high intensity electron events in the whole population of SEP events, but a larger fraction of low intensity proton events. However, the differences of the distributions between ICME and SoWi events are not significant: A chi-square test shows that the ICME and SoWi intensity distributions can be drawn from the same parent distribution with a probability of 67\% for the protons and 98\% for the electrons, respectively.

The two IMF categories of SEP events (ICME and SoWi) show no conspicuous difference in event-averaged composition ({\it e.g.}, electron-richness, iron-richness, etc., as identified for each event by \inlinecite{2010JGRA..11508101C}).

\subsection{Rise Time} %%%%%%%%%%%%%%
  \label{S-Rise_time}

In order to investigate if the particle transport is different in the two IMF categories, we evaluated the rise times of the intensity profiles\footnote{Estimating transport effects from rise times is valid under the assumption of a single, short in time particle injection.}. This is most easily done if the onset and peak of the profile are clearly defined. However, especially the intensity profiles of deka-MeV protons may be complex with fluctuations superposed upon a general rise or with a gradual flattening to a poorly defined maximum. The simple definition of the rise time as the time from start to maximum did not lead to consistent results for different instruments.  A detailed discussion of rise time determinations is given in \inlinecite{2007SpWea...505001P}.

 \begin{figure}[!t]
   \centerline{\hspace*{-0.02\textwidth}
               \includegraphics[width=0.53\textwidth,clip=]{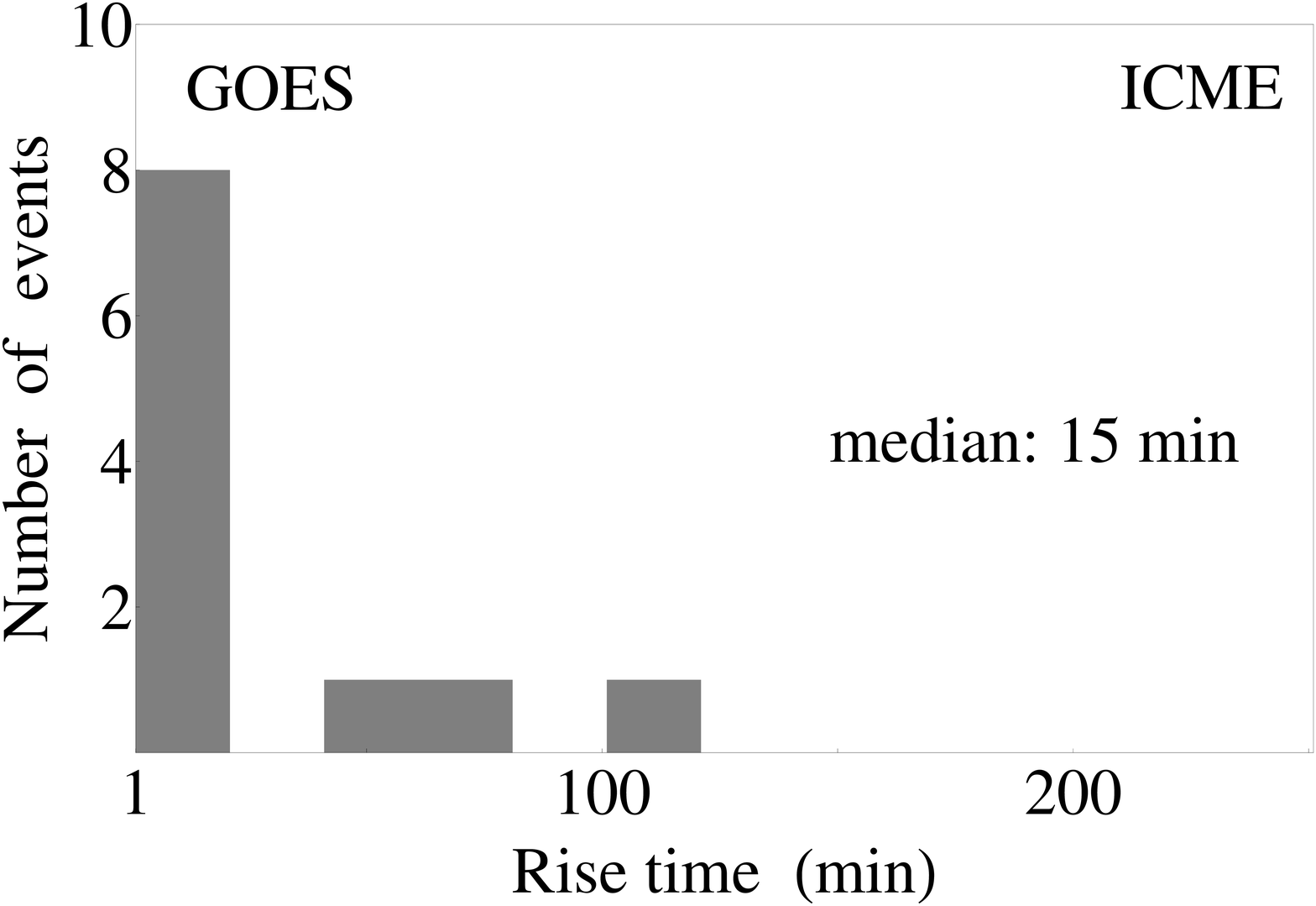}
               \hspace*{-0.07\textwidth}
               \includegraphics[width=0.53\textwidth,clip=]{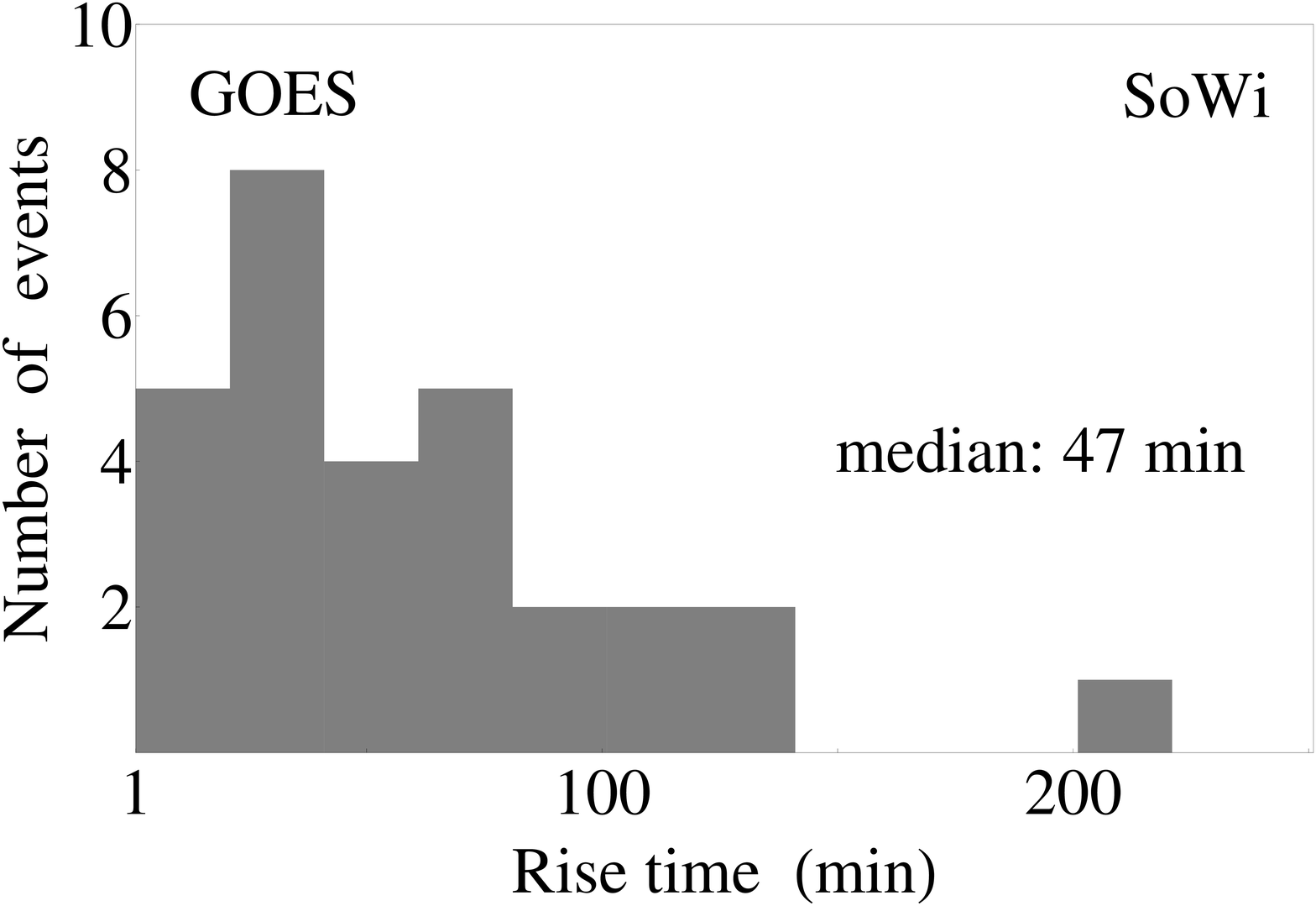}
               }
              \vspace*{-0.02\textwidth}
   \centerline{\hspace*{-0.02\textwidth}
               \includegraphics[width=0.53\textwidth,clip=]{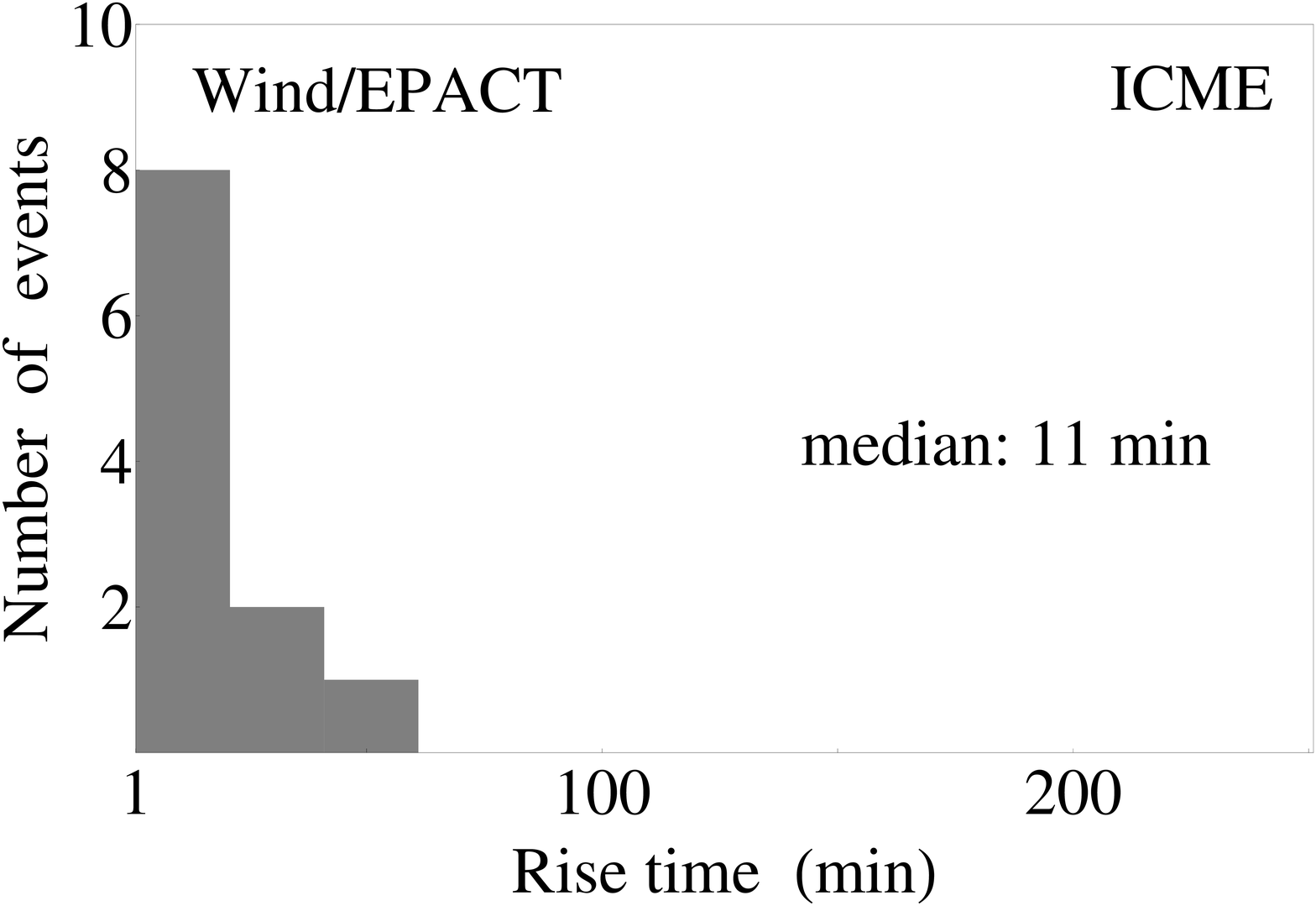}
               \hspace*{-0.07\textwidth}
               \includegraphics[width=0.53\textwidth,clip=]{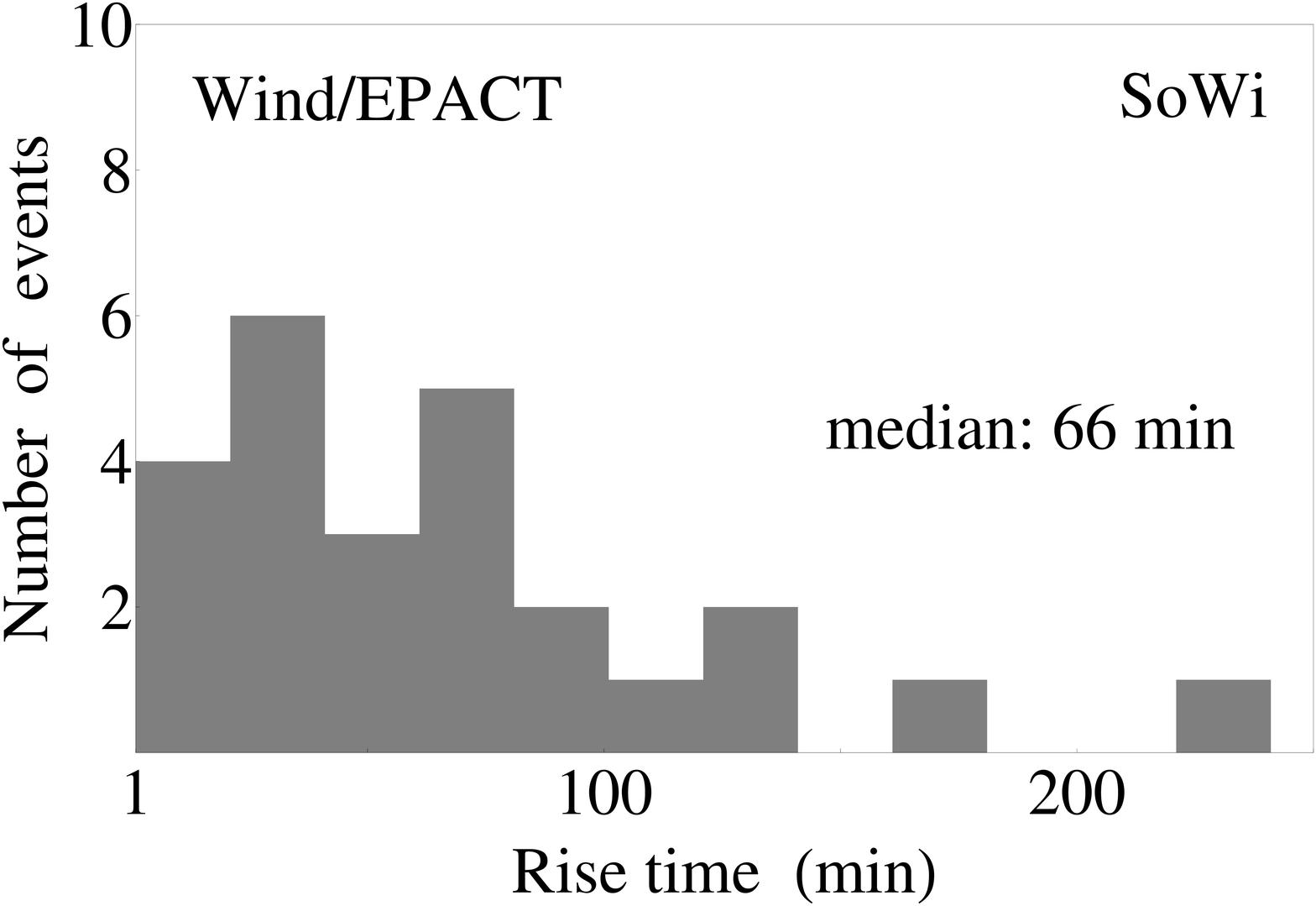}
              }
\caption{Distributions of the GOES and {\it Wind}/EPACT proton rise times for the different IMF categories of SEP events (ICME in the left and SoWi in the right panels). The number of the SEP events in each category is given by the length of the corresponding color bar in each bin.}
\label{F-rise_p}
   \end{figure}

 \begin{figure}[!b]
              \vspace*{-0.02\textwidth}
   \centerline{\hspace*{-0.02\textwidth}
               \includegraphics[width=0.53\textwidth,clip=]{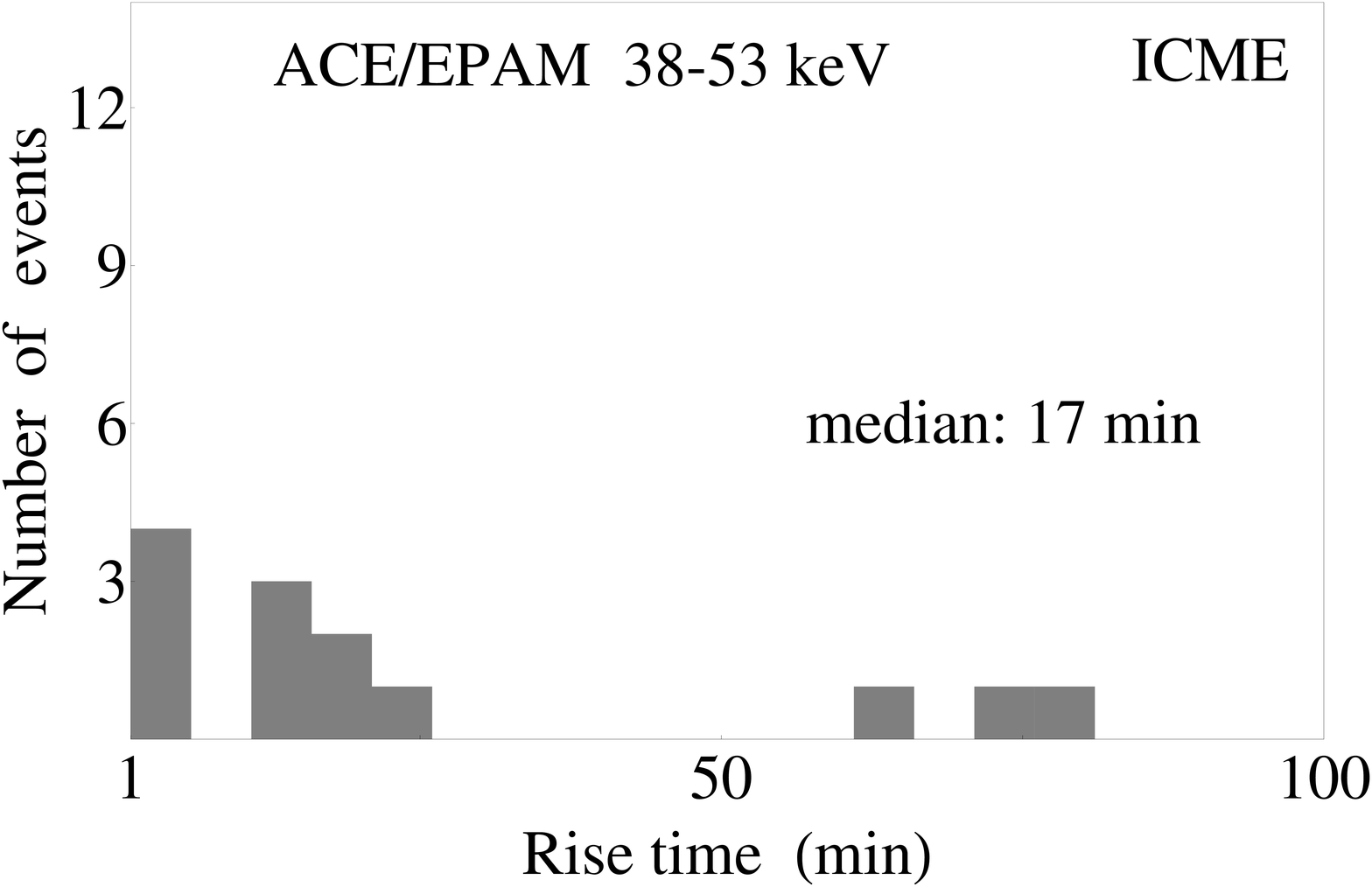}
               \hspace*{-0.04\textwidth}
               \includegraphics[width=0.53\textwidth,clip=]{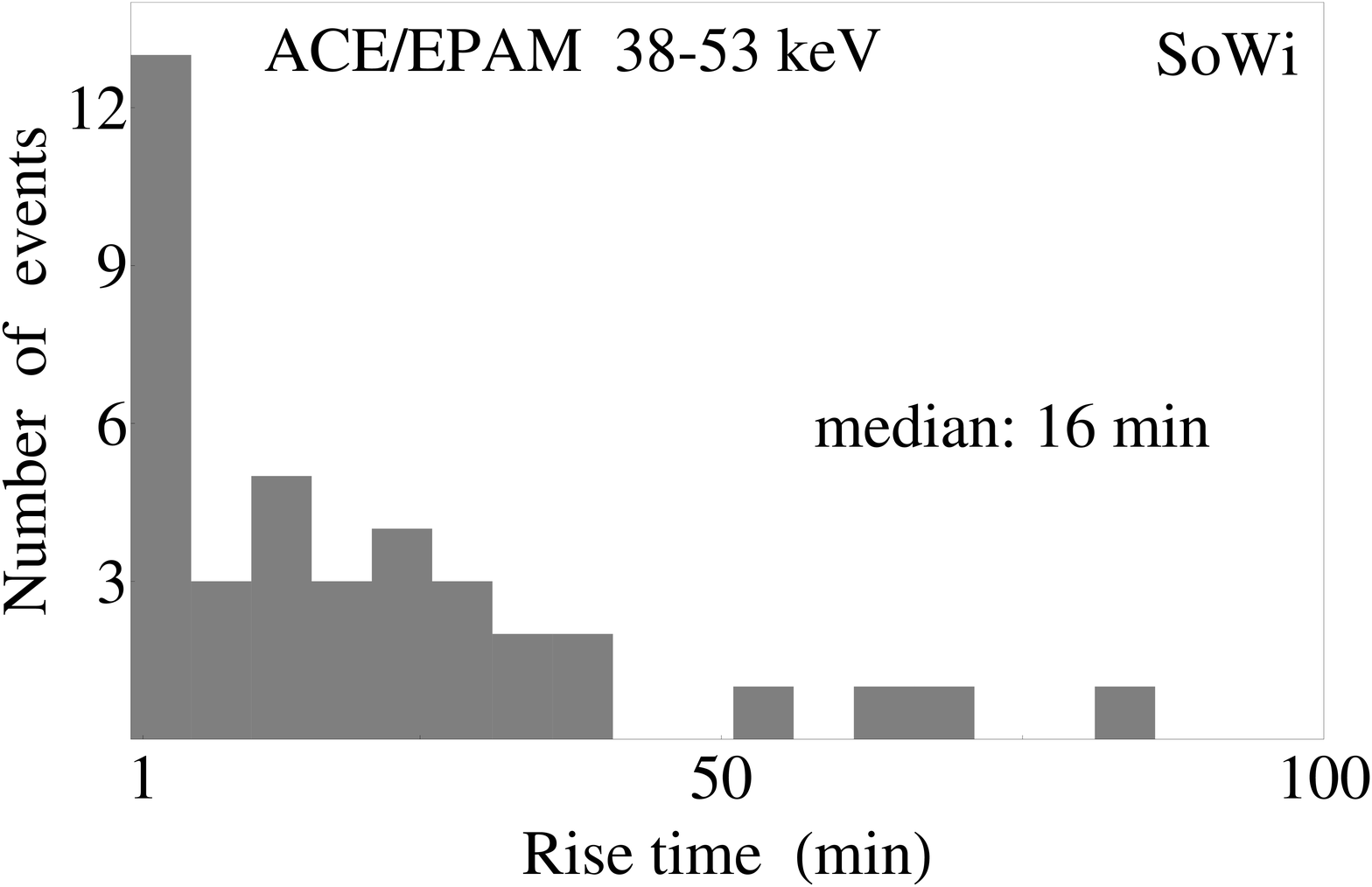}
              }
              \vspace*{-0.02\textwidth}
   \centerline{\hspace*{-0.02\textwidth}
               \includegraphics[width=0.53\textwidth,clip=]{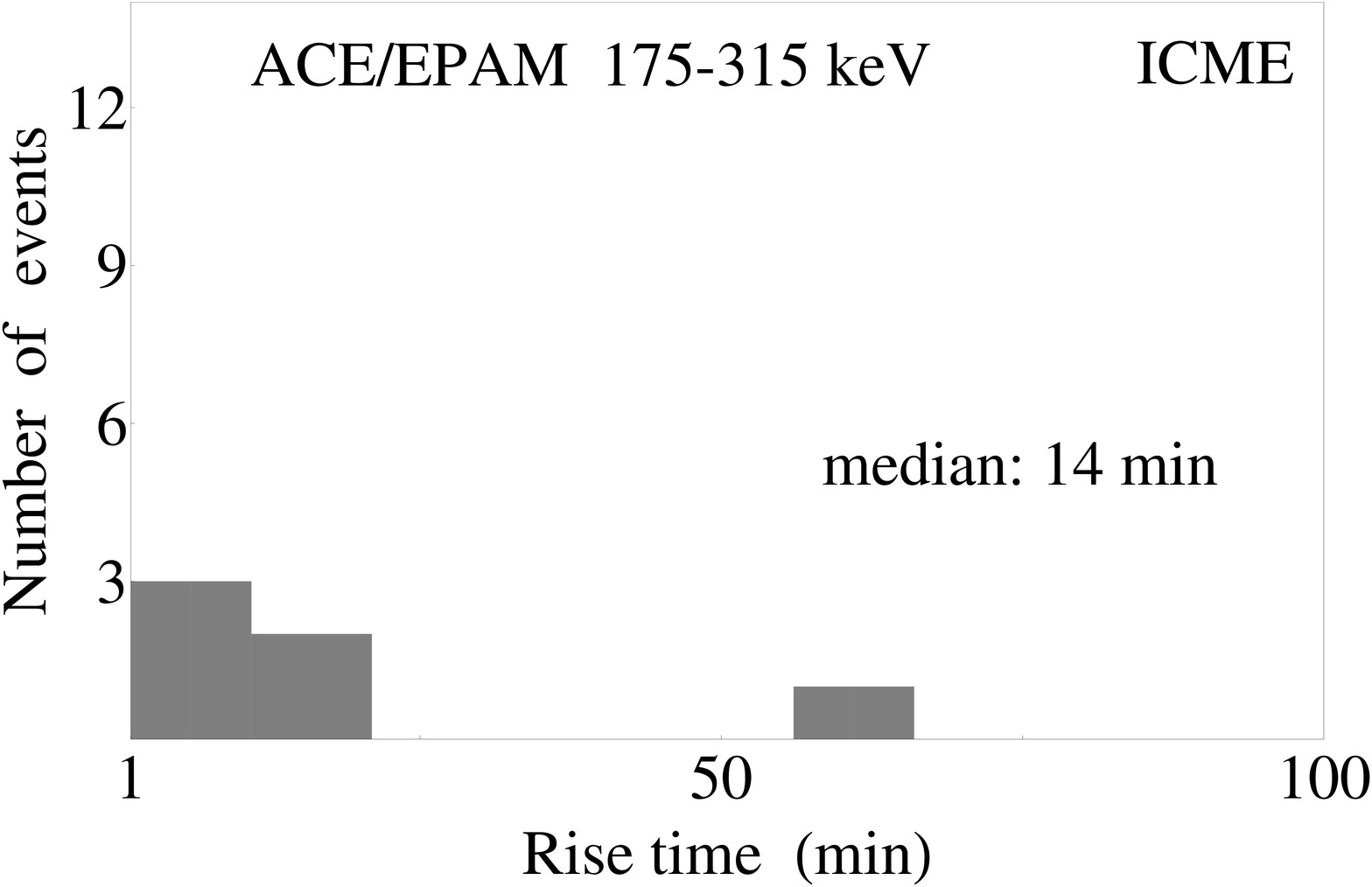}
               \hspace*{-0.04\textwidth}
               \includegraphics[width=0.53\textwidth,clip=]{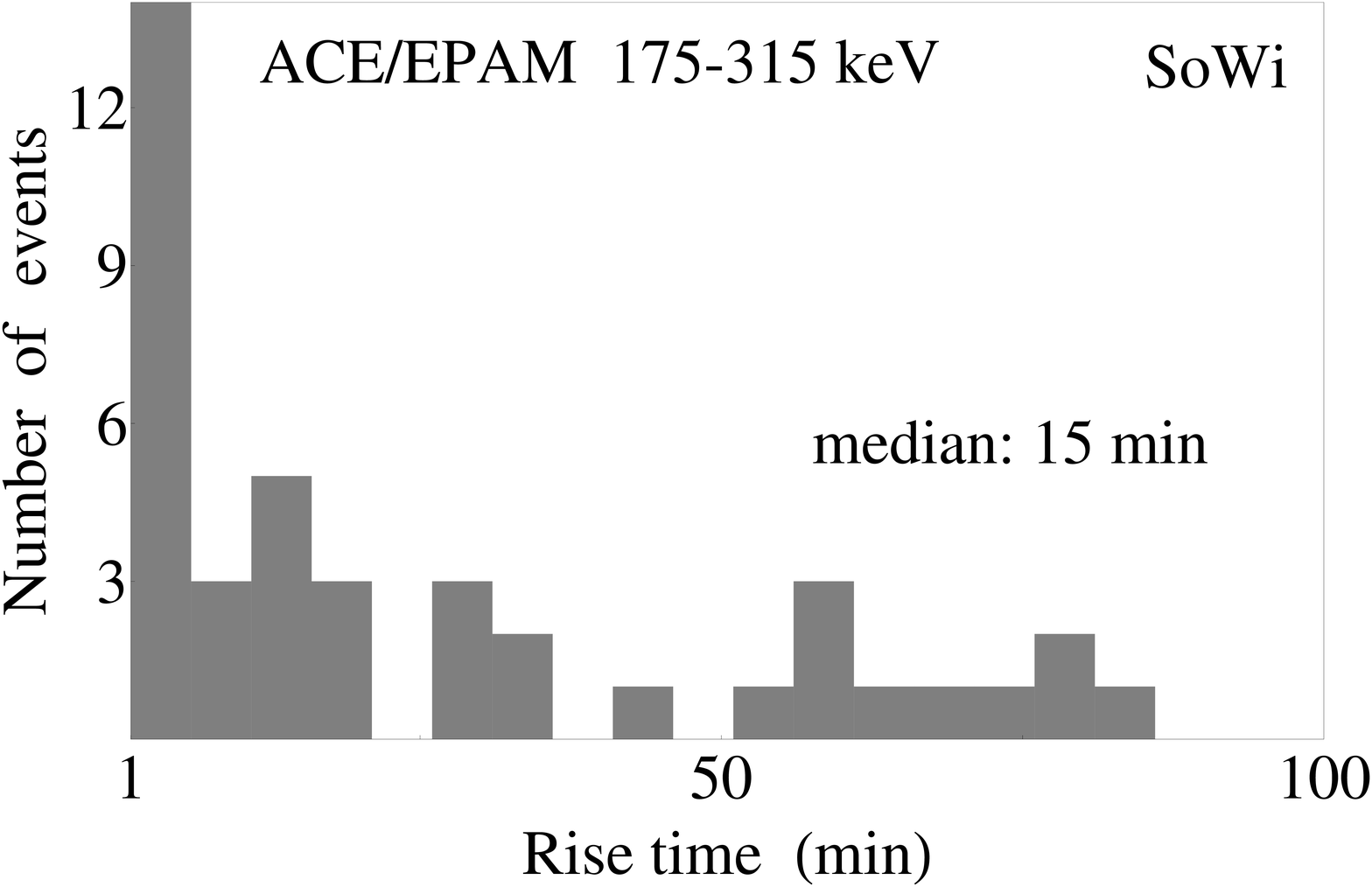}
              }
\caption{Distributions of the ACE/EPAM low and high energy channel electron rise times for the different IMF categories of SEP events (ICME in the left and SoWi in the right panels). The number of the SEP events in each category is given by the length of the corresponding color bar in each bin.}
\label{F-rise_e}
   \end{figure}

To determine the rise time of proton and electron time profiles, we used a method similar to \inlinecite{2012A&A...538A..32M}: The particle intensity was first divided by the pre-event background. The logarithm of the normalized background intensity is hence zero. This logarithmic time profile was then again normalized to the first identified maximum or to the value at the first significant break of the slope of the profile. Usually the rise slows down during the event. The time for the second normalization was hence chosen such as to identify the fastest  part of the rise phase. The estimated rise time is the time of the exponential rise of the profile as inferred from a linear least absolute deviation fit to the logarithmic profile between the levels 0.2 and 0.8. The results are given in column~(11) in Tables~\ref{T-ICME_events} and \ref{T-SoWi_events} in the Appendix.

We applied this method to the GOES and {\it Wind} proton data sets and obtained consistent results (given as histograms in Figure~\ref{F-rise_p}): Proton events in the ICME category have three times shorter rise times (median value and its deviation of 15$\pm 6$ min for GOES and 11$\pm 7$ min for {\it Wind} data) than SoWi events (median value of 47$\pm 25$ min for GOES and 66$\pm 35$ min for {\it Wind} data).

Rise times of the electrons, derived from the low-energy and high-energy channels of ACE/EPAM 5-min data, do not show a syste\-matic difference between ICME and SoWi events (Figure~\ref{F-rise_e}). The median values of the rise times are 17$\pm 13$~min (ACE/EPAM 38$-$53 keV) and 14$\pm 7$~min (ACE/EPAM 175$-$315 keV) in the ICME category and 16$\pm 12$~min and 15$\pm 14$~min in the SoWi category. Although there are more electron events with longer rise times in the SoWi category than in the ICME category, there are also numerous electron events propagating in the solar wind with rise times less than 5 min (given with one bin in the plots), which decreased the median value of the rise time. To optimize the presentation, only electron events with rise times shorter than 100 min are included in the histograms (but all events are used to calculate the median rise time).

\subsection{Connection Distance}
  \label{S-Conn_dist}

Inspection of column~(7) in Tables~\ref{T-ICME_events} and \ref{T-SoWi_events} suggests that the longitude distribution of the flares associated with the SEP events in the two IMF categories are different: The distribution is flat for the SoWi events, but peaks in the range $40^\circ$$-$$80^\circ$ for the ICME events. This difference can affect statistical relationships between the parameters of SEP events and the associated coronal activity in two ways. First, because of the broader distribution of projection angles, the projected CME speeds may be more strongly randomized in the SoWi sample than the ICME sample. Second, the distribution of angular distances between the IP field line through the Earth and the region of strongest particle injection in the corona is expected to induce a stronger dispersion of peak SEP intensities in the SoWi sample than the ICME sample. Projection effects on CME speeds will be addressed in Section~\ref{S-correlfc}. Here we evaluate the effect of the connection distance on the SEP detection and peak intensity.

Ideally the connection distance is the angular distance between the footpoint of the IMF line through the Earth on the solar wind source surface and the footpoint in the low corona of the field line onto which the bulk of the SEPs is injected. Here we use the Parker spiral as a proxy of the IMF line and the flare longitude as a proxy  of the coronal field line with the highest SEP intensity. We follow the common assumption of a spherical heliocentric source surface with radius $r_{\rm SS}=2.5\, R_\odot$. The longitude on the source surface of the Parker spiral through the GOES spacecraft at heliocentric distance $r_{\rm GOES} =214\, R_\odot$ is

\begin{eqnarray}
\phi_{\rm PS}=\frac{2\pi}{P\, V_{\rm SoWi}}(r_{\rm GOES}-r_{\rm SS}),
\label{E-parker_long}
\end{eqnarray}
with an averaged solar rotation period of $P=26$ days. The solar wind speed, $V_{\rm SoWi}$, is taken from the MTOF/PM sensor\footnote{\url{http://umtof.umd.edu/pm/crn/}} of the CELIAS instrument aboard SOHO \cite{1995SoPh..162..441H}, and is the averaged value over a 6-h period before the particle onset time (see Tables~\ref{T-ICME_events}$-$\ref{T-Other_events}). It is expected that SEPs are guided along field lines of different geometry within an ICME than in the solar wind. We nevertheless present the connection distance for the ICME category using the same formula as for the SoWi events, because the legs of the IP flux ropes are expected to be curved similarly to the Archimedian spiral by the combination of outward expansion and solar rotation \cite{1997...Marubashi,2002JGRA..107.1236V}. The results are given in the last column of Tables~\ref{T-ICME_events}$-$\ref{T-Other_events} in the Appendix.

 \begin{figure}[!t]
  \centerline{\hspace*{-0.01\textwidth}
               \includegraphics[width=0.6\textwidth,clip=]{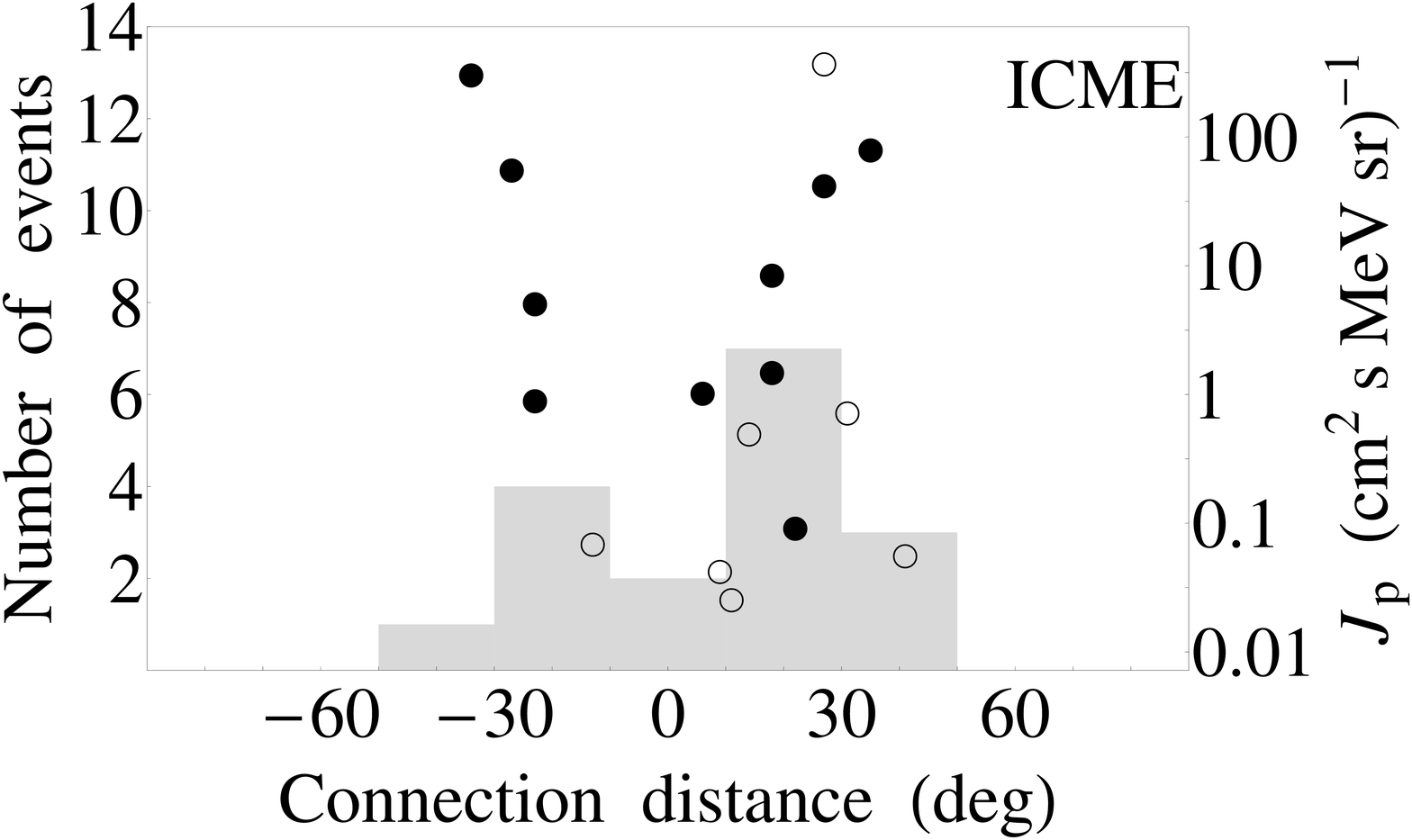}
               \hspace*{-0.1\textwidth}
               \includegraphics[width=0.6\textwidth,clip=]{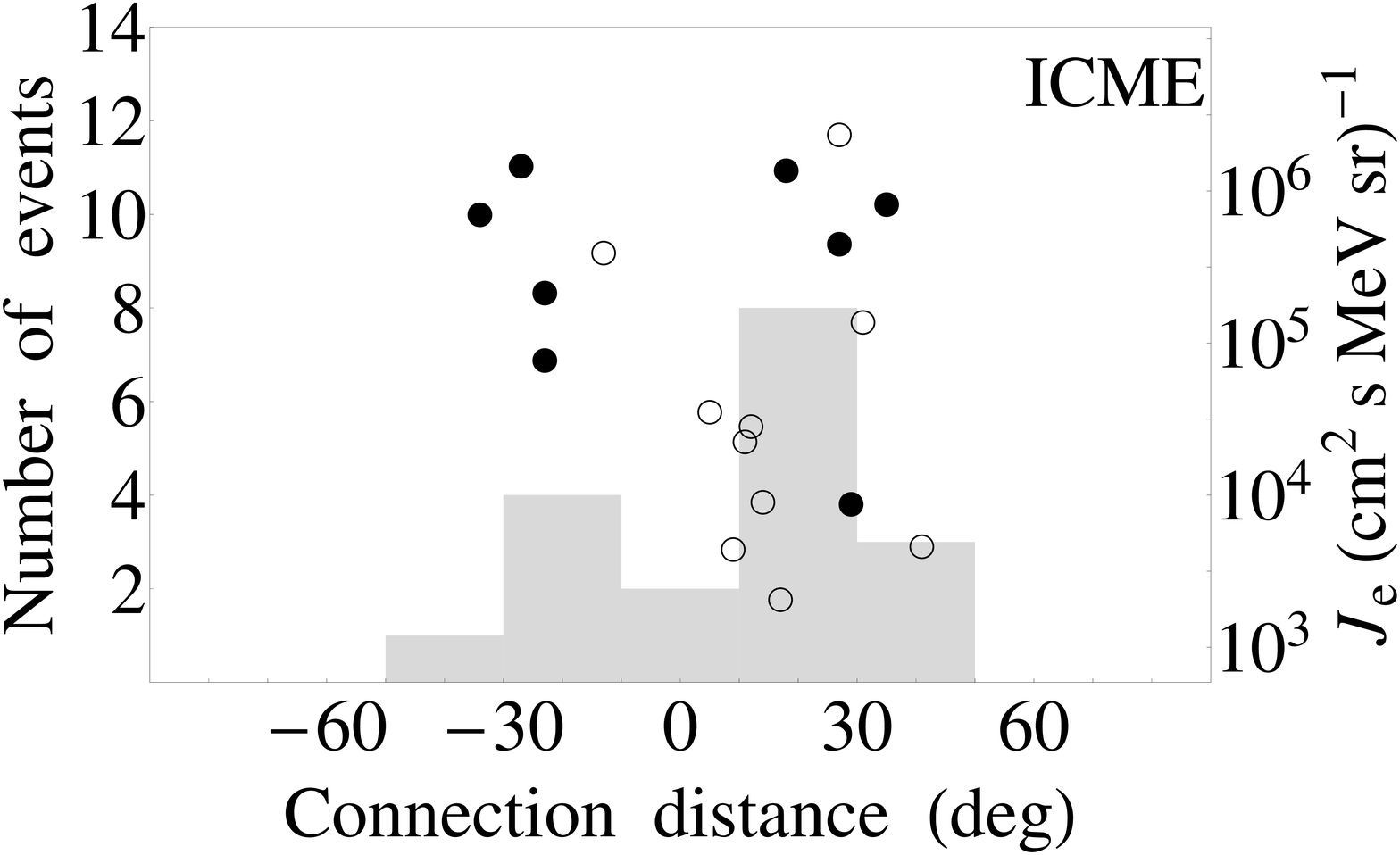}
               }
               \vspace*{-0.07\textwidth}
   \centerline{\hspace*{-0.01\textwidth}
               \includegraphics[width=0.6\textwidth,clip=]{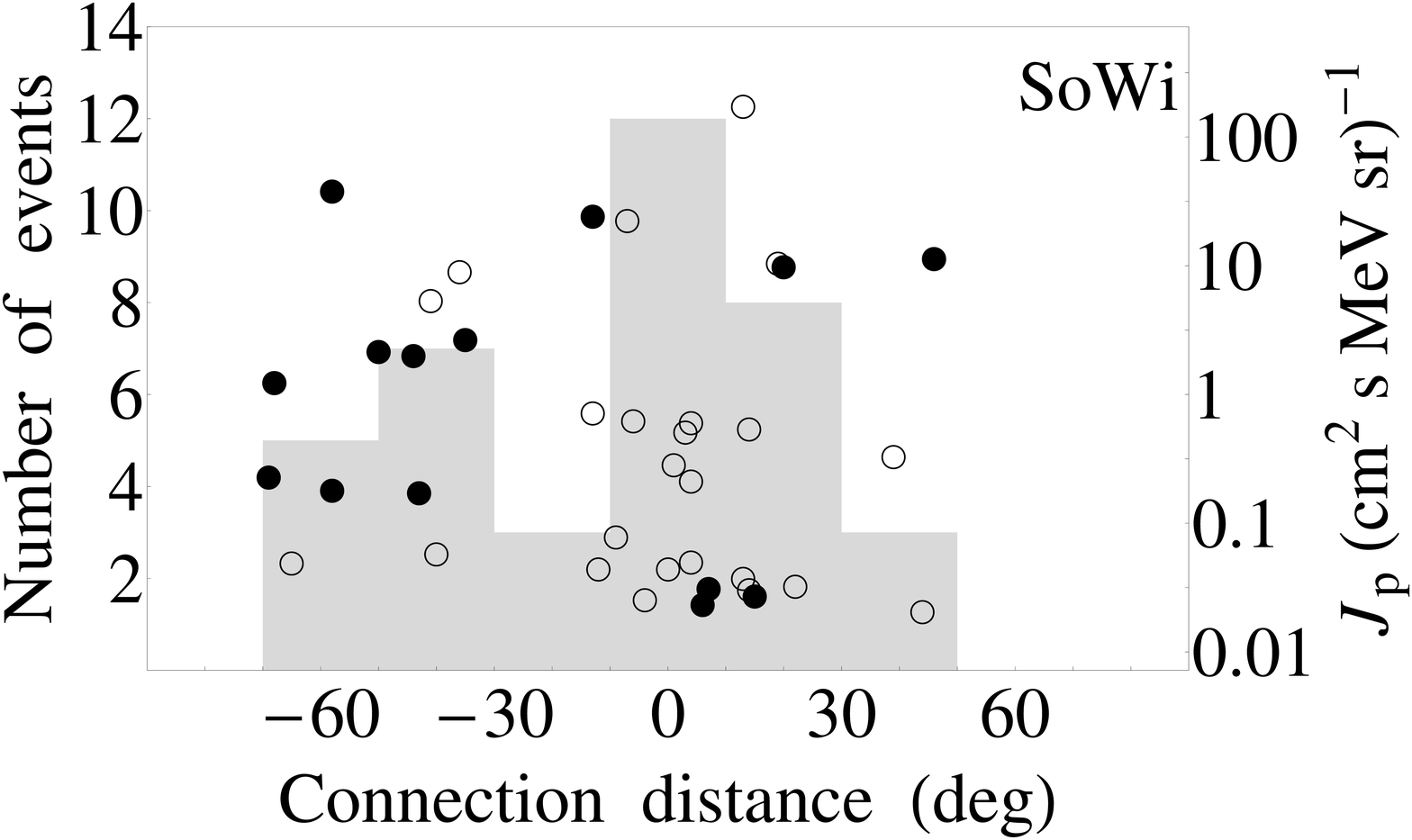}
               \hspace*{-0.1\textwidth}
               \includegraphics[width=0.6\textwidth,clip=]{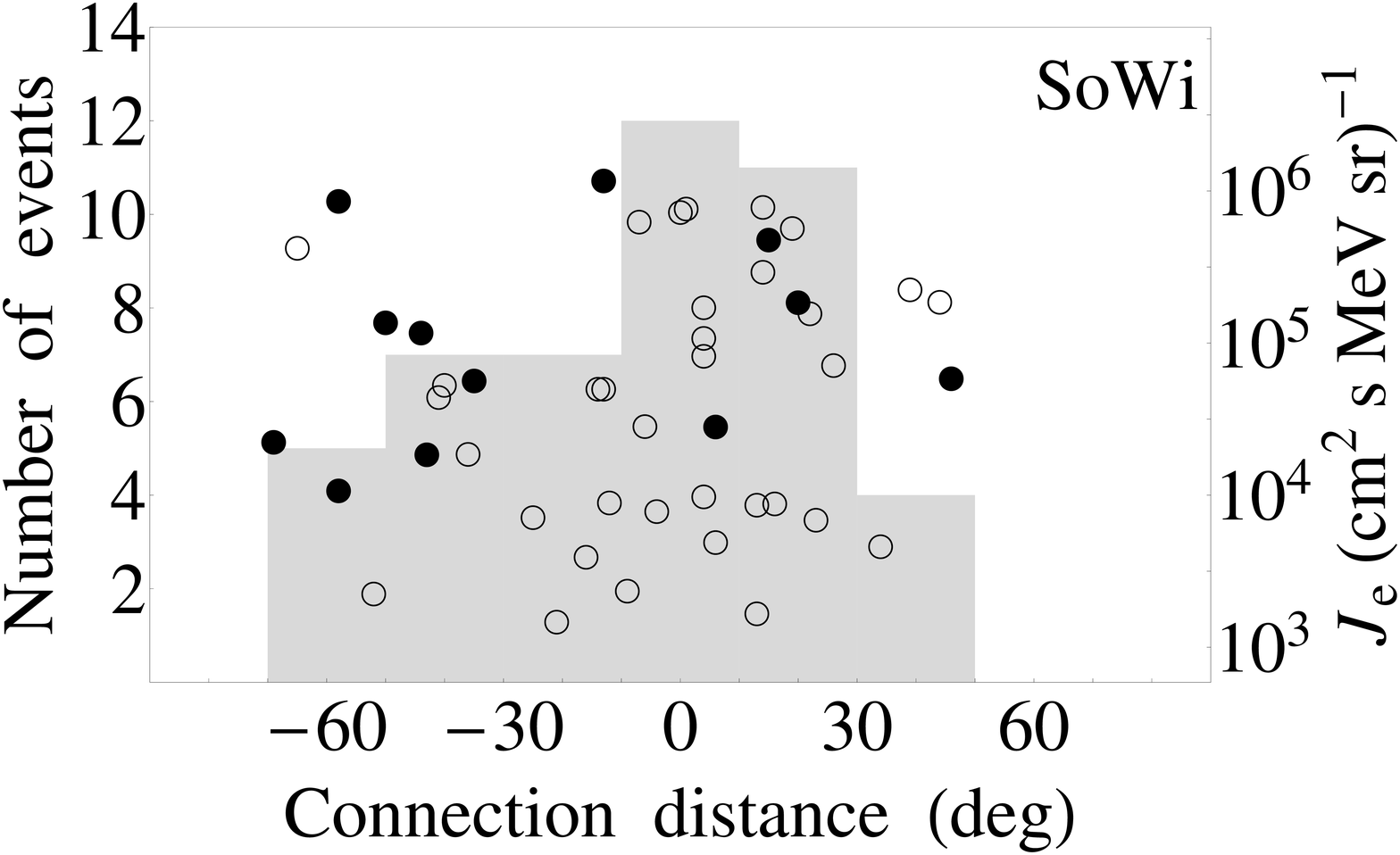}
               }
               \vspace*{-0.07\textwidth}
   \centerline{\hspace*{-0.01\textwidth}
               \includegraphics[width=0.6\textwidth,clip=]{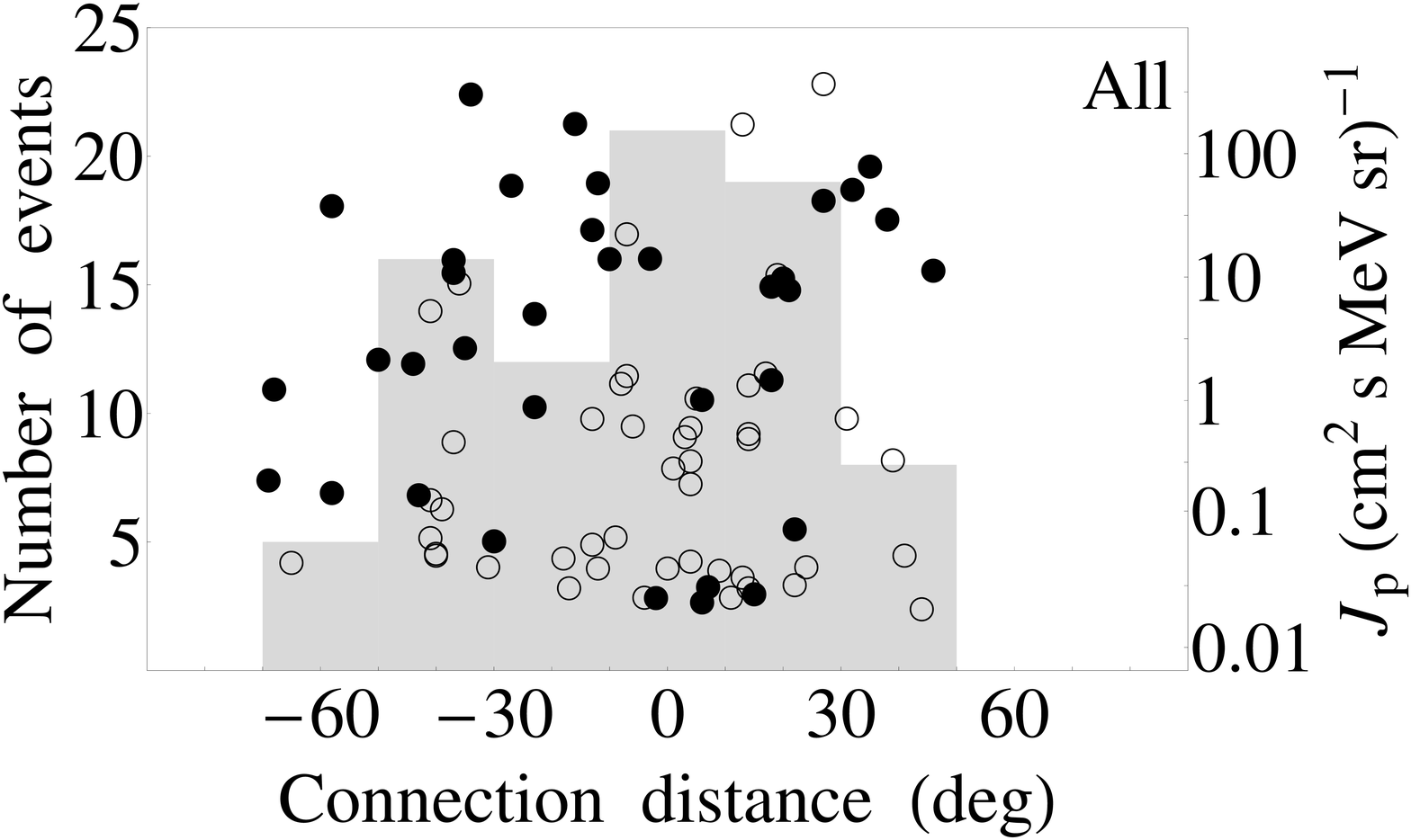}
               \hspace*{-0.1\textwidth}
               \includegraphics[width=0.6\textwidth,clip=]{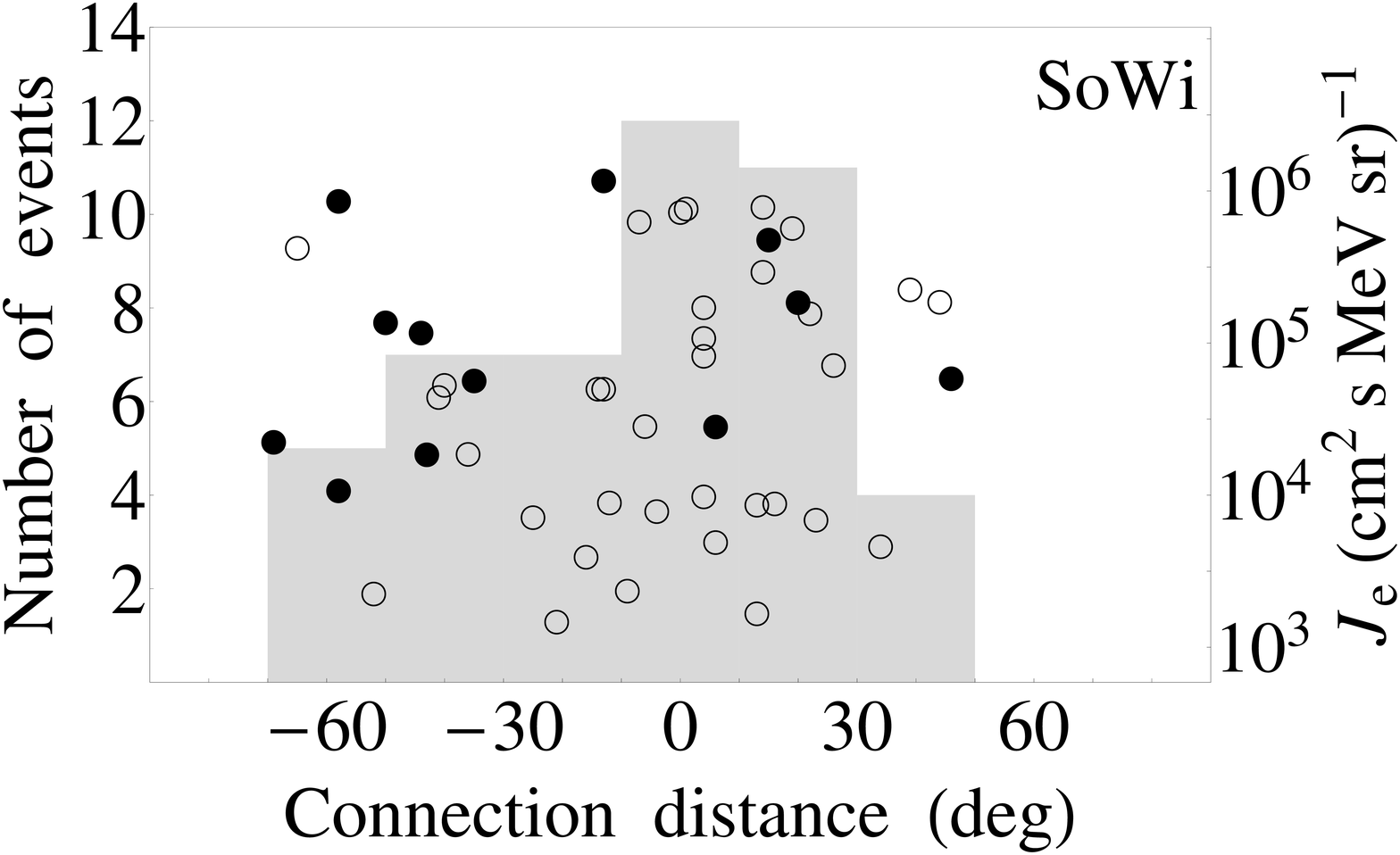}
               }
\caption{Distribution in connection longitude for different categories of SEP events. Overplotted are the SEP peak intensities for the GOES proton ($J_{\rm p}$, left) and ACE/EPAM electron ($J_{\rm e}$, right) data. Filled and open circles denote the SEP events associated with X-class and M-class flares, respectively.}
   \label{F-histo_cd}
   \end{figure}

The distributions of connection distances for the different IMF categories are shown in Figure~\ref{F-histo_cd} (histograms), together with the peak particle intensity (filled and empty circles for the SEP events associated with flares of classes X and M, respectively). The median values and deviations for the connection distance of GOES protons\footnote{For the {\it Wind}/EPACT proton data we obtain $22^{\circ} \pm 13^{\circ}$ ($20^{\circ} \pm 8^{\circ}$ in absolute values) for the ICME events and $-6^{\circ} \pm 20^{\circ}$ ($14^{\circ} \pm 11^{\circ}$ in absolute values) for the SoWi events.} are $14^{\circ} \pm 13^{\circ}$ ($23^{\circ} \pm 8^{\circ}$ in absolute values) for the ICME events and $-5^{\circ} \pm 19^{\circ}$ ($14^{\circ} \pm 11^{\circ}$ in absolute values) for the SoWi events. For the electron events (ACE/EPAM low energy channel) we obtain $13^{\circ} \pm 15^{\circ}$ ($23^{\circ} \pm 9^{\circ}$ in absolute values) for the ICME events and $-5^{\circ} \pm 20^{\circ}$ ($18^{\circ} \pm 14^{\circ}$ in absolute values) for the SoWi events, respectively. All distributions cover a wide range of distances ($\pm 60^{\circ}$) around the nominal connection longitude ($0^\circ$), but the scatter is more pronounced for the SoWi events, which have (for GOES data set) 26/38 events (68\%) at connection distances above 30$^\circ$, as compared to 13/17 ICME events (76\%).

At first glance there is no evidence for a systematic decrease of peak particle intensity with increasing connection distance. Both IMF categories comprise events which are intense, even though the associated flare is far from the nominal connection longitude. The connection distance distributions of the peak intensities are in general flat. We note, however, that the SEP events of the SoWi category with large connection distances ($>30^\circ$) are preferentially accompanied by strong (X-class) flares. So if, as we will show in Section~\ref{S-correlations}, stronger SXR bursts are associated with stronger SEP intensities, the over-representation of X-class flares in the remote SEP events does reveal a general trend of decrease of the peak particle intensity with increasing connection distance.

\subsection{SEP Events and Associated Coronal Activity}
  \label{S-SEP-coractivity}

Before investigating if and how the statistical relationship between SEP peak intensity and the parameters of the associated flare and CME depends on the IMF configuration, we address the correlation between the peak soft X-ray flux ($I_{\rm SXR}$) and the projected CME speed ($V_{\rm CME}$). This is necessary to see if the parent solar activity is similar in the two event categories or if selection effects are important. Some correlation is expected because of the general experience, termed the big flare syndrome \cite{1982JGR....87.3439K}, that solar events that are important with respect to one parameter are likely important with respect to other parameters, too.

\subsubsection{Correlations between Flare and CME Parameters in Different IMF Categories}
  \label{S-correlfc}

\begin{figure}[t!]
   \centerline{\hspace*{-0.05\textwidth}
               \includegraphics[width=0.55\textwidth,clip=]{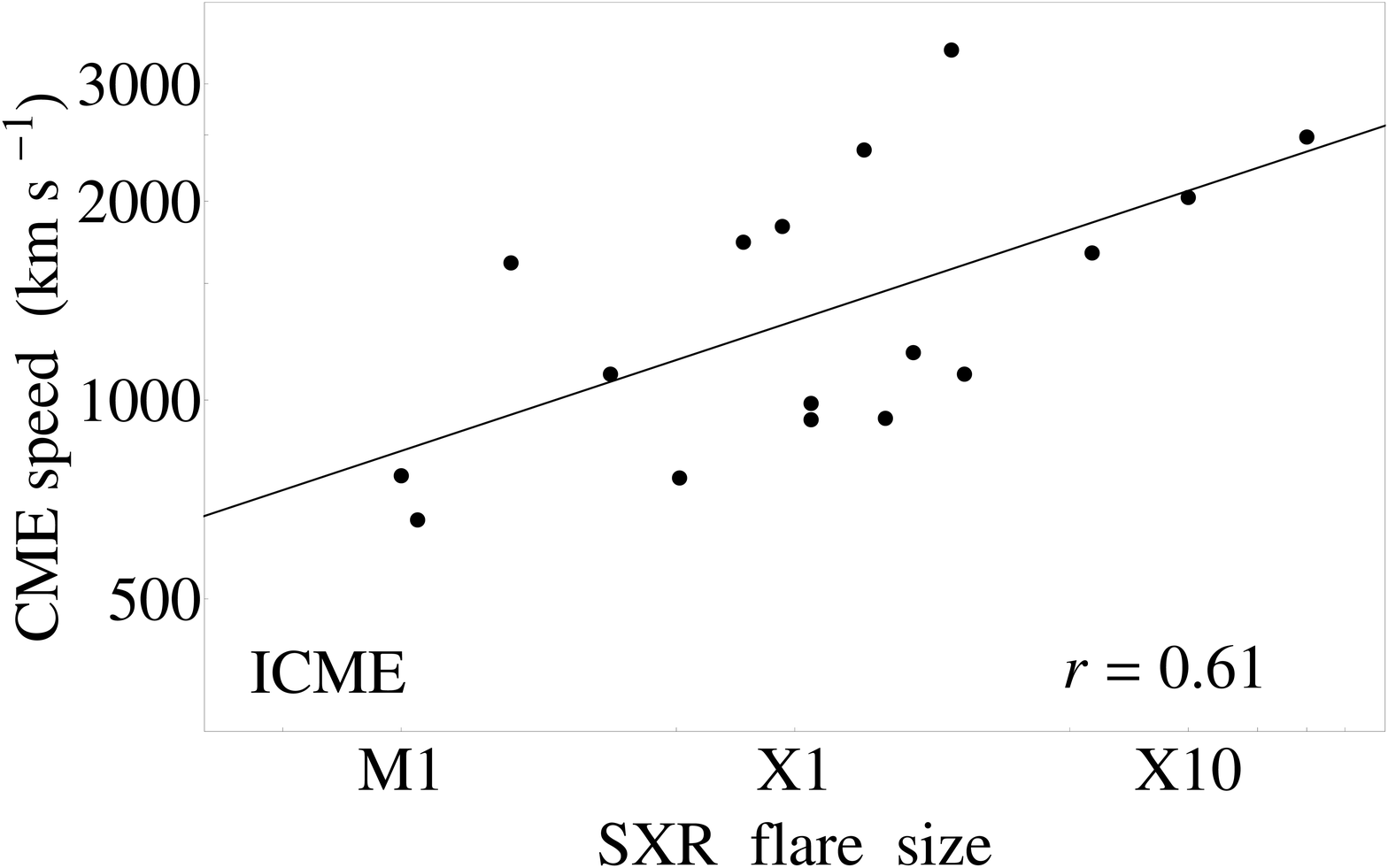}
               \hspace*{-0.05\textwidth}
               \includegraphics[width=0.55\textwidth,clip=]{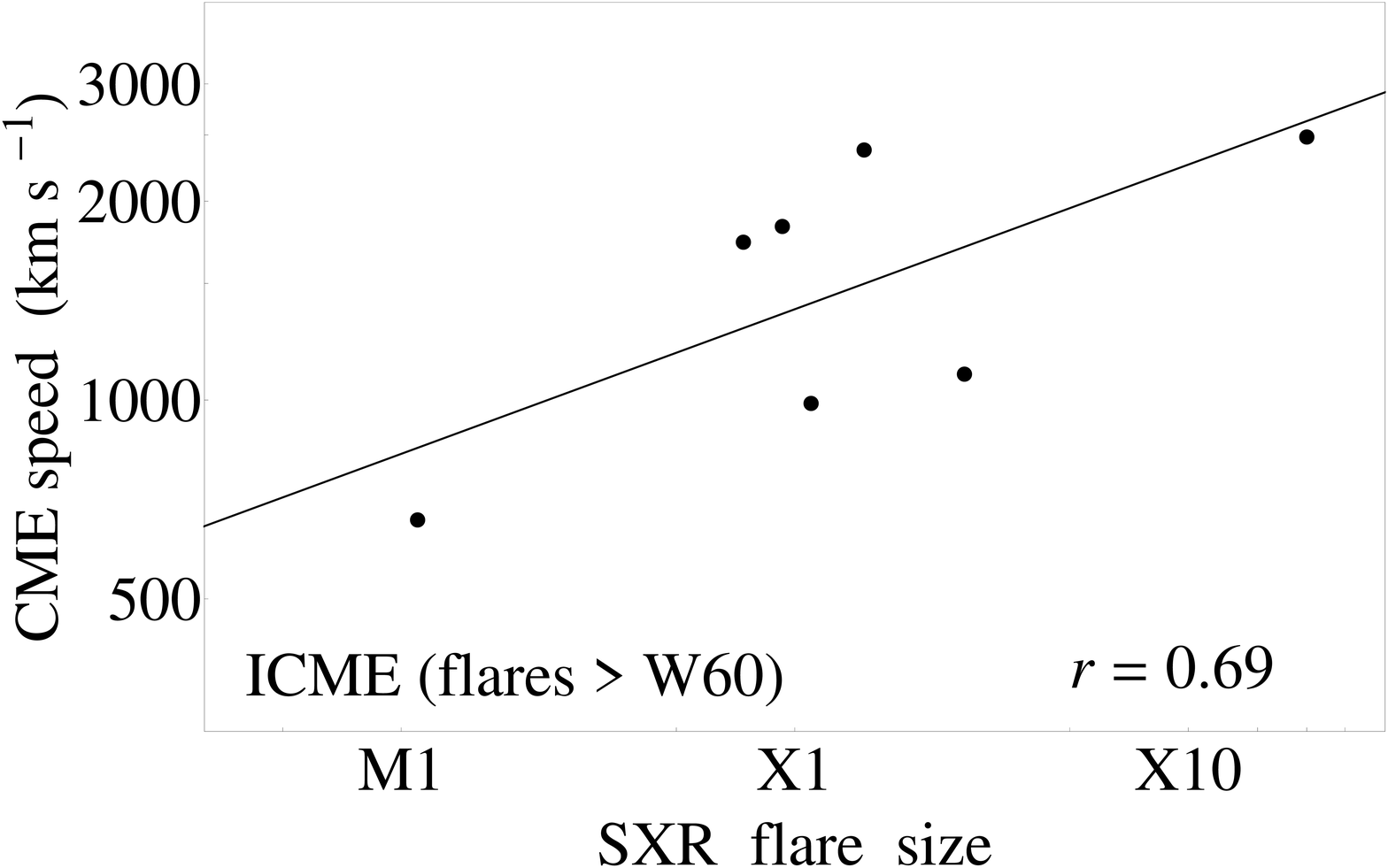}
               }
               \vspace*{-0.01\textwidth}
   \centerline{\hspace*{-0.05\textwidth}
               \includegraphics[width=0.55\textwidth,clip=]{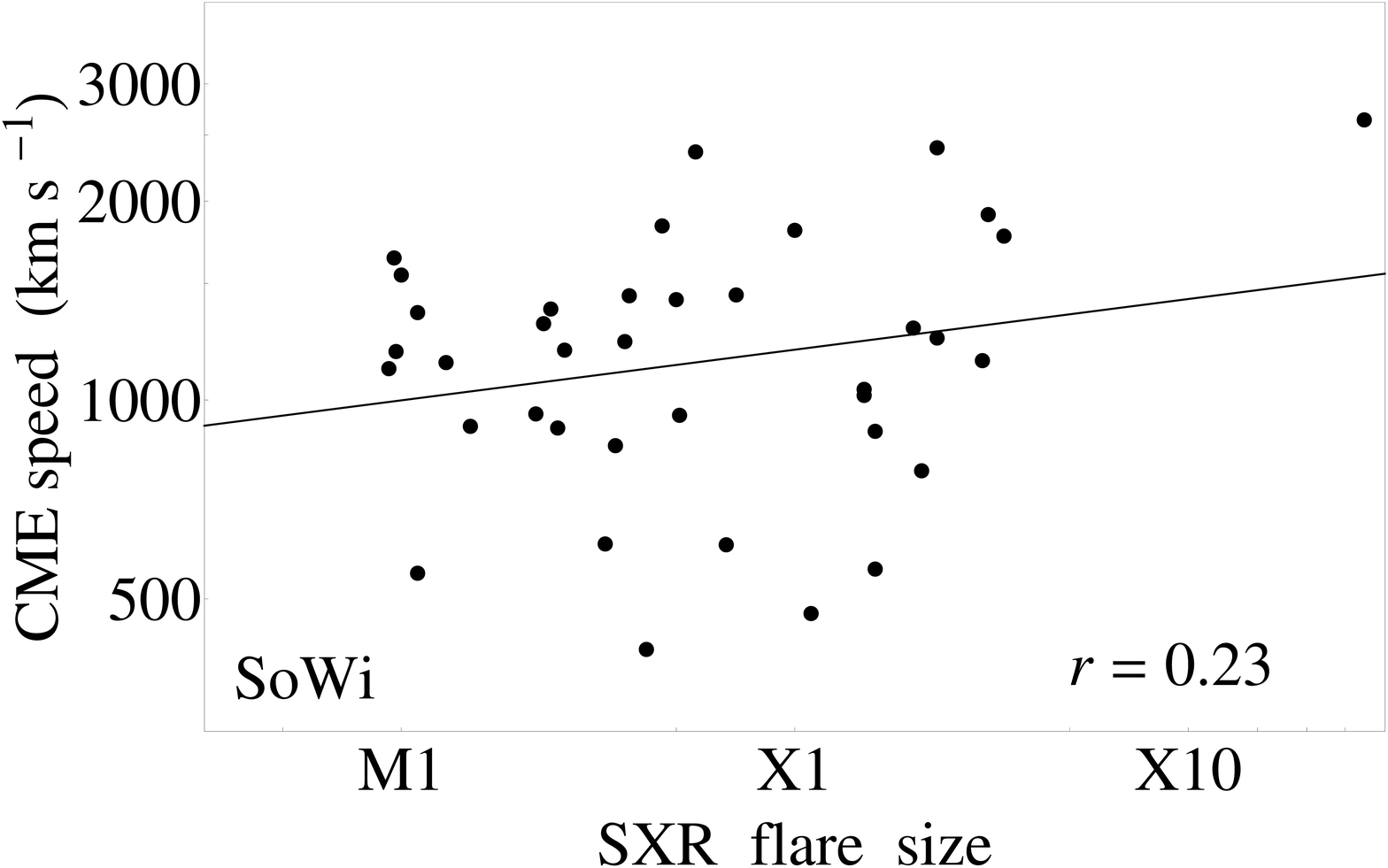}
               \hspace*{-0.05\textwidth}
               \includegraphics[width=0.55\textwidth,clip=]{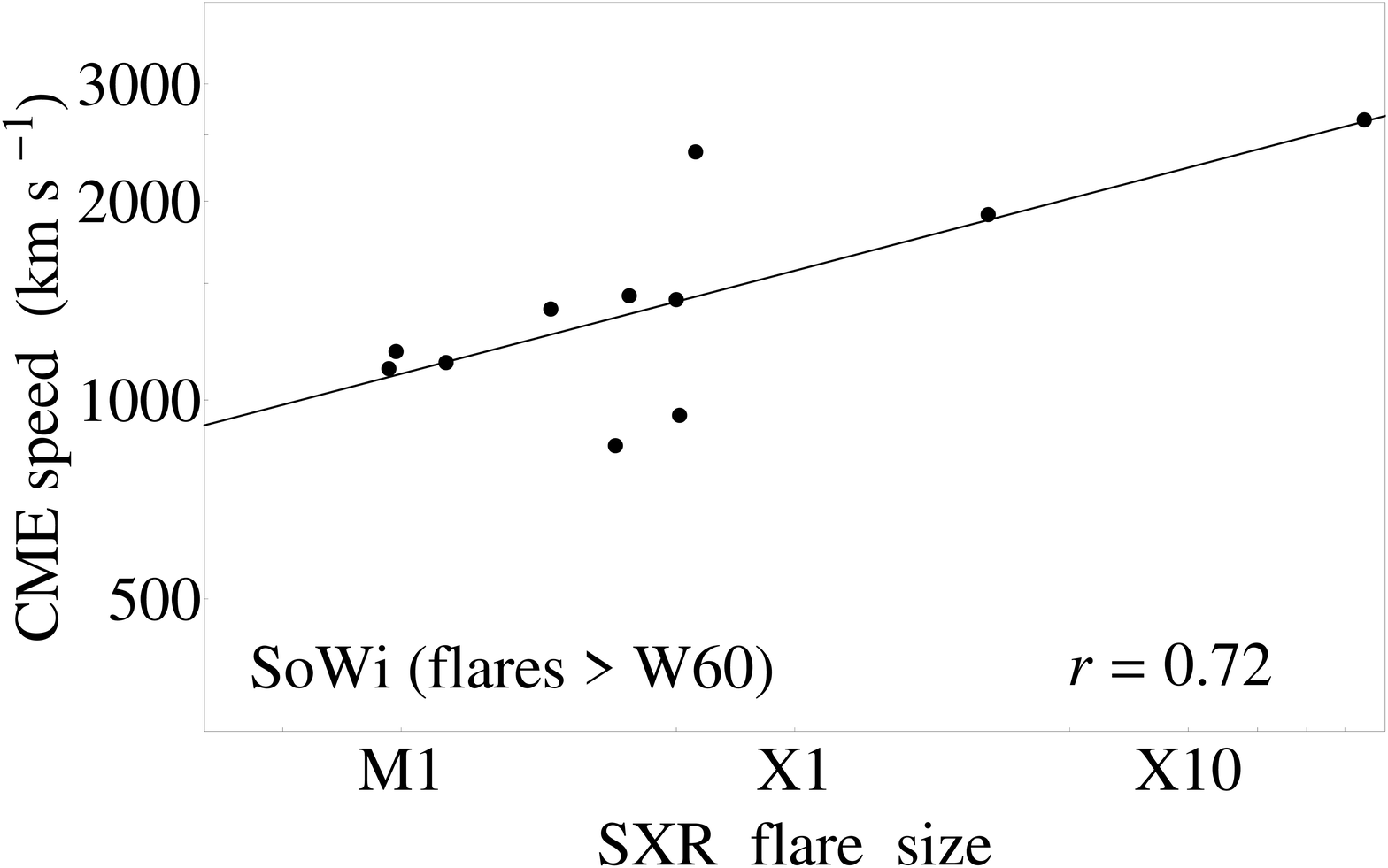}
               }
               \vspace*{-0.01\textwidth}
   \centerline{\hspace*{-0.05\textwidth}
               \includegraphics[width=0.55\textwidth,clip=]{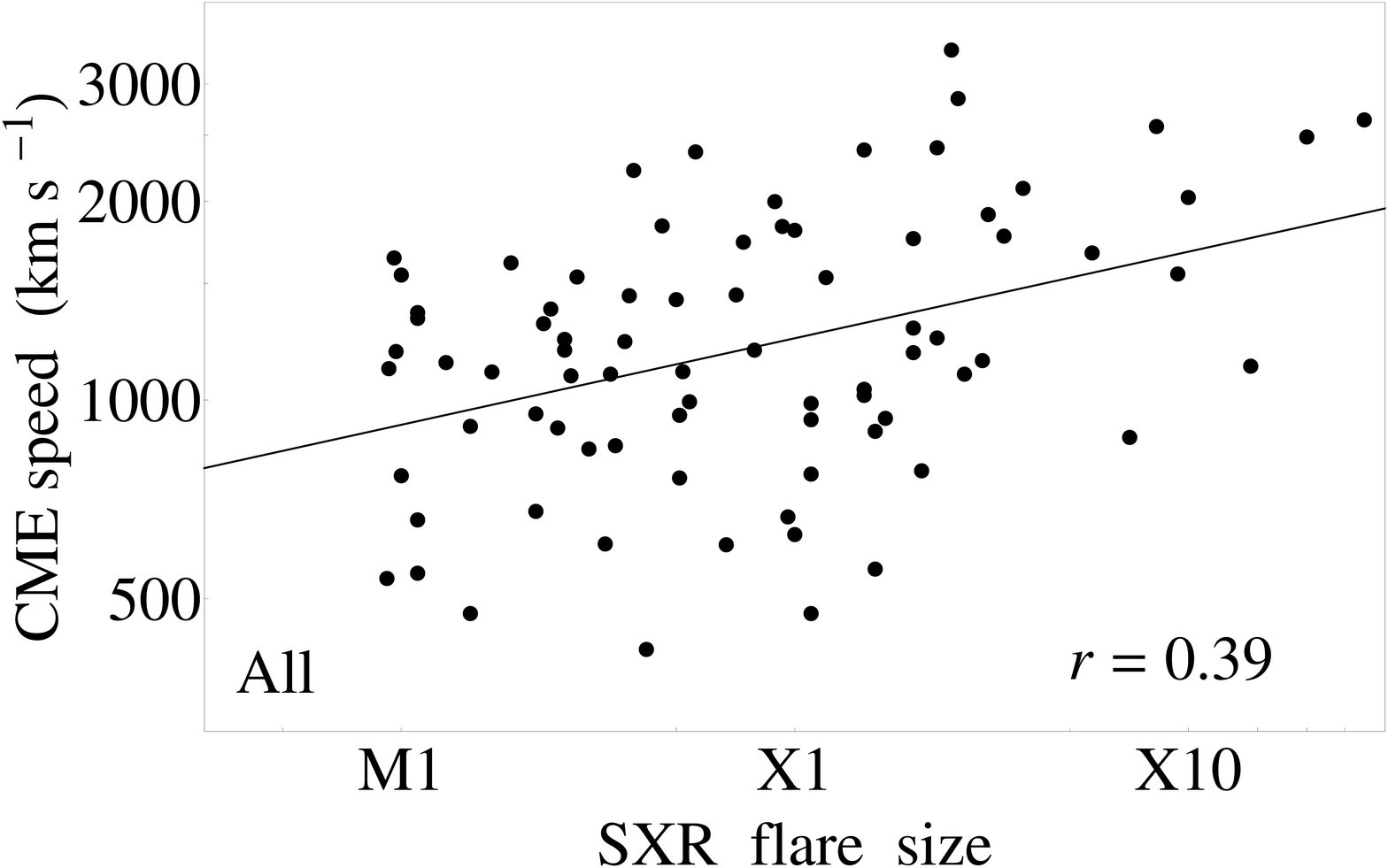}
               \hspace*{-0.05\textwidth}
               \includegraphics[width=0.55\textwidth,clip=]{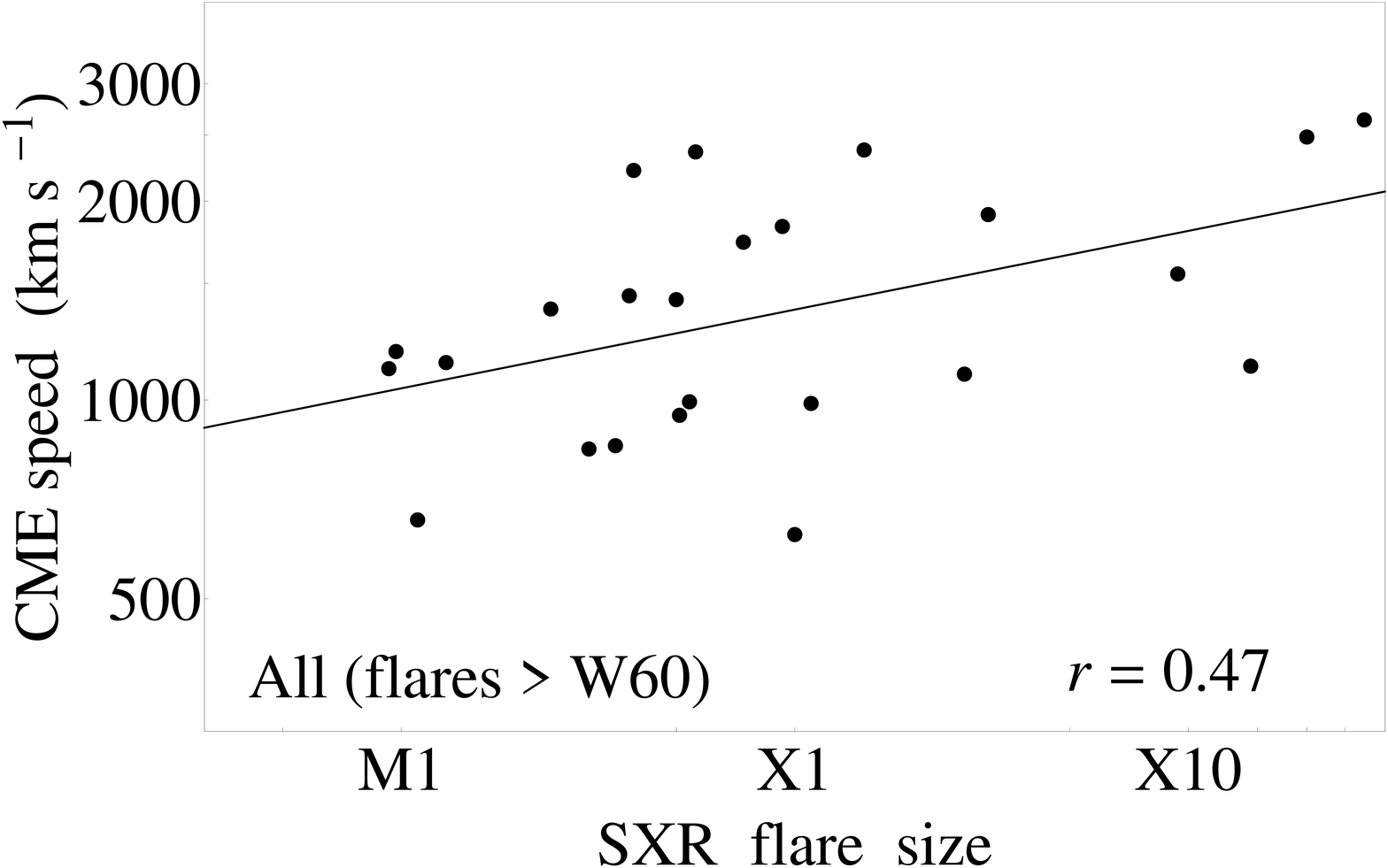}
               }
\caption{Scatter (log$-$log) plots of SXR flare size vs. CME speed for different IMF ca\-tegories of GOES proton events (left) and for a subset of these events associated with flares at longitudes $> {\rm W}60^{\circ}$ (right).}
   \label{F-scatter_plots_SXR_CMEp}
   \end{figure}

\begin{figure}[t!]
   \centerline{\hspace*{-0.05\textwidth}
               \includegraphics[width=0.55\textwidth,clip=]{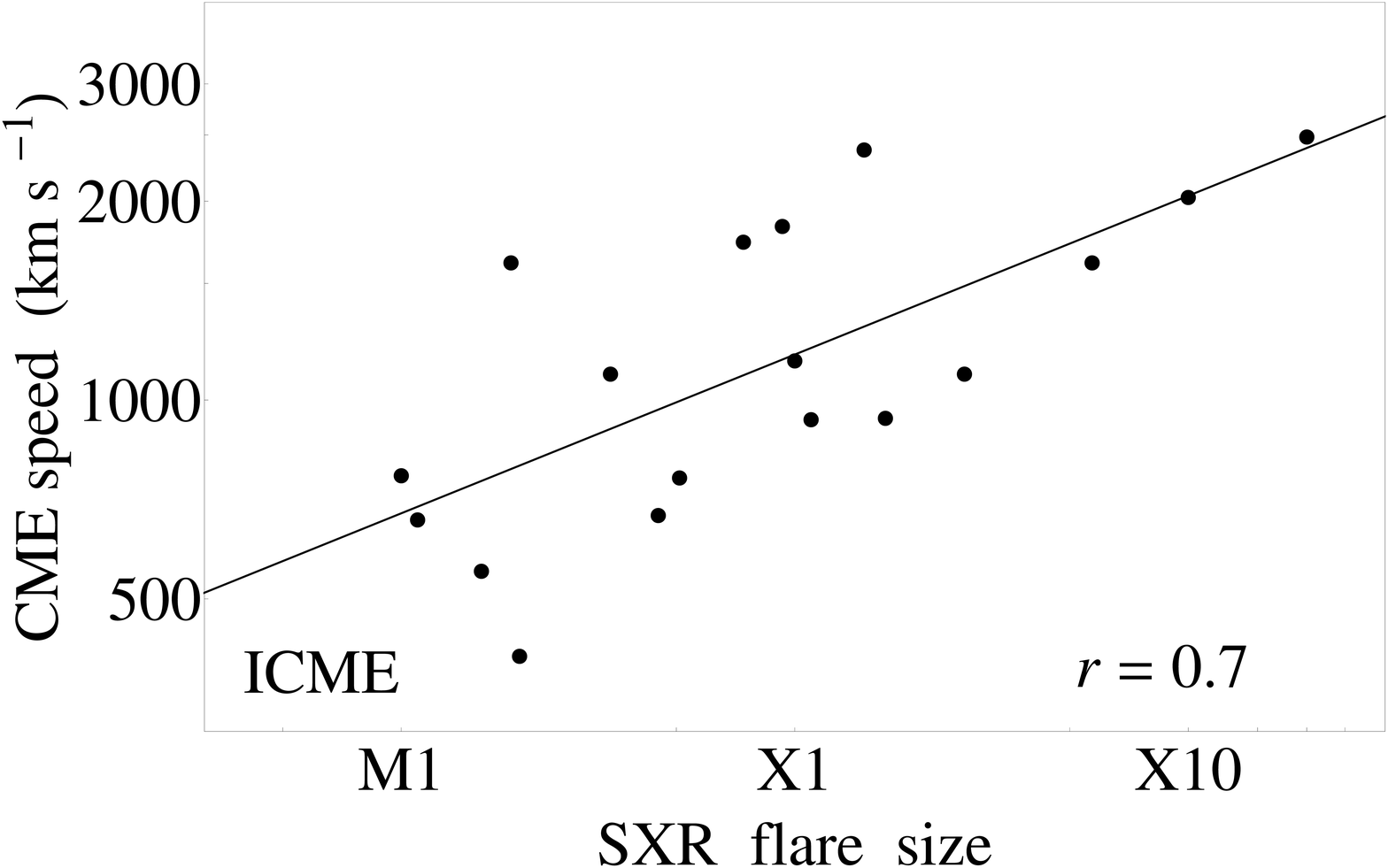}
               \hspace*{-0.05\textwidth}
               \includegraphics[width=0.55\textwidth,clip=]{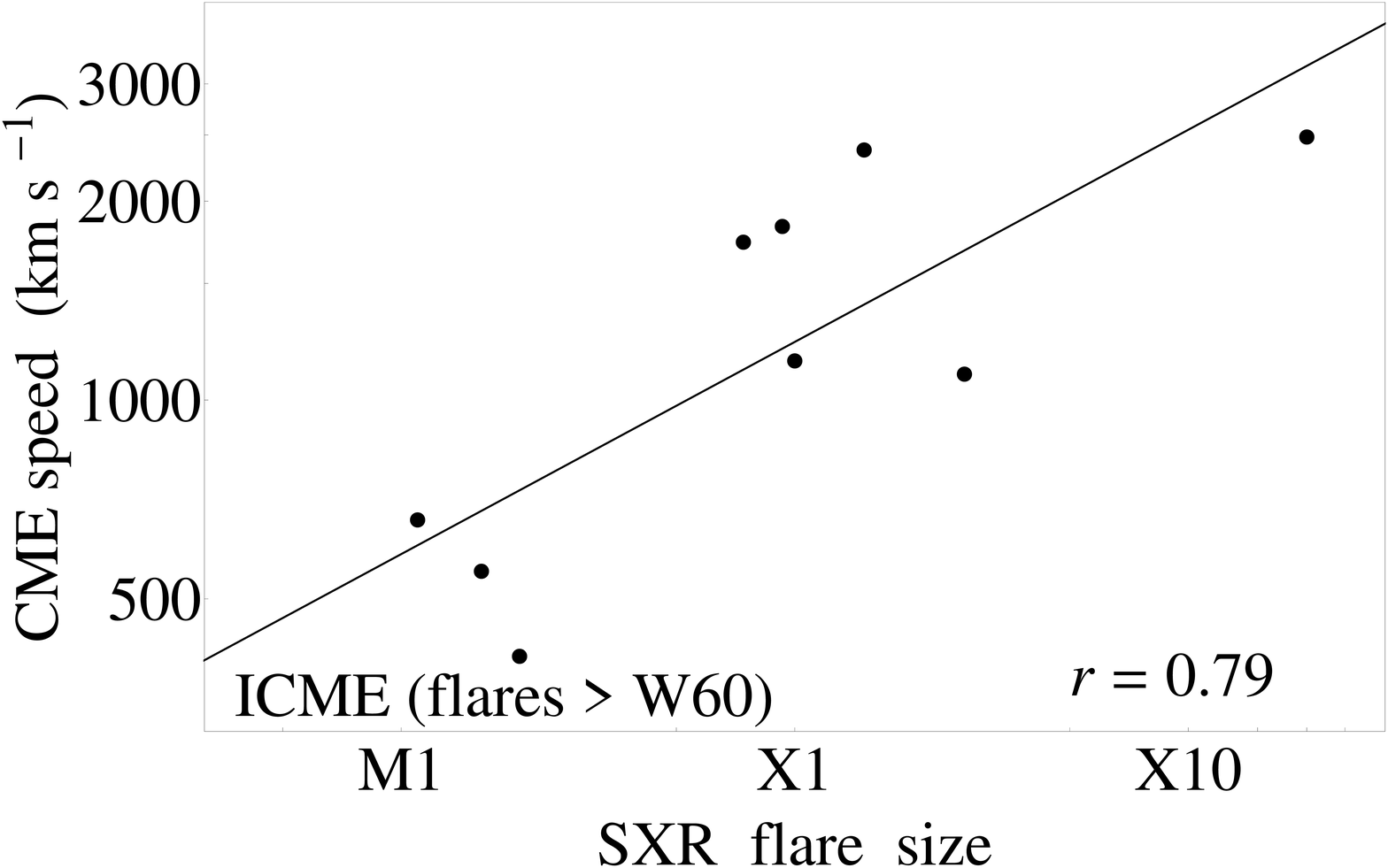}
               }
               \vspace*{-0.01\textwidth}
   \centerline{\hspace*{-0.05\textwidth}
               \includegraphics[width=0.55\textwidth,clip=]{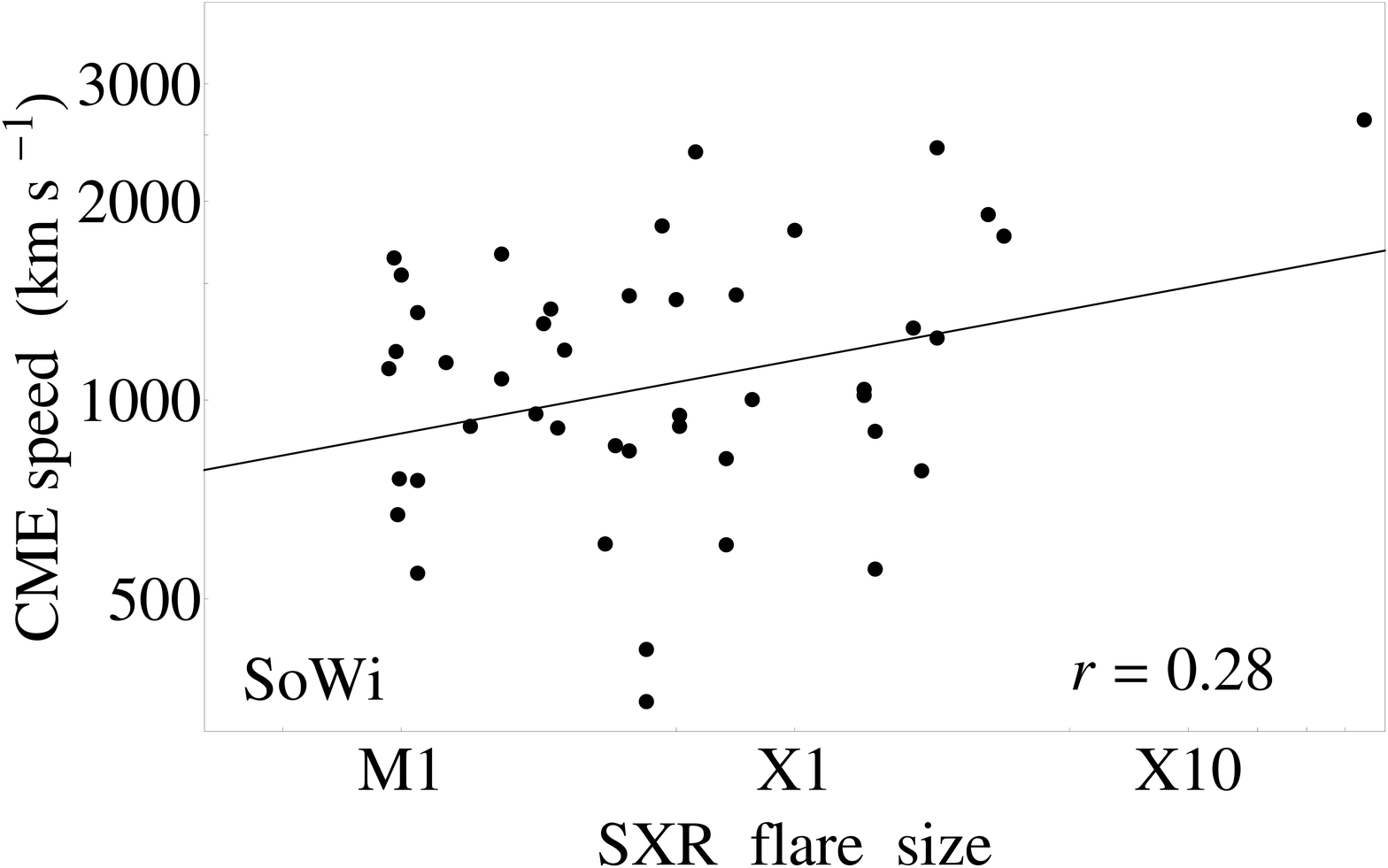}
               \hspace*{-0.05\textwidth}
               \includegraphics[width=0.55\textwidth,clip=]{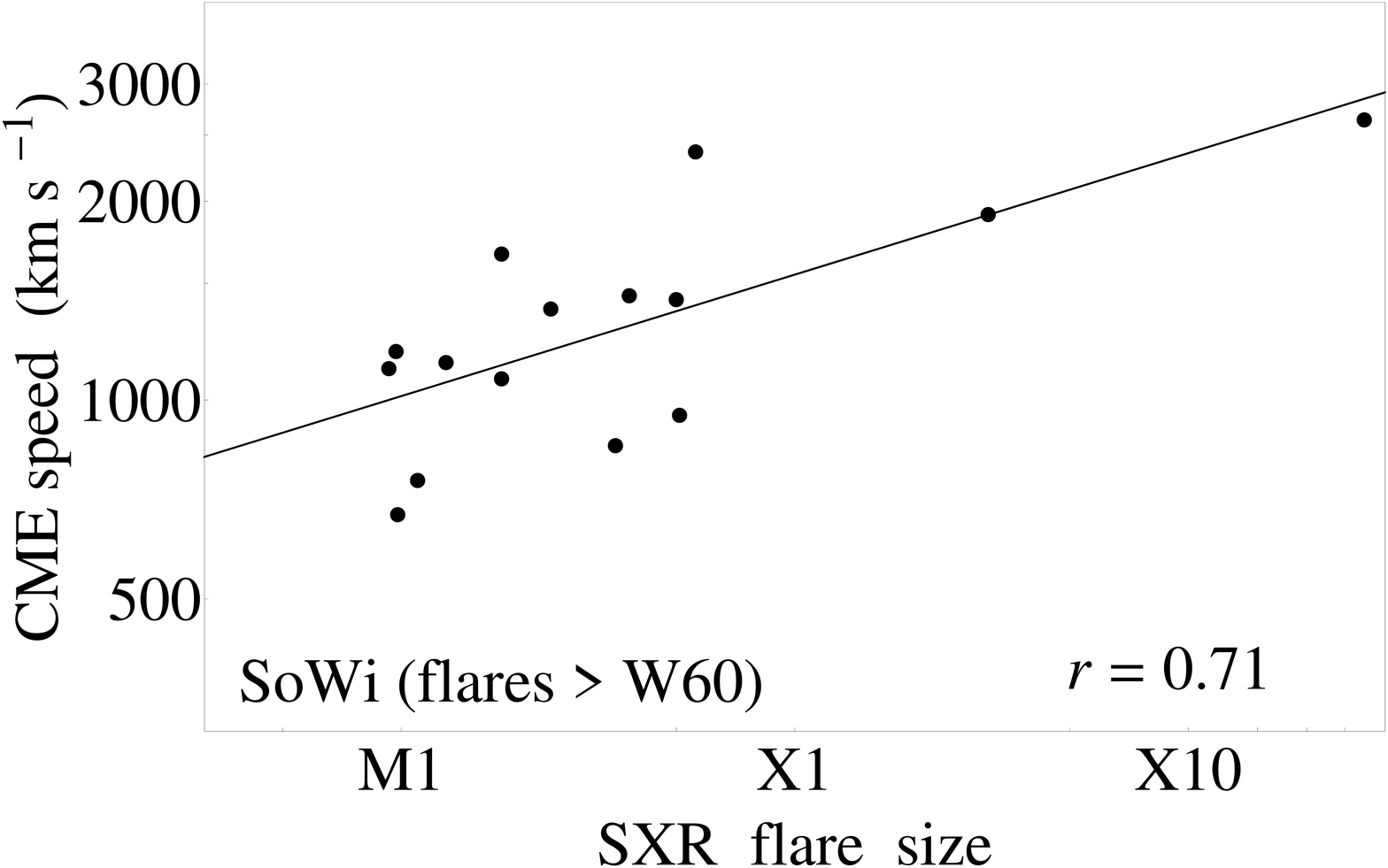}
               }
               \vspace*{-0.01\textwidth}
   \centerline{\hspace*{-0.05\textwidth}
               \includegraphics[width=0.55\textwidth,clip=]{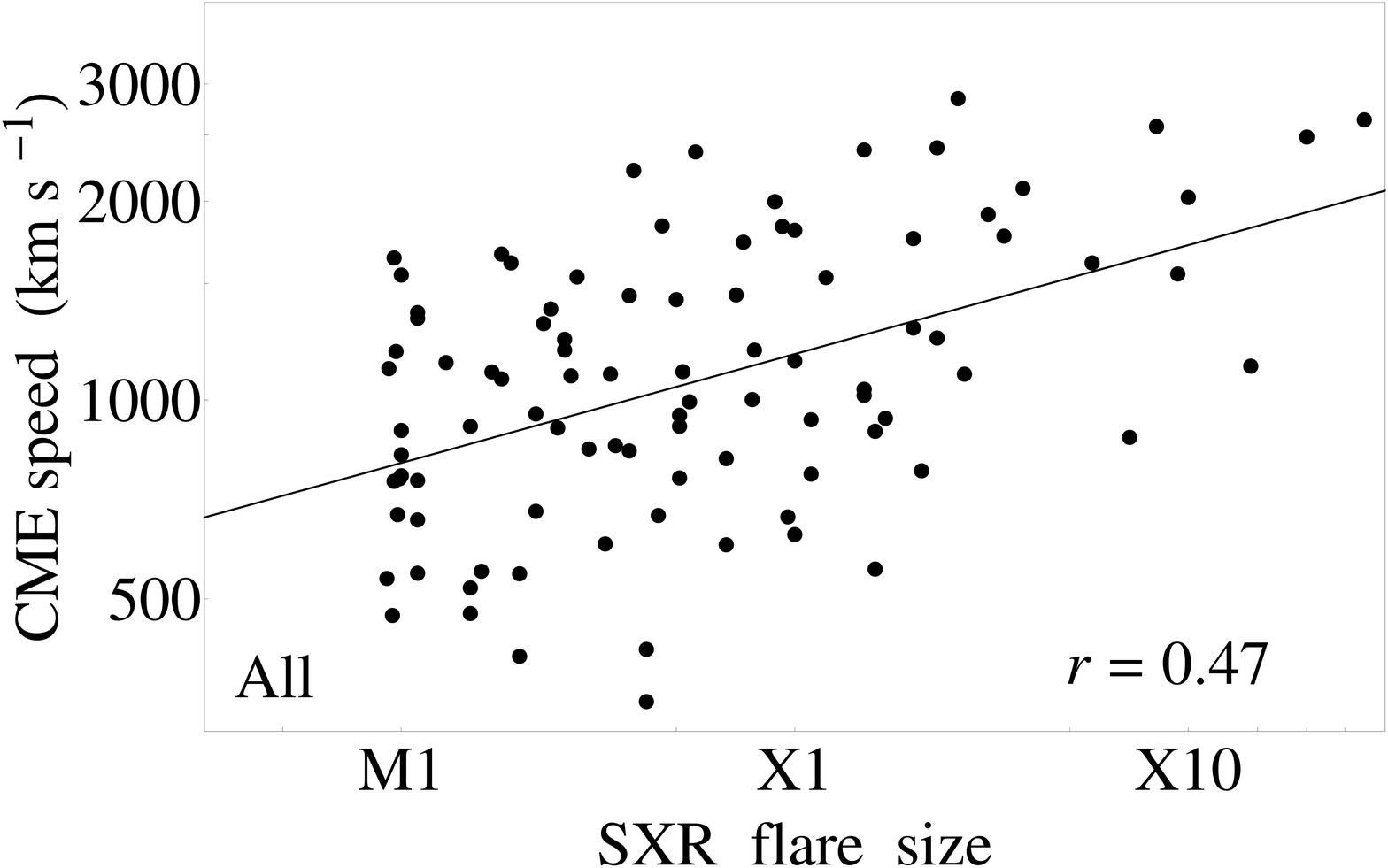}
               \hspace*{-0.05\textwidth}
               \includegraphics[width=0.55\textwidth,clip=]{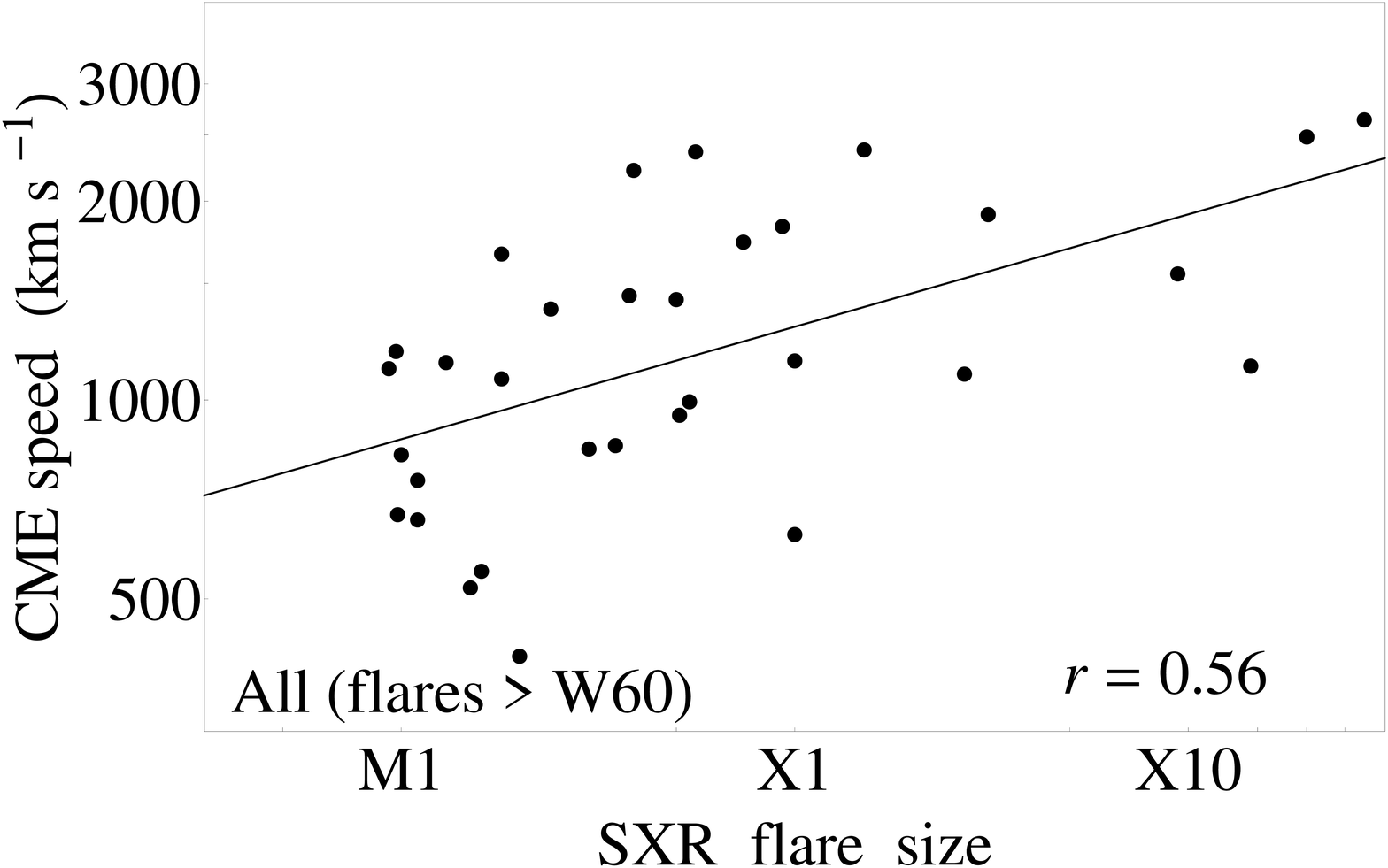}
               }
\caption{Scatter (log$-$log) plots of SXR flare size vs. CME speed for the different IMF ca\-tegories of ACE/EPAM electron events (left) and for a subset of these events associated with flares at longitudes $> {\rm W}60^{\circ}$ (right).}
   \label{F-scatter_plots_SXR_CMEe}
   \end{figure}

We present the log$-$log scatter plots of $I_{\rm SXR}$ vs. $V_{\rm CME}$ in different IMF categories. For the flare/CME events associated with proton events observed by GOES they are shown in Figure~\ref{F-scatter_plots_SXR_CMEp}, and for the events accompanied by electrons (ACE/EPAM low energy channel) in Figure~\ref{F-scatter_plots_SXR_CMEe}. There is an overall correlation for the entire event sample for both particle species, which shows that on average faster CMEs accompany stronger SXR bursts (correlation coefficient\footnote{ Note that the $r^2$-values of the correlation coefficients found in the present study (and also in earlier works) are usually $r^2 \lesssim 0.5$.} of 0.39$-$0.47). This was known before: \inlinecite{2005A&A...435.1149V} gave a value of 0.35 (with a statistical significance larger than 99.99\% by the $t$-test), and \inlinecite{2009IAUS..257..233Y} gave a value of 0.5.

In order to assess the significance of the correlations, we performed a simple error analysis using the `bootstrap' method \cite{2003psa..book.....W}. It consists in drawing repeatedly, out of a sample of $N$ events, $N$ events at random, and computing the correlation coefficient for this sample. This is repeated 1000 times, and the average value of the sample of correlation coefficients and its standard deviation are calculated. For all correlation coefficients calculated in the present work, the standard deviations will be given according to this method.

The left column of Figures~\ref{F-scatter_plots_SXR_CMEp} and \ref{F-scatter_plots_SXR_CMEe} suggests a difference between the behaviors of the SEP events in the two IMF categories, with a stronger correlation between $\log I_{\rm SXR}$ and $\log V_{\rm CME}$ in the ICME events, 0.61$-$0.70, as compared to the SoWi events (about twice as low, 0.23$-$0.28). The difference is statistically significant, see Table~\ref{T-SXR-CME}.

The difference in the two IMF categories is of course unexpected, because there is no physical reason why the correlation between two manifestations of coronal activity should depend on the IMF structure. We therefore look for observational biases. An obvious one is the projection effect \cite{2004JGRA..10903103B,2007A&A...469..339V}, which will decrease the CME speed, measured in projection onto the plane of the sky.

In order to test this, we calculated the correlation coefficients for different sub-samples (Table~\ref{T-SXR-CME}). A closer inspection of the left column of plots in Figures~\ref{F-scatter_plots_SXR_CMEp} and \ref{F-scatter_plots_SXR_CMEe} shows that the SoWi events associated with fast CMEs, as seen in the upper envelope of the scatter plot, seem to follow a similar ({\it i.e.} stronger) correlation as the ICME events. This trend is confirmed when we removed all SEP events from the sample associated with projected CME speed below $600\,$km$\,$s$^{-1}$. This brings the correlation coefficients slightly closer to each other (see the values in Table~\ref{T-SXR-CME}).

In order to avoid projection effects due to events close to the disc center\footnote{Note that five of the eight events in the SoWi category with projected speed below $600\,$km$\,$s$^{-1}$ occurred near the central meridian.} we then removed all SEP events associated with flares at longitudes below $60^{\circ}$. The right column of plots in Figures~\ref{F-scatter_plots_SXR_CMEp} and \ref{F-scatter_plots_SXR_CMEe} shows that this improves the correlations for all IMF categories of events. The largest increase in the correlation coefficient is for the SoWi events (compare the middle plots), but the standard deviation also increases due to the smaller number of events (Table~\ref{T-SXR-CME}). The change is less for the ICME events and only a slight increase is found for the entire data set (denoted with `no event restriction' in Table~\ref{T-SXR-CME}). \inlinecite{2004JGRA..10903103B}, considering the correlation between the kinetic energies of CMEs and SXR peak flux, also showed that the correlation coefficient increased if one considers only limb events.

\begin{table}[t!]
\caption{Linear correlation coefficients (with standard deviations) between the $\log I_{\rm SXR}$ and $\log V_{\rm CME}$ for proton and electron data set in the different IMF categories.}
\label{T-SXR-CME}
\begin{tabular}{lccc}
\hline
SEP event          & \multicolumn{3}{c}{IMF categories of SEP events}\\
sub-samples        & ICME & SoWi & All SEPs \\
\hline
Protons & \multicolumn{3}{c}{ GOES 15$-$40 MeV} \\
No event restriction               & 0.61$\pm 0.14$ &  0.23$\pm 0.16$  & 0.39$\pm 0.09$\\
$V_{\rm CME}> 600\,$km$\,$s$^{-1}$ & 0.61$\pm 0.14$ &  0.29$\pm 0.18$  & 0.38$\pm 0.09$ \\
Flares $>{\rm W}60^{\circ}$        & 0.69$\pm 0.36$ &  0.72$\pm 0.21$  & 0.47$\pm 0.15$ \\
\hline
Electrons & \multicolumn{3}{c}{ ACE/EPAM 38$-$53 keV} \\
No event restriction               & 0.70$\pm 0.12$ &  0.28$\pm 0.14$  & 0.47$\pm 0.08$   \\
$V_{\rm CME}> 600\,$km$\,$s$^{-1}$ & 0.65$\pm 0.16$ &  0.36$\pm 0.16$  & 0.40$\pm 0.09$ \\
Flares $> {\rm W}60^{\circ}$       & 0.79$\pm 0.14$ &  0.71$\pm 0.15$  & 0.56$\pm 0.10$ \\
\hline
\end{tabular}
\end{table}

In summary, the poor correlation between the projected CME speed and the peak SXR flux in events associated with SEP in the SoWi category can be ascribed to randomization of an existing relationship by the broad variety of flare longitudes, and to a lesser extent to events with slow CMEs. Hence projection effects play a role in the determination of CME speed. This role is apparently more important in the SoWi event sample, where the flares happen to be distributed over a broader range of longitudes than in the ICME sample. It does not seem to be possible to avoid this problem with the data sets at hand, since a restriction to limb events increases the correlation coefficient, but also its uncertainty. Attempts to empirically correct the CME speed for projection effects do not seem to increase the correlation with the SXR flux \cite{2005SoPh..229..313Y}. The actual correlations of SEP parameters with CME speeds are expected to be higher than the values we find from our data set. This applies especially to the SoWi events. On the other hand, the broader range of longitudes may also affect the peak particle intensity, in line with our previous finding that the range of connection distances is broader in the SoWi category than in the ICME events.

\subsubsection{Statistical Relationships between SEP Peak Intensities and Parameters of the Associated Solar Activity}
  \label{S-correlations}

 \begin{figure}[t!]
   \centerline{\hspace*{-0.05\textwidth}
               \includegraphics[width=0.6\textwidth,clip=]{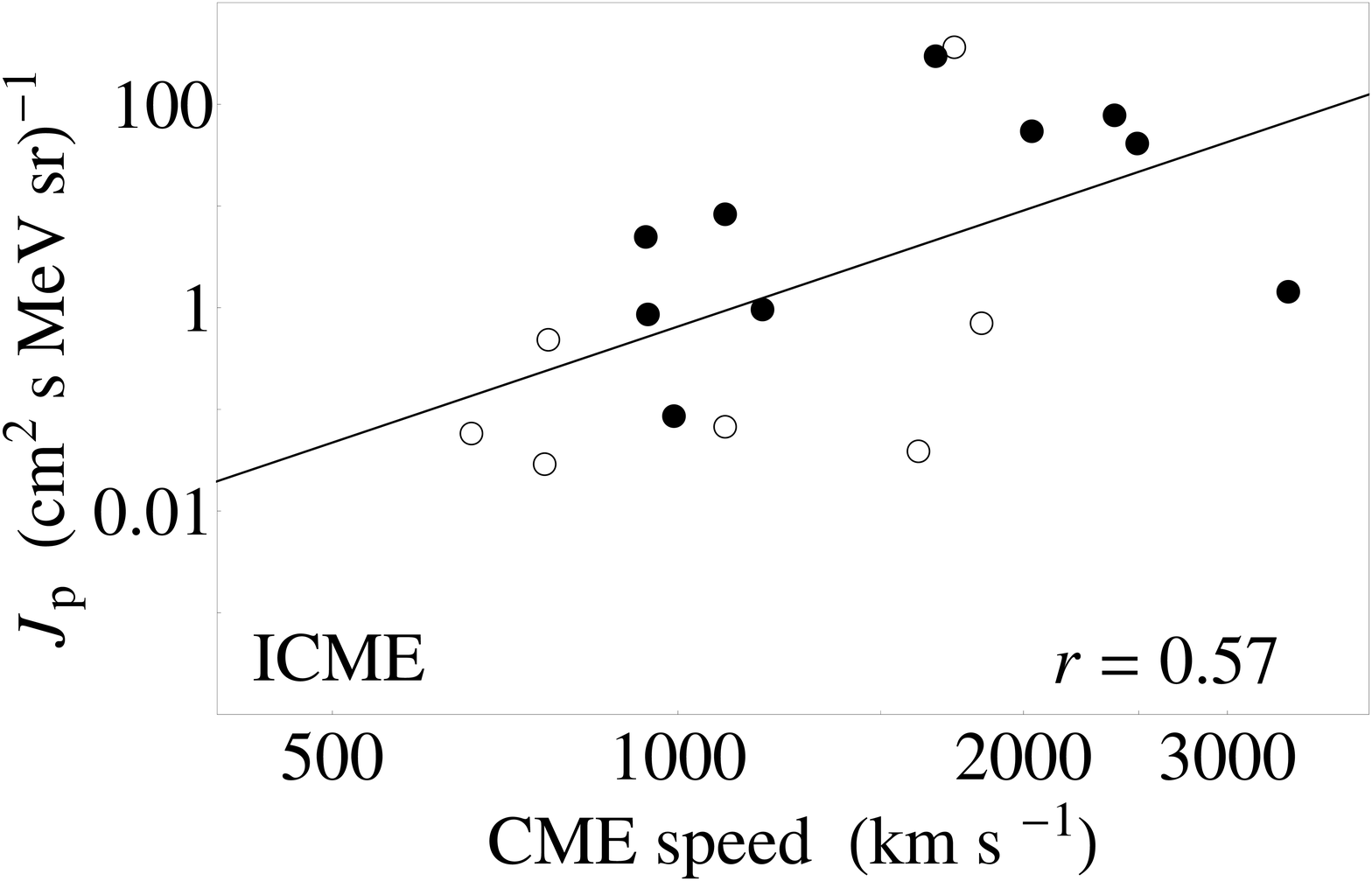}
               \hspace*{-0.1\textwidth}
               \includegraphics[width=0.6\textwidth,clip=]{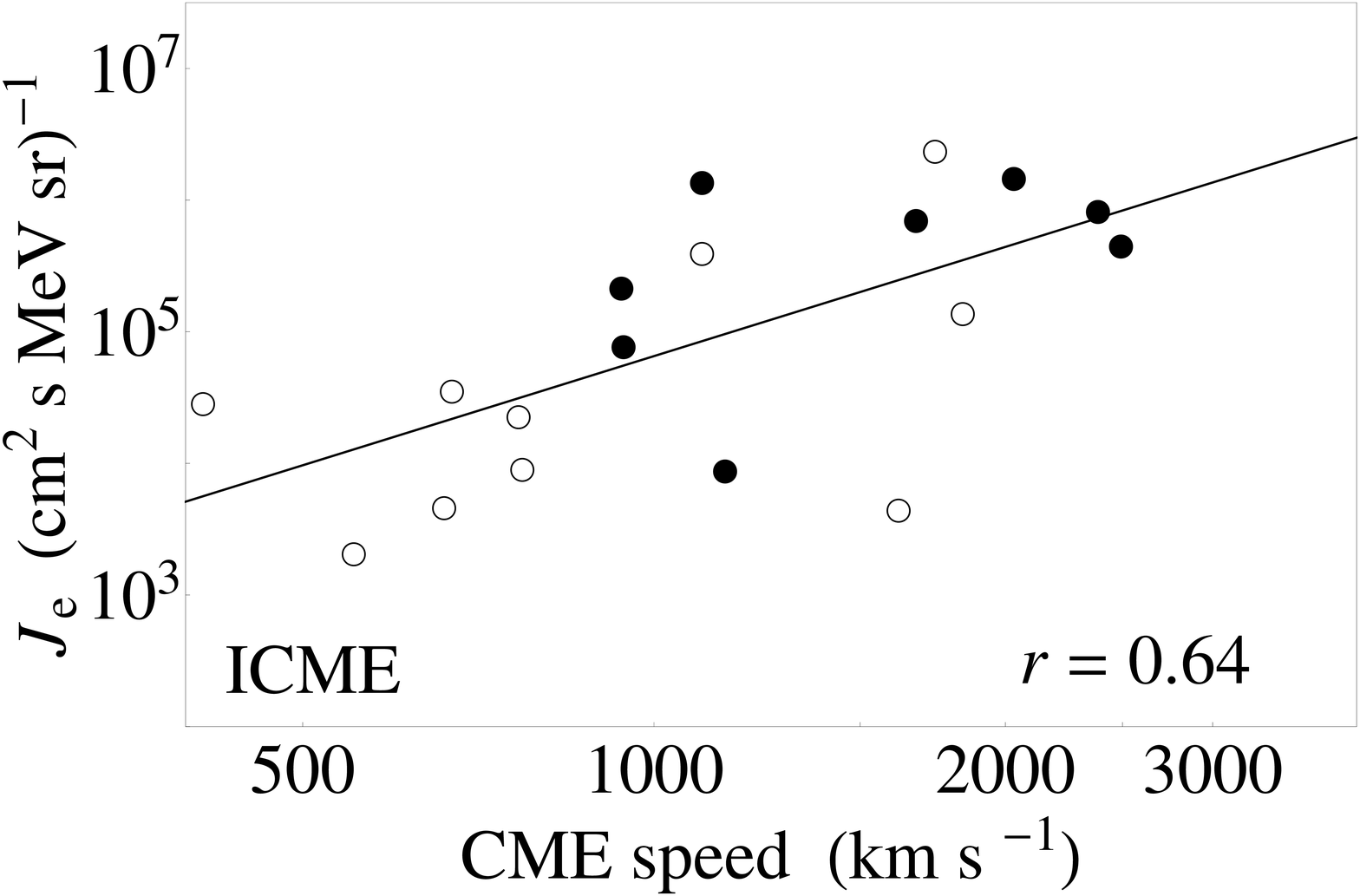}
               }
               \vspace*{-0.05\textwidth}
   \centerline{\hspace*{-0.05\textwidth}
               \includegraphics[width=0.6\textwidth,clip=]{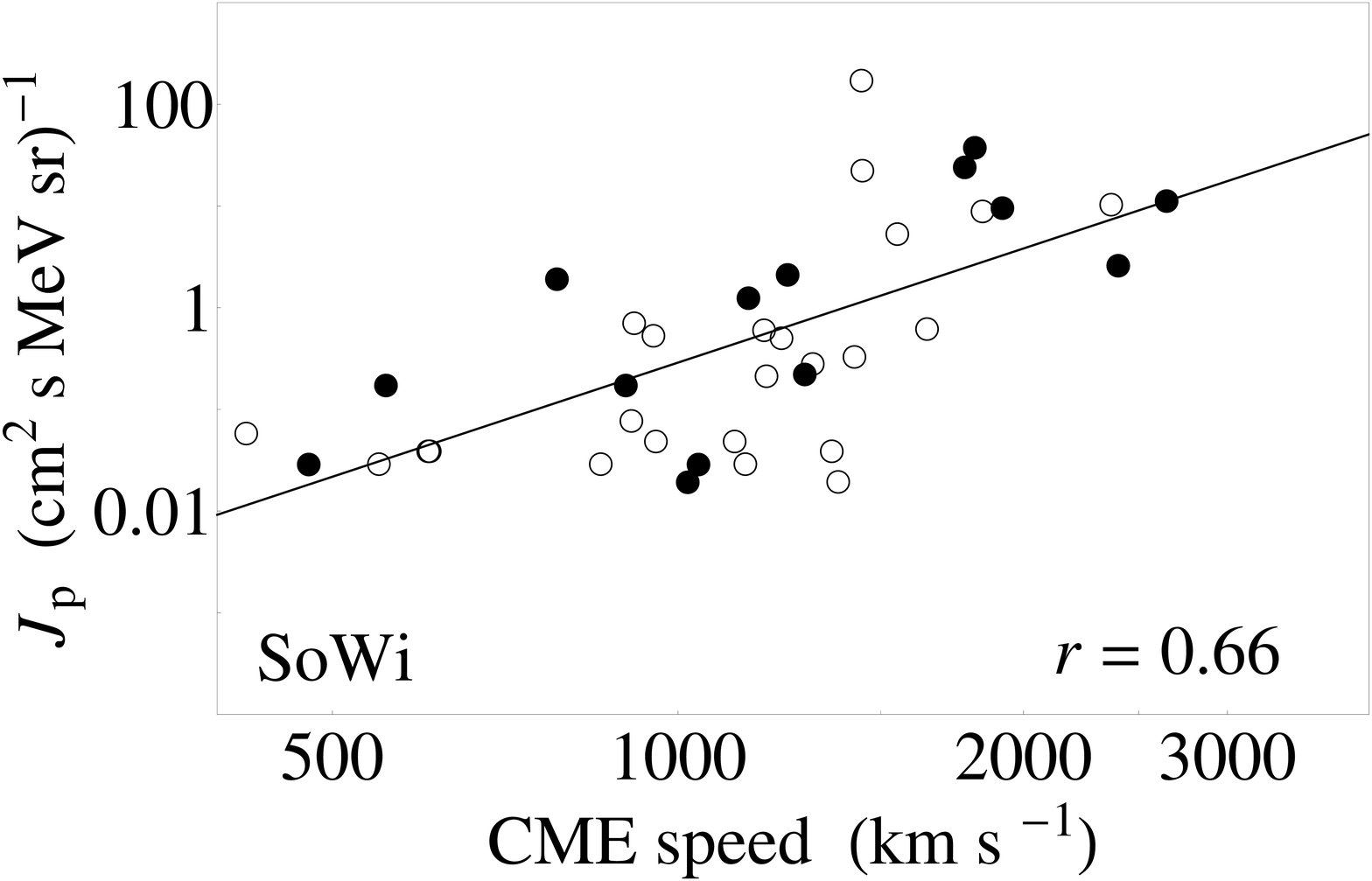}
               \hspace*{-0.1\textwidth}
               \includegraphics[width=0.6\textwidth,clip=]{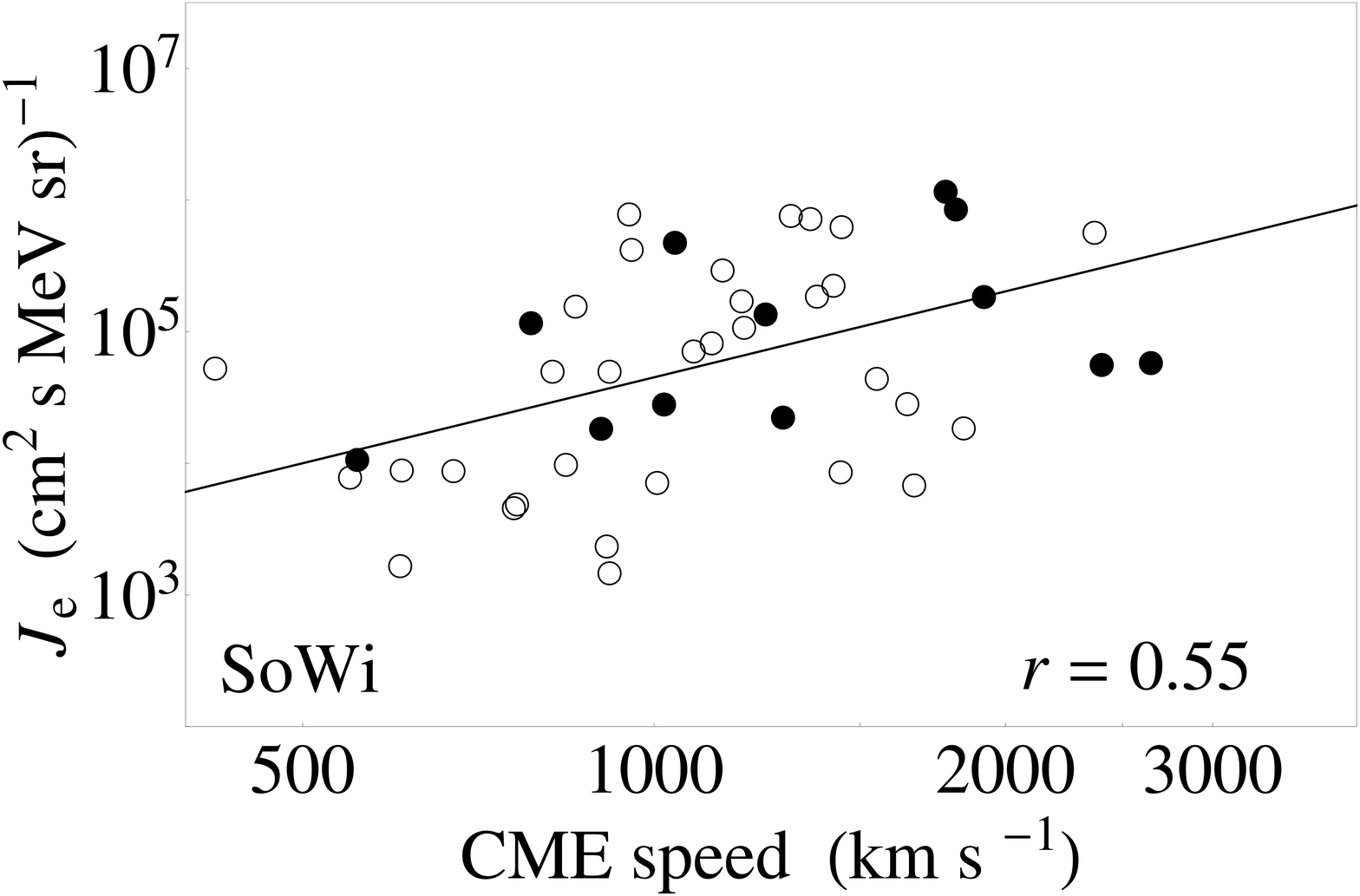}
               }
               \vspace*{-0.05\textwidth}
   \centerline{\hspace*{-0.05\textwidth}
               \includegraphics[width=0.6\textwidth,clip=]{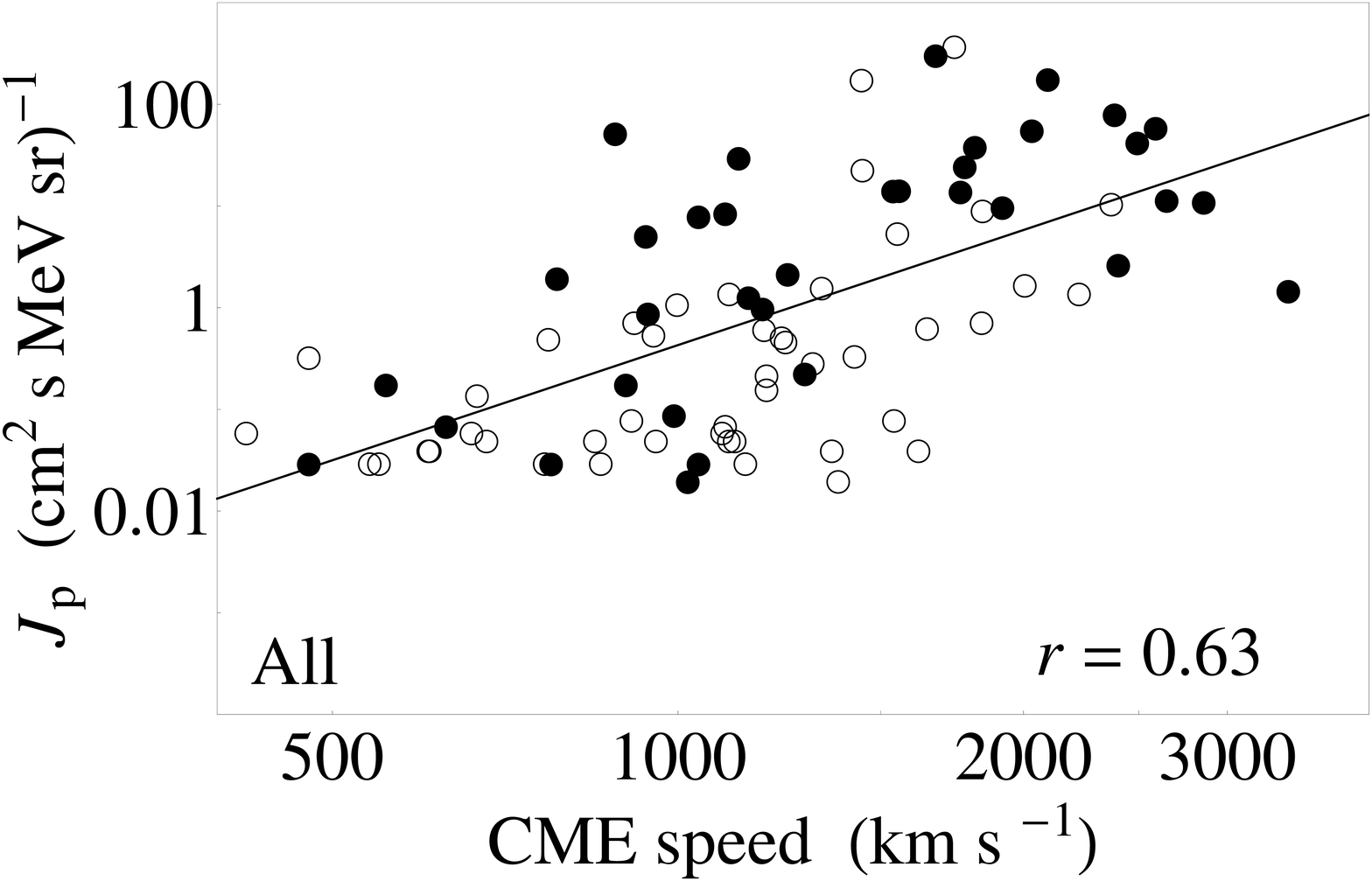}
               \hspace*{-0.1\textwidth}
               \includegraphics[width=0.6\textwidth,clip=]{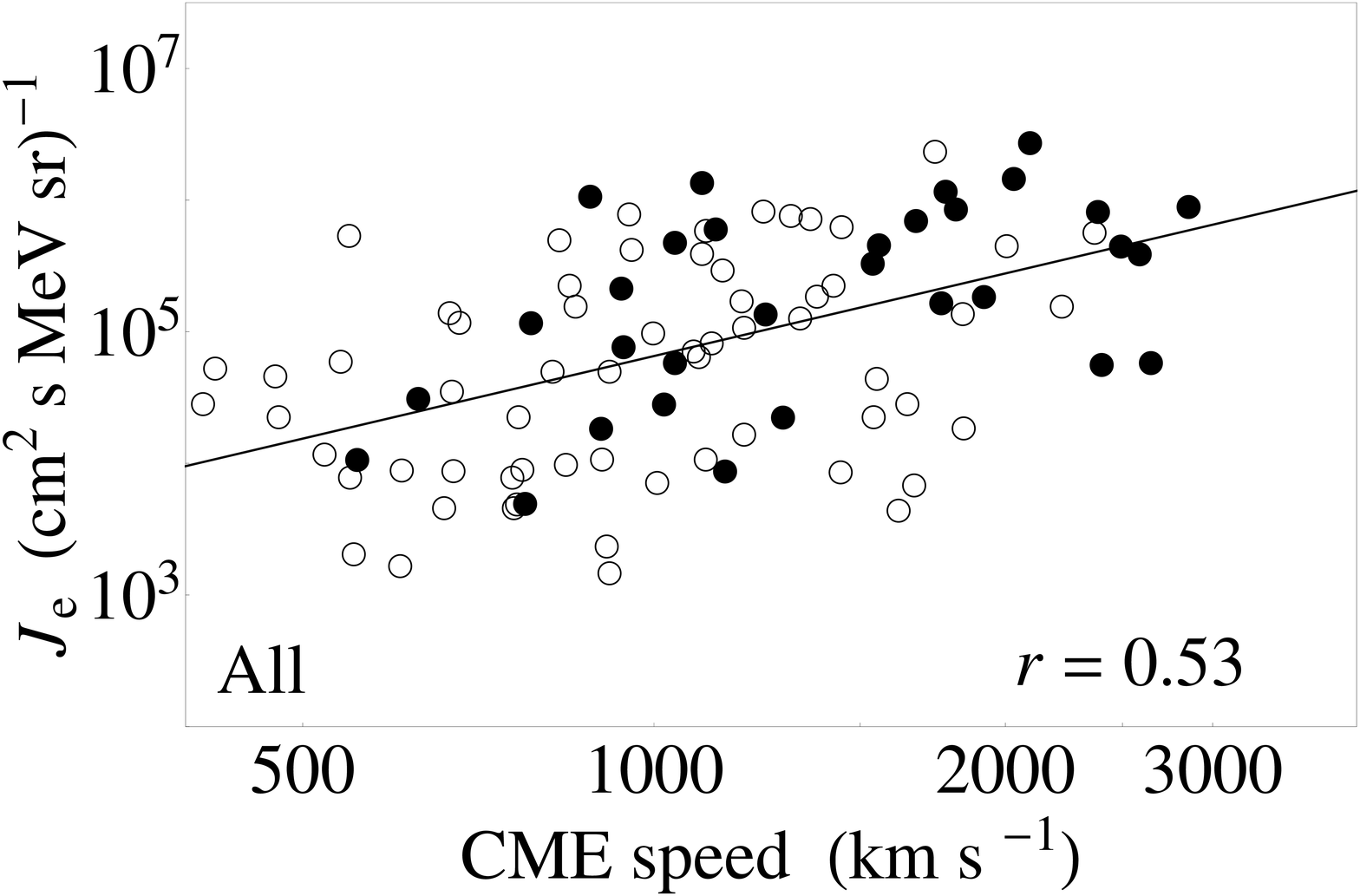}
               }
\caption{Scatter (log$-$log) plots of the peak proton ($J_{\rm p}$, left) and electron ($J_{\rm e}$, right) intensity vs. the CME speed for different IMF categories of SEP events.}
   \label{F-scatter_plots_CME}
   \end{figure}

 \begin{figure}[t!]
   \centerline{\hspace*{-0.04\textwidth}
               \includegraphics[width=0.55\textwidth,clip=]{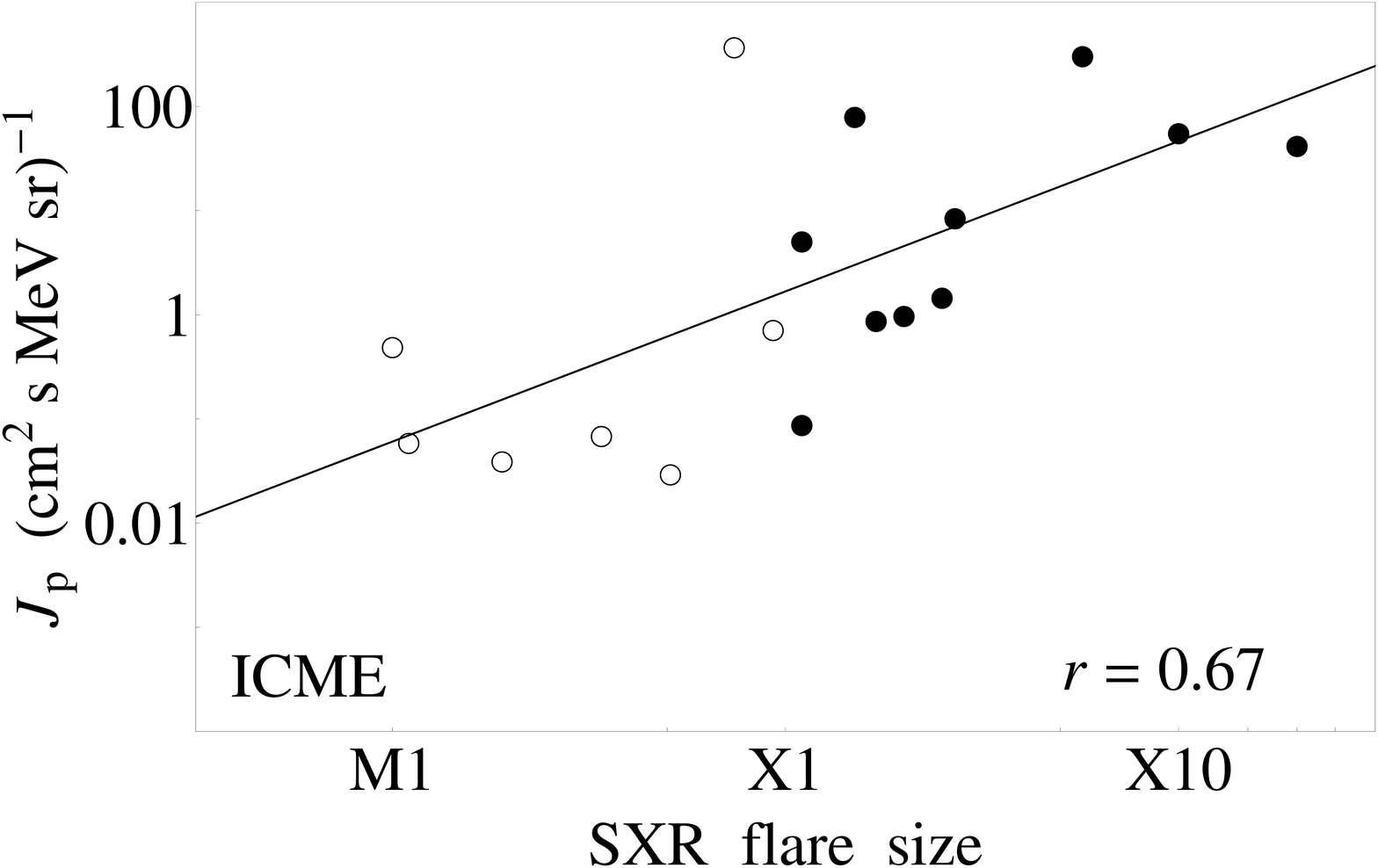}
               \hspace*{-0.04\textwidth}
               \includegraphics[width=0.55\textwidth,clip=]{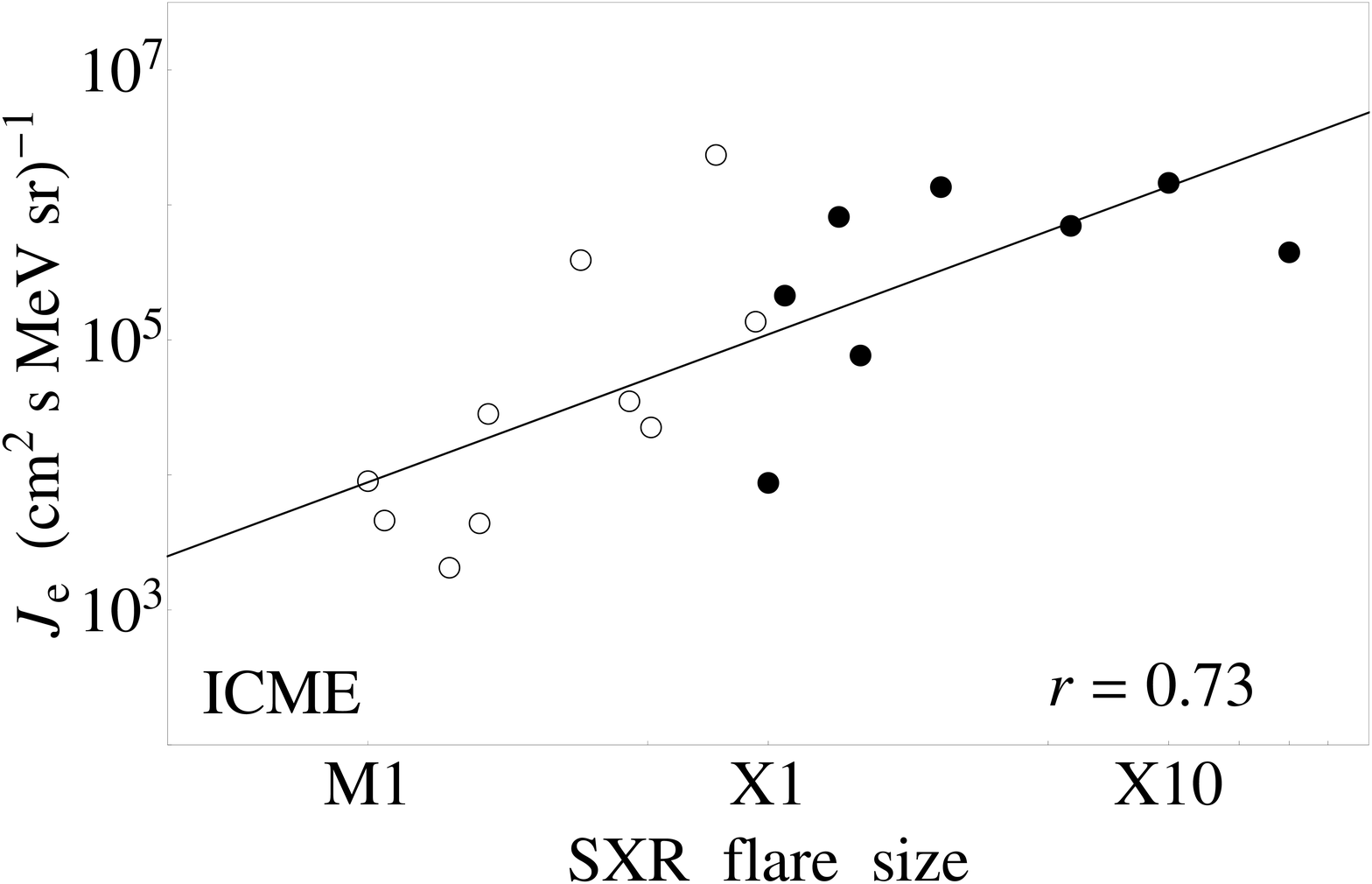}
               }
               \vspace*{-0.01\textwidth}
   \centerline{\hspace*{-0.04\textwidth}
               \includegraphics[width=0.55\textwidth,clip=]{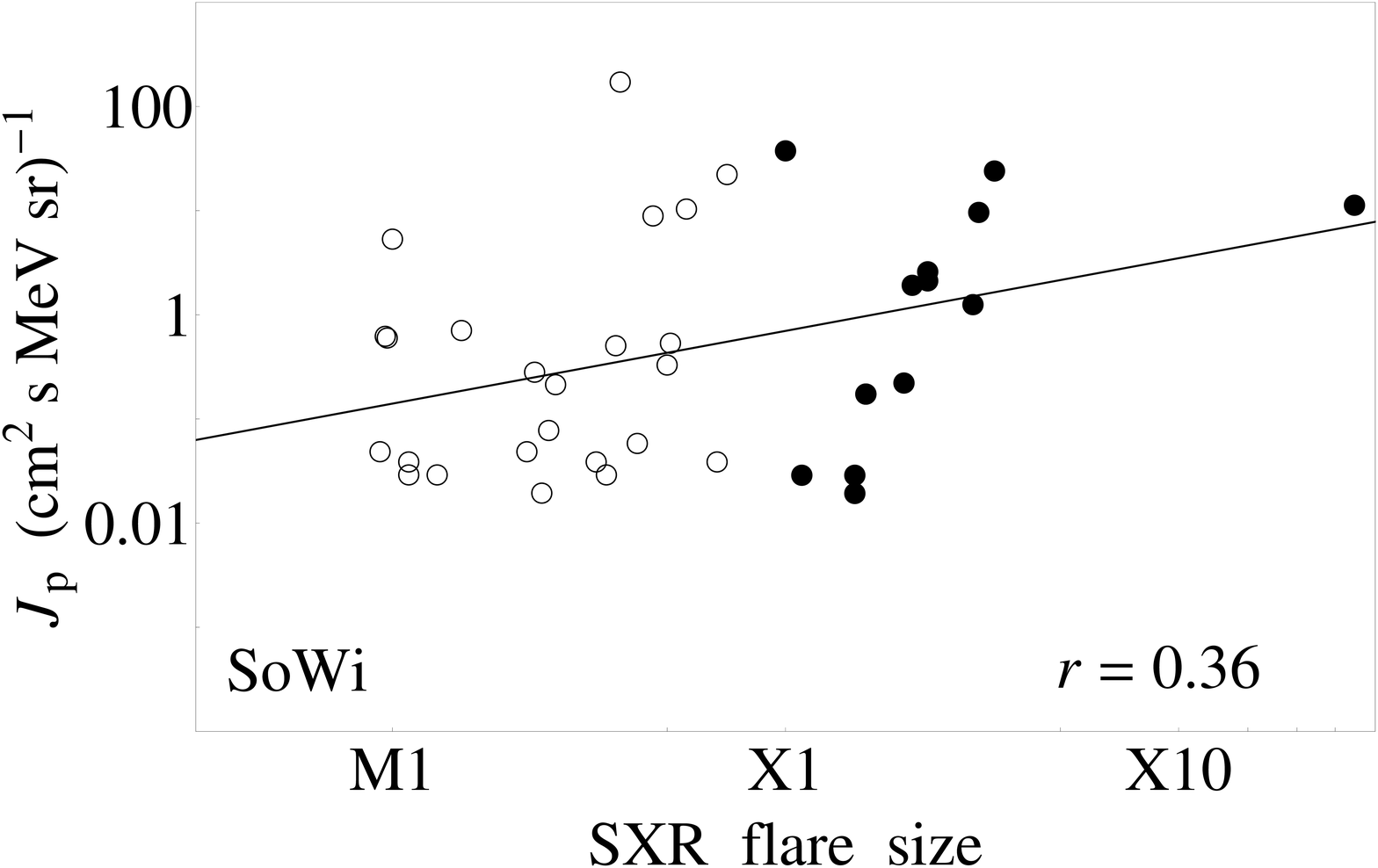}
               \hspace*{-0.04\textwidth}
               \includegraphics[width=0.55\textwidth,clip=]{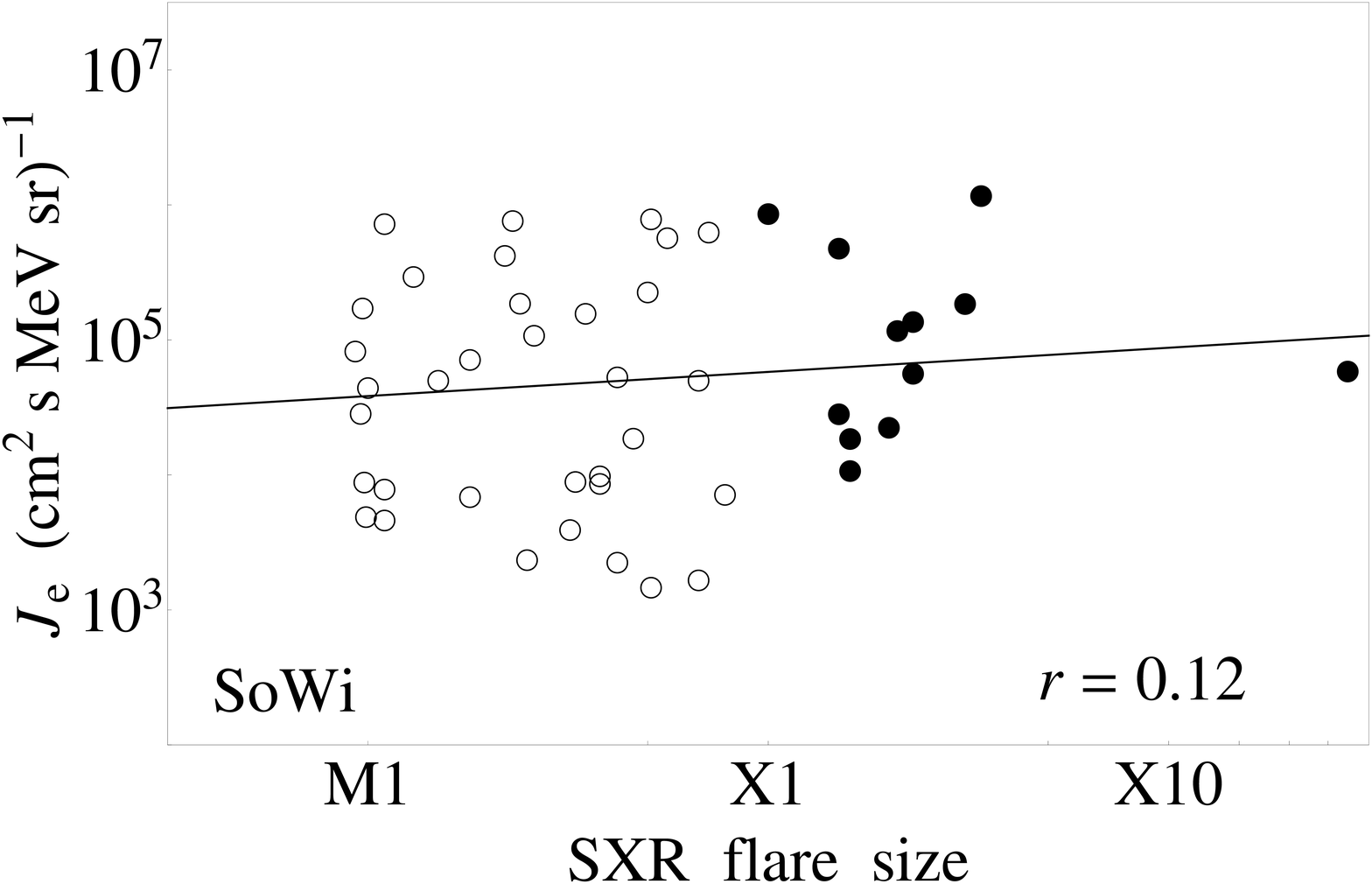}
               }
               \vspace*{-0.01\textwidth}
   \centerline{\hspace*{-0.04\textwidth}
               \includegraphics[width=0.55\textwidth,clip=]{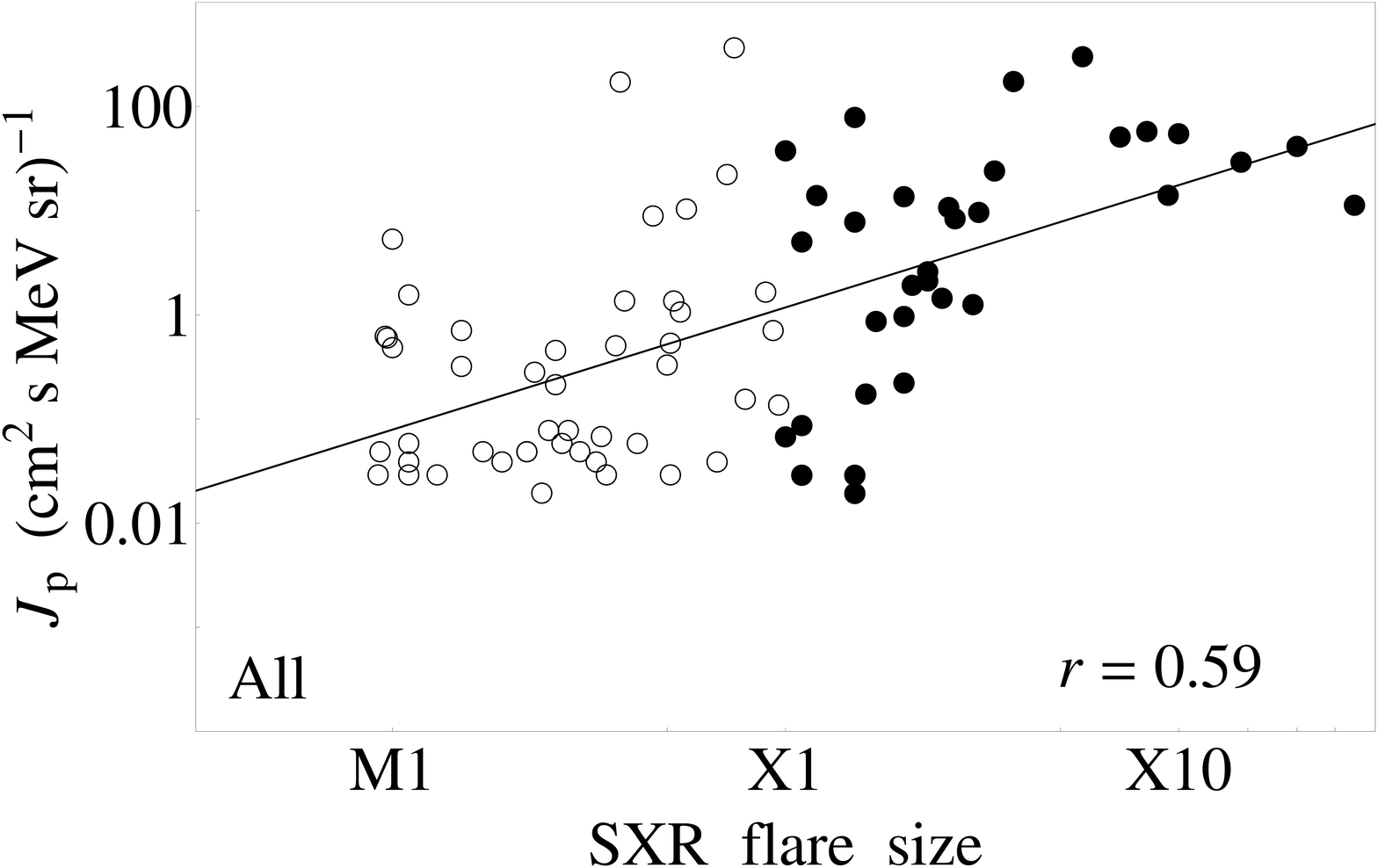}
               \hspace*{-0.04\textwidth}
               \includegraphics[width=0.55\textwidth,clip=]{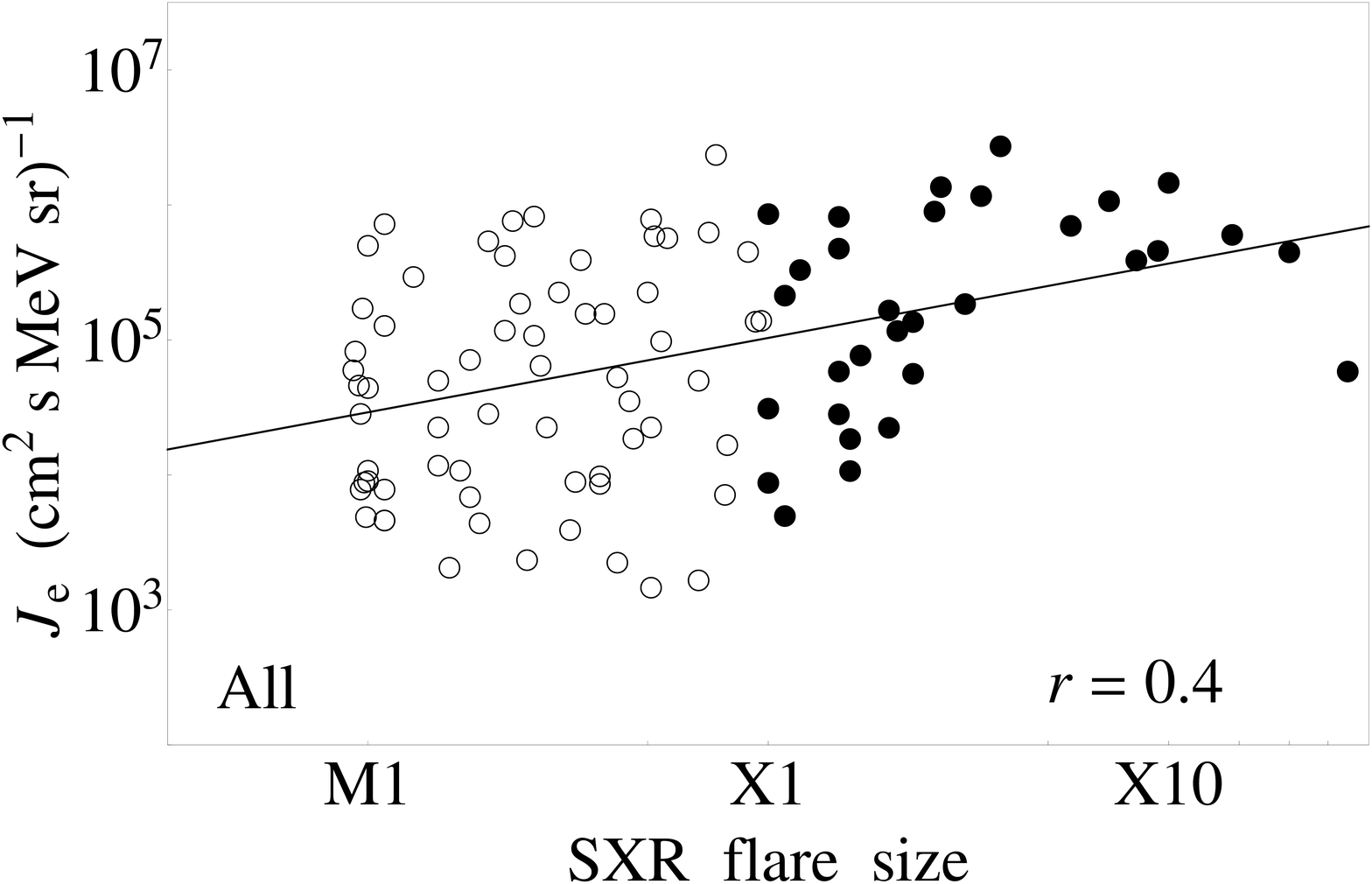}
               }
\caption{Scatter (log$-$log) plots of the peak proton ($J_{\rm p}$, left) and electron ($J_{\rm e}$, right) intensity vs. the soft X-ray flare size for different IMF categories of SEP events.}
   \label{F-scatter_plots_SXR}
   \end{figure}

\begin{table}[h!]
\caption[]{Linear correlation coefficients of $\log J_{\rm max}$ and $\log I_{\rm SXR}$, $\log V_{\rm CME}$ and angular width (AW) for the different IMF categories.}
\label{T-correlations}
\small
\begin{tabular}{lccc}
\hline
Correlation      & \multicolumn{3}{c}{IMF categories of SEP events}     \\
$\log J_{\rm max}$ with:   & ICME  & SoWi  & All SEPs \\
%(1)              & (2)   &  (3)  &  (4) \\
\hline
%\hline
Protons & \multicolumn{3}{c}{ GOES 15$-$40 MeV} \\
$\log I_{\rm SXR}$     & 0.67$\pm 0.13$ & 0.36$\pm 0.13$  &  0.59$\pm 0.07$ \\
$\log V_{\rm CME}$     & 0.57$\pm 0.15$ & 0.66$\pm 0.07$  &  0.63$\pm 0.05$ \\
AW                     & 0.52$\pm 0.18$ & 0.23$\pm 0.13$  &  0.29$\pm 0.10$ \\
Number                 & 17    & 38    & 81 \\
\hline
Protons & \multicolumn{3}{c}{ {\it Wind}/EPACT 19$-$28 MeV}  \\
$\log I_{\rm SXR}$     & 0.71  & 0.41 &  0.62 \\
$\log V_{\rm CME}$     & 0.73  & 0.68 &  0.68 \\
AW                     & 0.58  & 0.28 &  0.37 \\
Number                 & 22    & 43   &  96   \\
\hline
Protons & \multicolumn{3}{c}{ Cane {\it et al.} (2010) $>$ 25 MeV}\\
$\log I_{\rm SXR}$     & 0.86  & 0.31  &  0.61 \\
$\log V_{\rm CME}$     & 0.68  & 0.55  &  0.61 \\
AW                     & 0.69  & 0.22  &  0.33 \\
Number                 & 20     & 52    & 104 \\
\hline
Electrons & \multicolumn{3}{c}{ ACE/EPAM 38$-$53 keV} \\
$\log I_{\rm SXR}$     & 0.73$\pm 0.10$  & 0.12$\pm 0.11$    & 0.40$\pm 0.08$ \\
$\log V_{\rm CME}$     & 0.64$\pm 0.14$  & 0.55$\pm 0.09$    & 0.53$\pm 0.07$ \\
AW                     & 0.48$\pm 0.14$  & $-$0.09$\pm 0.12$ & 0.09$\pm 0.10$ \\
Number                 & 18    &  46   & 96 \\
\hline
Electrons & \multicolumn{3}{c}{ ACE/EPAM 175$-$315 keV} \\
$\log I_{\rm SXR}$     & 0.74 & 0.31 &  0.57 \\
$\log V_{\rm CME}$     & 0.63 & 0.63 &  0.59 \\
AW                     & 0.56 & 0.05 &  0.22 \\
Number                 & 18    &  46   & 95\\
\hline
\end{tabular}
\end{table}

In this section we search for statistical relationships between the peak SEP intensity ($J_{\rm max}$) and parameters of the associated coronal activity, namely the peak SXR flux and the CME projected speed. Only the scatter plots for the GOES proton (Figure~\ref{F-scatter_plots_CME}) and ACE/EPAM electron (Figure~\ref{F-scatter_plots_SXR}) data sets are shown, for the ICME, SoWi events and the entire sample. Linear correlation coefficients $r$ between the logarithmic quantities are given in each frame.

The logarithms of $J_{\rm p}$ and $J_{\rm e}$ show similar correlation coefficients with $I_{\rm SXR}$ and $V_{\rm CME}$ in the range 0.40$-$0.63 for the entire event sample (see the bottom panels of Figures~\ref{F-scatter_plots_CME} and \ref{F-scatter_plots_SXR}). Differences arise when the two IMF configurations are considered: While the correlations of $\log J_{\rm max}$ with $\log V_{\rm CME}$ are similar for ICME and SoWi events (Figure~\ref{F-scatter_plots_CME}), the correlations with $\log I_{\rm SXR}$ differ (Figure~\ref{F-scatter_plots_SXR}). Peak particle intensities are more strongly correlated with $I_{\rm SXR}$ than average in the ICME events, and are only weakly correlated in SoWi events. The difference is about a factor of two for the protons, and six for the electrons. The same results are found when the start-to-peak fluence, {\it i.e.}, the time-integral of the background-subtracted SEP intensity until the peak time, is used instead of the peak intensities.

As done previously (Section~\ref{S-correlfc}) the statistical uncertainties of the correlation coefficients are evaluated using the `bootstrap'  method. As a consistency check, we calculated the average value of each correlation coefficient, based on 1000 random selections. This value differs only less than 5\% from the value reported in Figures~\ref{F-scatter_plots_CME} and \ref{F-scatter_plots_SXR}, drawn from the complete set of measured points. It is the latter correlation coefficients that are listed in Table~\ref{T-correlations} for the GOES and ACE/EPAM low energy channel data sets, together with the error estimates. The number of events in each SEP group is also given. Within the estimated errors the correlation coefficient between $J_{\rm max}$ and $V_{\rm CME}$ is the same for ICME and SoWi events. It is also similar to the correlation between $J_{\rm max}$ and $I_{\rm SXR}$ for the ICME events. All these correlation coefficients are statistically higher than that between $J_{\rm max}$ and $I_{\rm SXR}$ for the SoWi events. The weak correlation of $\log J_{\rm max}$$-$$\log I_{\rm SXR}$ in the SoWi category is statistically significant result compared to the higher correlation found in the ICME category. The correlation coefficients drawn from the different data sets are consistent (see Table~\ref{T-correlations}).

\begin{figure}[t!]
   \centerline{\hspace*{-0.01\textwidth}
               \includegraphics[width=0.57\textwidth,clip=]{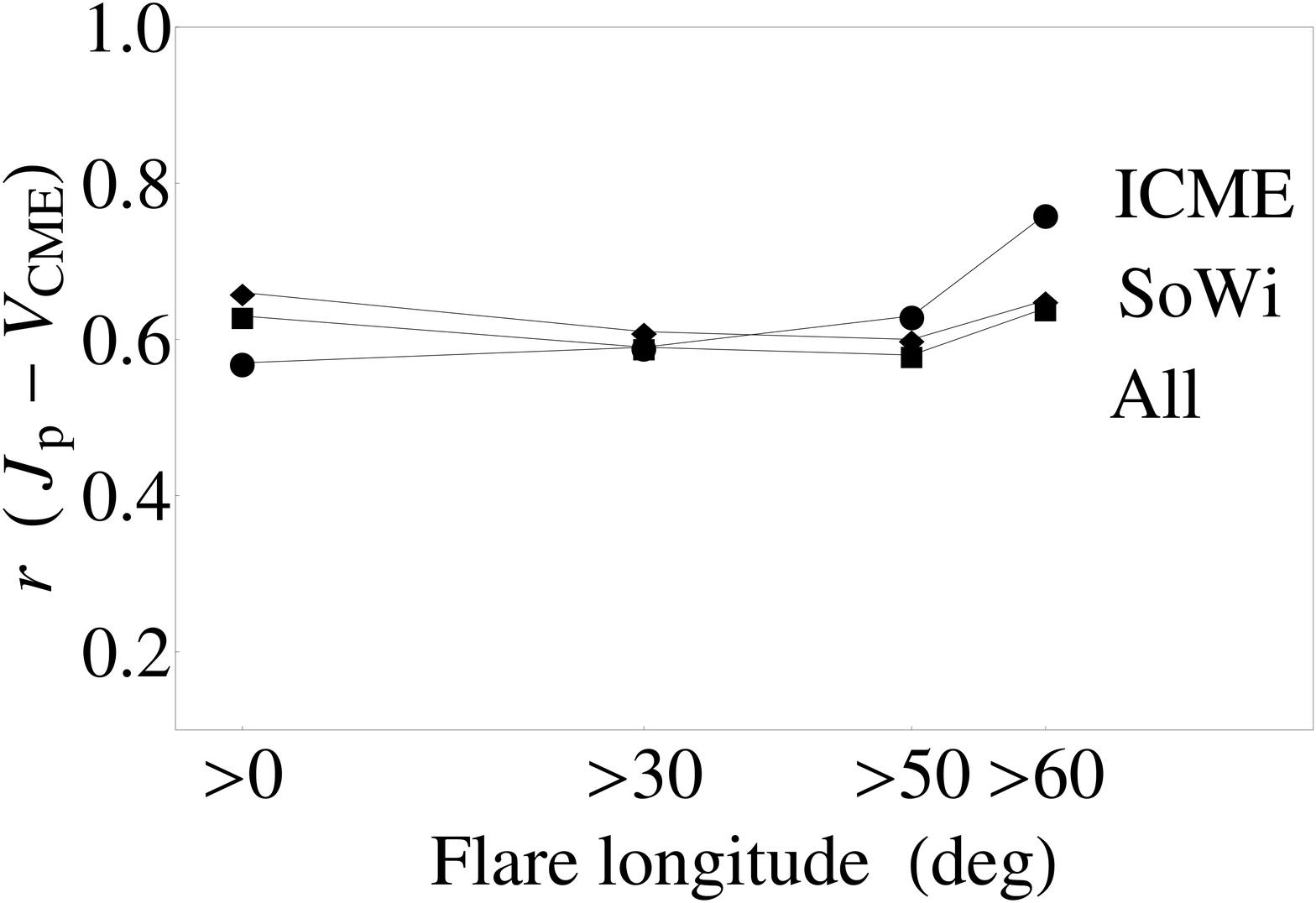}
               \hspace*{-0.1\textwidth}
               \includegraphics[width=0.57\textwidth,clip=]{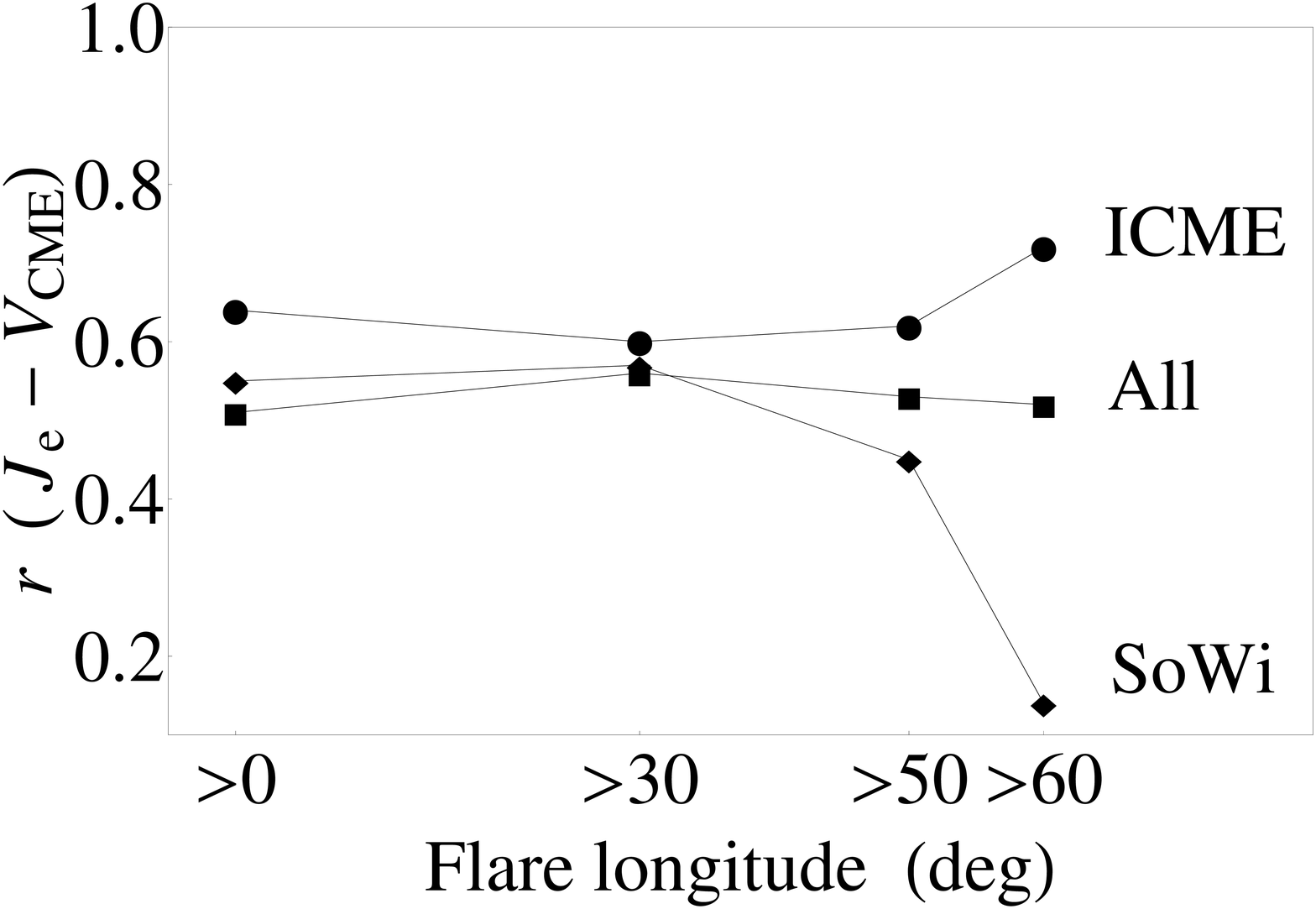}
               }
               \vspace*{-0.05\textwidth}
   \centerline{\hspace*{-0.01\textwidth}
               \includegraphics[width=0.57\textwidth,clip=]{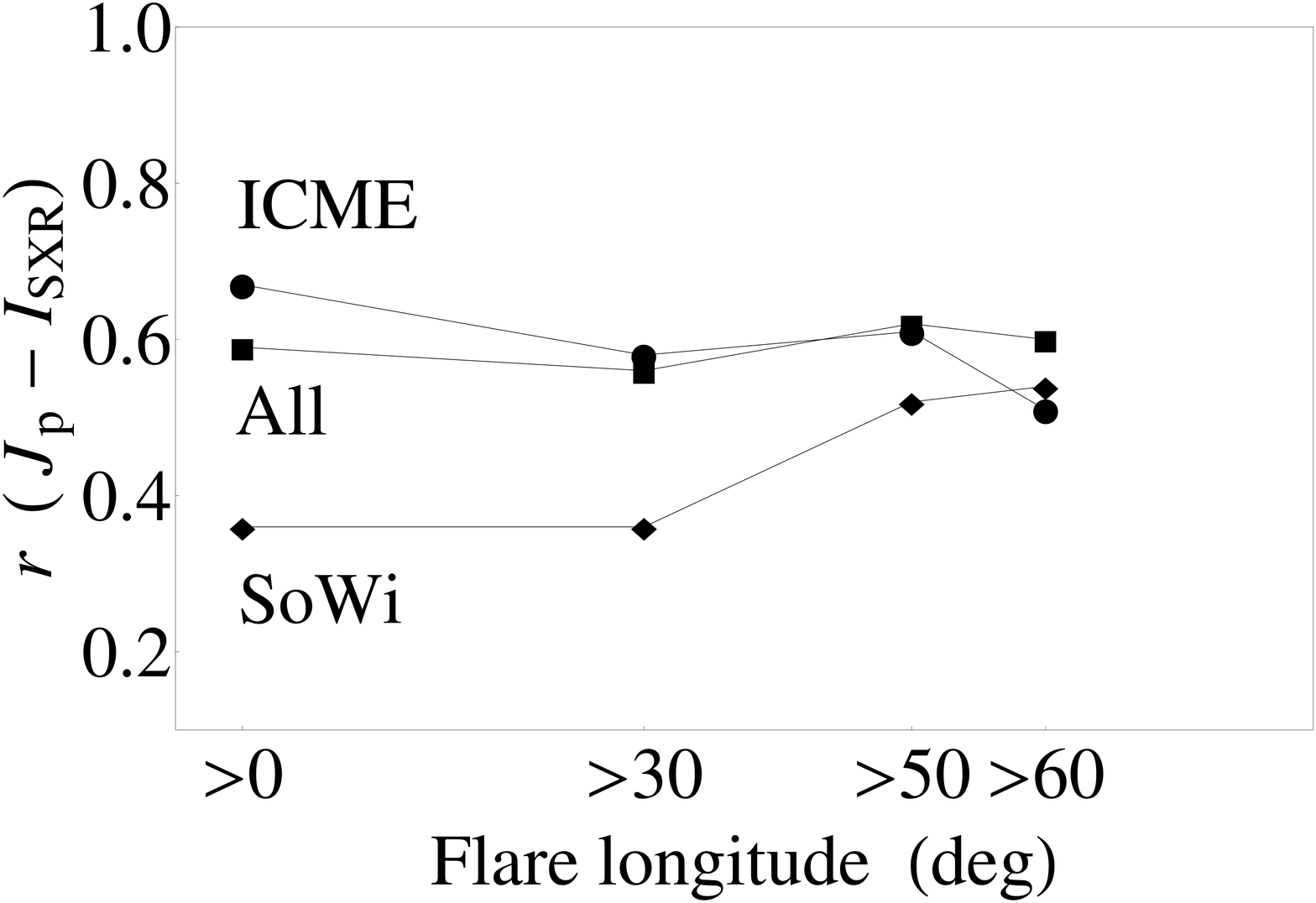}
               \hspace*{-0.1\textwidth}
               \includegraphics[width=0.57\textwidth,clip=]{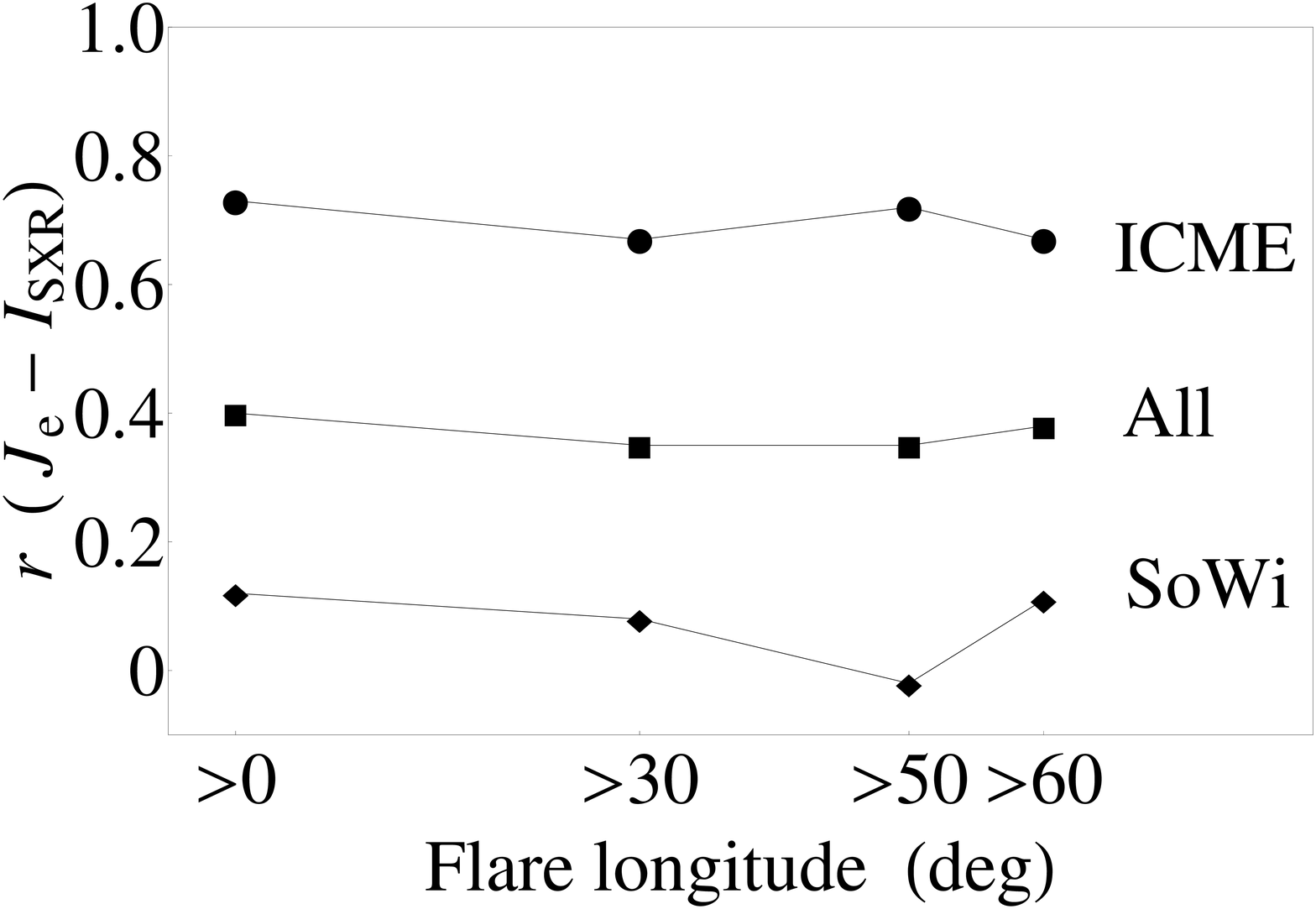}
               }
\caption{Trends for the correlation coefficients of the particle intensity ($J_{\rm max}$) with $V_{\rm CME}$ and $I_{\rm SXR}$ for sub-samples of SEP events associated with flares at longitude above certain threshold. Dots, diamonds, and squares denote ICME events, SoWi and All SEP events, respectively.}
   \label{F-r-trend-flare}
   \end{figure}

Given the discussion of the $I_{\rm SXR}$$-$$V_{\rm CME}$ correlation in Section~\ref{S-correlfc}, different correlations in the ICME and SoWi event categories may be biased due to different longitude distributions of the flare/CME events in the two categories. We therefore investigate if and how this bias also influences the correlations of $J_{\rm max}$ and the parameters of the coronal activity. For this we calculated the corresponding correlation coefficients of $J_{\rm max}$ with $I_{\rm SXR}$ and $V_{\rm CME}$ in the different IMF categories and for particle events associated with flares beyond certain longitude threshold. The results are summarized in Table~\ref{T-J-SXR-CME} (in the Appendix), where the first row in each section is taken from Table~\ref{T-correlations} to facilitate the comparison. The trends for the different correlation coefficients are presented in Figure~\ref{F-r-trend-flare}. The restriction to limb events ($>60^{\circ}$), however, reduces the number of events in each category to less than half of the original size and increases the uncertainty. Hence, due to the small number of events for this sub-sample, selection effects may play a role.

In general, the correlation $\log J_{\rm max}$$-$$\log V_{\rm CME}$ stays the same within the uncertainties when we consider different subsets of limb events (compare the upper panel of plots in Figure~\ref{F-r-trend-flare}). For the GOES protons, associated with flares at longitude $>60^{\circ}$, we find no statistically significant difference in $\log J_{\rm p}$$-$$\log V_{\rm CME}$ (see the upper left plot in Figure~\ref{F-r-trend-flare}): The increase of the correlation coefficient for ICME events to 0.76$\pm 0.22$ (from the previous value of 0.57$\pm 0.15$) is within the uncertainties, and there is practically no change in the correlation for the SoWi events (0.65$\pm 0.14$ compared to the previous value of 0.66$\pm 0.07$). However, we obtain a statistically lower correlation $\log J_{\rm e}$$-$$\log V_{\rm CME}$ in the SoWi category, 0.14$\pm 0.28$ (see the upper right plot), but only when limb events $>60^{\circ}$ are considered (and not for $>50^{\circ}$ for example), so selection effects may be responsible for this result.

For the correlation $\log J_{\rm max}$$-$$\log I_{\rm SXR}$ we find different trends for the two particle species (lower panel of plots in Figure~\ref{F-r-trend-flare}). For the protons, the factor of two difference in the $\log J_{\rm p}$$-$$\log I_{\rm SXR}$ correlation between the two IMF categories is no longer present for the limb events. Namely, we obtain 0.51$\pm 0.34$ for the ICME events and 0.54$\pm 0.17$ for the SoWi events (see the lower left plot in Fig.~\ref{F-r-trend-flare}). For the ACE/EPAM low energy channel electron data, however, we still find statistically significant difference in the correlations $\log J_{\rm e}$$-$$\log I_{\rm SXR}$ between the two IMF categories (see the lower right plot in Figure~\ref{F-r-trend-flare}). Namely, the correlation coefficient in the SoWi category, 0.11$\pm 0.28$, is lower than that in the ICME category, 0.67$\pm 0.18$, within the uncertainties.

We conclude that SEP events associated with flares close to disc centre could be the main observational artifact that weakens the correlation of the proton intensity with the $I_{\rm SXR}$ for the SoWi events. For the electron events propagating in the solar wind another explanation is needed.

Finally, we consider the correlation between $\log J_{\rm max}$ and the linear value of the angular width (AW) in the two IMF categories. As can be seen from Table~\ref{T-correlations}, the correlation with the angular width is similar in behavior to that of the $I_{\rm SXR}$, and not with the $V_{\rm CME}$. Namely, the correlation coefficients for the protons (electrons) for the SoWi events are roughly twice (five times) as weak compared to those for the ICME events.

\subsubsection{A Closer Look at the ICME Events} %%%%%%%%%%%%%%
  \label{S-ICME-category}

We considered so far ICME events as a well-defined category (in total we have 22 events; see Table~\ref{T-ICME_events}). But one must actually envisage two physical scenarios: If particles accelerated in the corona are released into the legs of the ICME connected to the Earth, the SEP propagation is expected to be favored. This is probably the case when the flare and CME associated  with the SEP event occur close to the footpoints of the ICME. If they occur far away, the ICME may actually shield the Earth from the SEP. In this case the ICME should weaken the relationship between the SEP peak intensity and the parameters of the coronal activity. In order to check which of these scenarios applies, we compared the active region (AR) associated with the SEP event with the AR of the flare associated with the CME that gave rise to the ICME.

The number of ICMEs that come clearly from the same AR or at least from the same group of ARs as the SEP event is 16/22, where 10/16 of the SEPs are X-class associated events and 6/16 are related to M-class flares. There are three uncertain associations and three cases where the ARs of the SEP associated flare and the ICME associated flare are not the same. Nevertheless, the majority of the SEP events, arriving inside (and presumably also propagating along the field lines of) the ICME, come from the same AR (13/22) or from the same group of ARs (3/22) as the ICME. It is hence to be expected that the SEP are directly released into the magnetic flux tube of the ICME, which is still magnetically connected to that AR.

Restricting the ICME sample to these 16 cases where the ICME and the SEP associated activity come from the same AR (or group of ARs), and using {\it Wind}/EPACT data that comprise more events than the GOES data set, we obtained (log$-$log) correlation coefficients between $J_{\rm p}$ and $I_{\rm SXR}$ of 0.81$\pm 0.08$. There is a slight increase compared with 0.71$\pm 0.10$ for the complete set of ICME events in Table~\ref{T-correlations}, but it is within the uncertainty estimate. The correlation coefficient of $J_{\rm p}$ with $V_{\rm CME}$ does not show a significant improvement, {\it i.e.}, for the 16 ICME events we obtained correlation coefficient of 0.77$\pm 0.08$ (compared to 0.73$\pm 0.08$ for the whole ICME group). Only 12 of these 16 ICME events from the same AR or cluster of ARs have electron intensities detected by ACE/EPAM. The corresponding (log$-$log) correlation coefficients of $J_{\rm e}$ are 0.64$\pm 0.14$ with $I_{\rm SXR}$ (compared to 0.73$\pm 0.10$ for the whole sample). The difference is not statistically significant, and there is no change in the correlation coefficient with $V_{\rm CME}$, 0.65$\pm 0.15$ (compared with 0.64$\pm 0.14$).

In summary, we find an increase in the $J_{\rm p}$$-$$I_{\rm SXR}$ correlation, a decrease in the $J_{\rm e}$$-$$I_{\rm SXR}$ correlation, both within the error margins, and practically no change with the $V_{\rm CME}$ for both particle species. Due to the large uncertainties, the difference of the subset of ICME events and the entire ICME group is not statistically significant. Hence, proton propagation could be favored in the ICME, a trend that is supported by the faster rise times found for ICME events. Electrons from the same AR as the ICMEs show no trend of increased correlation with the parent activity. They also have similar rise times in both IMF configurations.

\subsubsection{The Effect of the Connection Distance on the Correlations} %%%%%%%%%%%%%%
  \label{S-Corr_conn_dist}

We now compare the correlation coefficients for particle events around the best connection longitude (Section~\ref{S-Conn_dist}), referred to as inner events, with those farther away, outer events, for both IMF categories. The ICME and SoWi events are divided into inner and outer by the absolute value of the median connection distance for each IMF category. For the {\it Wind}/EPACT protons we have $20^{\circ}$ for the ICME and $14^{\circ}$ for the SoWi events, whereas for the ACE/EPAM electrons we have $23^{\circ}$ and $28^{\circ}$, respectively. The results (see Table~\ref{T-conn_dist} in the Appendix) have large error bars. The only significant effect is a better correlation of $J_{\rm e}$ with $V_{\rm CME}$ for the inner SoWi electrons than for the outer. No such trend is found for the protons. The observed difference in the correlation coefficients of $J_{\rm max}$ with $I_{\rm SXR}$ between the ICME and SoWi events cannot be explained by a mere longitude spread (inner/outer events) around the best nominal connection within the same IMF category. Another effect needs to be present.

\section{Discussion}
\label{S-disc}

\subsection{Summary of Observational Findings}

The results of the statistical study of SEP events of solar cycle 23 associated with flares of class M or X in the western solar hemisphere are summarized as follows:
\begin{enumerate}
\item A significant number of SEP events (about 20\% $-$ 17/81 GOES proton events and 18/96 ACE/EPAM electron events) were detected while the Earth was immersed in an interplanetary coronal mass ejection. This means that these deka-MeV protons and near-relativistic (tens to hundreds of keV) electrons were guided along transient interplanetary field lines, instead of the  Parker spiral of the nominal solar wind.
\item  SEP events are relatively more often detected within ICMEs when the associated flare is of class X (10/35, 29\%) than of class M (7/46, 15\%).
\item  The peak intensities of electrons and protons cover a similar range for SEP events detected within ICMEs and in the standard solar wind. Indications of different peak intensity distributions in the two event categories are seen for protons, but they are not statistically significant. No such indication is seen for electrons.
\item The proton profiles have a median rise time shorter (by a factor of 3) within ICMEs than in the solar wind. 
\item Contrary to protons, electrons have similar distributions of rise times in ICMEs and in the solar wind. 
\item The longitudes of the parent active regions of the SEP events cluster around the nominal Parker spiral with some scatter for the ICME events, but have a very broad distribution for the events detected in the solar wind. There is no evident dependence of the peak intensity of protons or electrons on the connection distance.
\item The underlying relationship between flares and CMEs ({\it i.e.}, between peak SXR flux and CME speed) is stronger for ICME events ($r\,$=$\,$0.61$-$0.70) and weaker for SoWi events ($r\,$=$\,$0.23$-$0.28). The weak correlation is ascribed to the randomization of the projected CME speed in the SoWi category. We conclude that there is a stronger intrinsic correlation between SXR peak flux and CME speed in the two IMF categories. The correlation for the entire event sample ($r\,$=$\,$0.39$-$0.47) is comparable to previous reports.
\item The correlation of SEP peak intensities $J_{\rm max}$ with the peak flux of the associated soft X-ray burst, $I_{\rm SXR}$, and the speed of the associated CME, $V_{\rm CME}$, depends on the IMF configuration:
\begin{itemize}
\item  On average over all events, the correlation coefficients of $\log J_{\rm max}$$-$$\log I_{\rm SXR}$ and $\log J_{\rm max}$$-$$\log V_{\rm CME}$ are comparable, with values in the range 0.6$-$0.7 for the protons and 0.4$-$0.6 for the electrons.
\item The correlation $\log J_{\rm p}$$-$$\log I_{\rm SXR}$ is twice as high in SEP events detected within ICMEs as for SEP events in the solar wind. The difference disappears when the sample is restricted to limb events.
\item The correlation $\log J_{\rm e}$$-$$\log I_{\rm SXR}$ is about six times higher for the entire ICME sample and also for limb events as for SoWi sample.
\item The correlation $\log J_{\rm max}$$-$$\log V_{\rm CME}$ is similar in both event categories and particle species.
\end{itemize}
\end{enumerate}

\subsection{The IMF Configuration of SEP Events}

The occasional detection of SEP events within ICMEs is a well-known phenomenon. ICMEs provide an evident explanation why fast rising SEP events are occasionally observed in association with activity in the eastern solar hemisphere \cite{1991JGR....96.7853R}. These authors estimated that about 15\% of SEP events from the eastern solar hemisphere have a rapidly rising intensity profile and showed that this may be due to propagation within an ICME. The percentage found in the present study for the SEP events from the western solar hemisphere ({\it i.e.}, 20\% arrive within ICMEs) is similar.

The fraction of GOES SEP events detected within ICMEs increases with the flare size. Such trend is also found by a relativistic proton event study. \inlinecite{2012A&A...538A..32M} showed that 7/10 relativistic SEP events of solar cycle 23 occurred within or in the vicinity of ICMEs. The associated flares ranged from X5.7 to X14 (with the exception of one behind-the-limb event). The numbers are not directly comparable, because the events in the neighborhood of ICMEs are excluded from the ICME events in the present study. Nonetheless there appears to be a trend that the more energetic the flare associated with a particle event, the greater the likelihood to detect the SEP within or in the vicinity of an ICME. This probably reflects the fact that on the one hand the ICME rate is enhanced in periods of high activity \cite{2010SoPh..264..189R}, and that on the other hand strong flares \cite{1987ApJ...314..795B,2000ApJ...540..583S} and fast CMEs \cite{2008ApJ...680.1516W} preferentially occur in a small number of highly active regions.

We found no significant evidence that the SEP intensity distributions differ between ICME events and SoWi events. However, ICME events display on average faster rises in the proton profiles than SoWi events. This is consistent with the reported long scattering mean free paths of energetic protons in ICMEs \cite{1987JGR....92....6T,2004ApJ...600L..83T}. No such effect is seen for the electrons. This is the first time, to our knowledge, that such a comparison is carried out.

\subsection{IMF Configuration and the Correlation between SEP Parameters and Solar Activity}

A number of studies in the literature report overall, but noisy, correlations between the logarithms of SEP proton ($J_{\rm p}$) intensity and the logarithms of SXR peak flux and/or CME speed. The correlation coefficients between $\log I_{\rm SXR}$ and $\log J_{\rm p}$ at deka-MeV energies were found near 0.5 (36 events in 1973$-$1979) from \inlinecite{1982ApJ...261..710K} or 0.4 (25 events in 1996$-$2001) from \inlinecite{2003GeoRL..30lSEP3G}. Higher correlation coefficients were reported with $\log V_{\rm CME}$: 0.7 (71 events in 1986$-$2000) from \inlinecite{2001JGR...10620947K} or 0.6 from \inlinecite{2003GeoRL..30lSEP3G}. \inlinecite{2010JGRA..11508101C} report the same value of 0.6 for the correlation with $\log I_{\rm SXR}$ and $\log V_{\rm CME}$ ($\approx$100 events in 1997$-$2006). No estimate of the uncertainty of these correlation coefficients was given.

The overall correlations found in the present study between the logarithms of peak SEP intensity and SXR peak flux on the one hand, CME speed on the other, are comparable, with correlation coefficients in the range of 0.4$-$0.7 for both electrons and protons and a statistical uncertainty of about $\pm 0.07$. The entire event sample does not support the claim \cite{2003GeoRL..30lSEP3G} of a higher correlation coefficient of SEP peak intensities with CME speed than with soft X-ray flux. Our result agrees in this respect with that of \inlinecite{2010JGRA..11508101C} who reported a value of 0.6.

A finer distinction between the two IMF categories $-$ the standard solar wind and ICMEs $-$ is subject to caution, because the solar wind sample is more strongly affected by projection effects on CME speeds (due to larger variety in longitudes of the associated flare) and variable connections between the parent solar activity and the Earth-connected IMF line. We find no difference between the correlation coefficients $\log J_{\rm max}$$-$$\log V_{\rm CME}$ in the two IMF categories, but a marked difference for the correlation $\log J_{\rm max}$$-$$\log I_{\rm SXR}$. This is true for both protons and electrons. For electron events propagating in the solar wind there is virtually no correlation between  $\log J_{\rm e}$ and $\log I_{\rm SXR}$, while it is significant in ICME events. The behavior of protons is less clear: They show low correlation between $\log J_{\rm p}$ or peak fluence and $\log I_{\rm SXR}$ for the entire SoWi sample (as the electrons), but the difference disappears when the sample is restricted to limb events. However, it is uncertain if the latter relationship points to a real physical effect due to the small event sample involved and large error bars on the correlation coefficients. We conclude that the IMF structure likely affects the correlation between peak particle intensity and peak SXR flux, and that this difference comes mostly from the fact that the ICME events covers a narrower range of flare longitudes and connection distances than the SoWi events.

\subsection{A Tentative Interpretation}

The correlation between $\log J_{\rm max}$ on the one hand, and $\log V_{\rm CME}$ or $\log I_{\rm SXR}$ on the other, is not suited to discriminate clearly between CME-related and flare-related SEP acceleration processes. Statistical relationships do exist, but they appear to be strongly dependent on the location of the parent activity in the corona and the observer. Furthermore, the statistical relationship between SXR peak flux and CME speed is rather strong itself. This agrees with the recent findings of close relationships between CME kinematics and energy release in flares \cite{Zha:al-01,Bei:al-12}. Irrespective of the acceleration agent and mechanism, particles are released from the acceleration region in a direction that can cover a range of angular orientations and also be event-dependent. If it scatters around the normal to the photosphere, the direction of maximum particle intensity will be in some range around the longitude and latitude of the flare, and may well vary during the event ({\it cf.} \opencite{2012SoPh..276..199M}). Since they are detected at the Earth, the particles reach the magnetically well-connected IMF line, which is likely a Parker spiral on average, but again with a broad scatter ({\it e.g.}, \opencite{Smi-08book}). During their interplanetary travel the particles are also subject to pitch-angle scattering due to the magnetic inhomogeneities. All these effects will introduce additional blurring in the correlation between the particle intensity and the parameters of the parent activity. We consider in more details the IP transport and the connection distance.

The estimation of rise time in the present study is used as a proxy for the IP transport. We find the rise times of proton profiles scatter over a much broader range in SoWi events than in ICME events. This suggests a broader variation of mean free paths with respect to pitch angle scattering in the SoWi events. One then expects that a given peak intensity at injection will be smeared out into a range of peak intensities at 1~AU, depending on the scattering conditions encountered. The much lesser dispersion of rise times in the ICME events is consistent with the long mean free paths found in some earlier studies. Then, any existing correlation between the peak SEP intensity and the flare strength will be better preserved in ICME events than in SoWi events. But while this argument works for protons, it does not for electrons: Electrons, like protons, show stronger correlation with SXR flux in ICME events than in SoWi events, but have similar (median values for the) rise times in the two IMF categories. So the interplanetary propagation of the SEP cannot be the only reason for the different correlations with peak SXR flux.

The other difference between the ICME and SoWi samples is in the connection distance. We found that both electrons and protons tend to come from activity that is closer to the optimal magnetic connection with the Earth in ICME events than in SoWi events.

Flare-related particle acceleration occurs in small volumes within or around an active region. The coronal magnetic field guiding the particles from the acceleration site to the magnetically well connected field lines is essential for the detection of flare-related particles. The broad scattering of connection distances observed in SoWi events then suggests a stronger blurring than in ICME events, and an ensuing loss of correlation between $J_{\rm max}$ and $I_{\rm SXR}$. The broad range in connection distances combined with a spread in injection angles of the SEPs may also be responsible for the flat distribution of the particle intensity with connection distances (see Figure~\ref{F-histo_cd}).

A CME shock is a rather extended accelerator and is expected $-$ at least in simple scenarios $-$ to inject particles into a broad cone of interplanetary field lines. This makes the correlation of the peak SEP intensity with the CME speed less sensitive to the connection distance than the correlation with the SXR peak flux.

At the present time such interpretation is speculative. Irrespective of the physical relationships, the usefulness of either the CME speed or the SXR peak flux in empirical schemes of SEP prediction is confirmed by the present study, with a greater sensitivity to the angular connection if the peak SXR flux is used.

\begin{acks}
The authors acknowledge D.~Boscher (ONERA Toulouse) for making the IPODE database of GOES particle measurements available to us. We also thank T. Dudok de Wit, M. Temmer, G. Trottet, H. Reid, and A. Veronig for helpful discussions and the referee for her/his comments. RM acknowledges a post-doctoral fellowship by Paris Observatory. The CME catalog is generated and maintained at the CDAW Data Center by NASA and The Catholic University of America in cooperation with the Naval Research Laboratory. SOHO is a project of international cooperation between ESA and NASA.
\end{acks}

\appendix
\label{Appendix}

Tables~\ref{T-ICME_events}$-$\ref{T-Other_events} summarize all data used in the paper, organized in different IMF categories, namely ICME, SoWi and SEP events in the vicinity of an ICME (Section~\ref{S-IP_field}). The events in each table are listed chronologically: The event date is given in column~(1). The proton and electron peak intensities (with their onset time) follow in columns (2)$-$(5). The next four columns give the SXR peak flux (with the onset time), the flare position on the western (W) hemisphere, the projected CME speed and the angular width (AW), as reported in catalogs or from previous works. The data sources are explained in detail in the footnotes under each Table. In column~(10) we give the temporal offset between the GOES SEP start (or at {\it Wind} at 1~AU) and the nearest-in-time boundary of the ICME (shifted at GOES orbit or as observed at 1~AU). This value is used as a confidence check for the identification of the IMF category. Although we used exclusively the timings of the ICME boundaries as reported in \inlinecite{2010SoPh..264..189R}, differences might exist with other ICME lists due to different definition used for an ICME, variation in the IMF data from different satellites and also due to the subjectivity of the observer. In the ICME category (Table~\ref{T-ICME_events}) two events are relatively close (about 2 h) to the reported ICME onset and may change category after a detailed analysis. All other events in this category are well within the body of the ICME. Similarly for the last SEP category (Table~\ref{T-Other_events}), some SEP events might be propagating in quiet solar wind conditions, although many are in the sheath region of the ICME or occur only several hours before or after the ICME boundary. Rise times are given in column~(11) in Tables~\ref{T-ICME_events} and \ref{T-SoWi_events}. Finally, in the last two columns in each table the solar wind speed (averaged values) and the connection distance are given.

\begin{sidewaystable}[h]
\caption[]{ICME solar energetic particle events.\\}
\label{T-ICME_events}
\tiny
\vspace{0.5cm}
\begin{tabular}{crrrr|rrrr|rrrr}
\hline
Event &\multicolumn{4}{c|}{Particle intensity$\quad$ (cm$^2$ s sr MeV)$^{-1}$} &\multicolumn{2}{c}{$\quad$Flare} & \multicolumn{2}{c|}{$\quad$CME} & \multicolumn{2}{c}{$\quad$SEP} &  & \\
date  &  GOES   & {\it Wind}/EPACT &  \multicolumn{2}{c|}{ACE/EPAM  ($\times 10^4$)} & Peak SXR      & Long. &       &      &       & Rise time & SoWi & Conn.\\
yymmdd& 15$-$40 & 19$-$28    &  38$-$53     & 175$-$135                        & flux         & W    & speed & AW  & offset &  (2)/(3)/(4)/(5)  & speed  & dist.\\
      &   [MeV]   &  [MeV]       &    [keV]       & [keV]                              & [W$\,$m$^{-2}$] & [deg]   & [km/s]  & [deg] & [hrs] &  [min] & [km/s]  & [deg] \\
(1)   & (2)     & (3)        &  (4)         & (5)                              & (6)           & (7)   &  (8)  &  (9)  & (10) & (11) & (12) & (13) \\
\hline
98 05 02 & 5.2 (14:00)  & 1.5 (14:00)            & 22 (13:55)   & 1    & X1.1 (13:31) & 15 & 938  & 130 & $-$8.2 & 8/18/19/5      & 619 & $-$23\\
98 05 06 & 8.7 (08:30)  & 5.7 (08:30)            & 140 (08:10)  & 4.1  & X2.7 (07:58) & 65 & 1099 & 90  & +39.2  & 9/11/4/$-$     & 506 & ~18\\
99 12 28 & $-$          & 0.006$^w$ (02:00)      & 3.6 (01:45)  & 0.11 & M4.5 (00:39) & 56 & 672  & 60  & +2     & $-$/$-$/62/18  & 465 & 5\\
00 06 25 & 0.04 (12:30)$^{d}$& 0.03 (12:00)$^{d}$& 0.45 (08:00) &0.0032& M1.9 (07:17) & 55 & 1617 & 70  & +20.3  & $u$            & 508 & 9\\
00 07 14 & 312 (10:30)  & 88 (10:40)             & 72 (10:40)   & 12   & X5.7 (10:03) & 7  & 1674 & 360 & +5.2   & 11/6/5/6       & 579 & $-$34\\
00 08 12 & 0.06 (11:00) & 0.05$^{u}$ (11:00)     & 0.47 (10:55) &0.0046& M1.1 (09:45) &(79)& 662  & 60  & $-$5.3 & $u$            & 617 &(41)\\
00 09 09 & $-$          & 0.02 (10:00)           & 0.21 (10:00) &0.0077& M1.6 (08:28) & 67 & 554  & 70  & $-$22  & $-$/$-$/25/58  & 468 & 17 \\
00 09 19 & 0.03 (14:00)$^{d}$& 0.02 (13:00)$^{d}$& 2.3 (09:15)  & 0.019& M5.1 (08:06) & 46 & 766  &60   & +34.7  & $u$/$u$/81/65  & 672 & 11 \\
00 11 08 & 377 (23:30)  & 179$^{u}$ (23:00)      & 240 (23:00)  & 32   & M7.4 (22:42) &[78]& 1738 &120  & $-$9.6 & 6/4/4/2        & 460 &[27]\\
01 03 29 & 0.9 (12:30)  & 0.66 (12:00)           & 7.9 (10:37)  & 0.35 & X1.7 (09:57) & 19 & 942  & 360 & $-$18.7& 69/45/74/20    & 566 & $-$23\\
01 04 02 & 0.09 (12:00) & 0.08 (13:00)           & $-$          & $-$  & X1.1 (10:58) &(62)& 992  & 50  & +28.8  & 119/6$^u$/$-$/$-$ & 585 & (22)\\
01 04 02 & 43 (23:00)   & 36$^m$ (23:00)         & 46 (22:05)   & 2.2  & X20 (21:32)  &[70]& 2505 & 100 & +16.7  & 18/4/14/9      & 543 & [27]\\
01 04 12 & 1$^{m}$ (11:30) & 0.75$^{m}$ (12:00)  & $-$          & $-$  & X2.0 (09:39) & 43 & 1184 & 120 & $-$12.9& $u$/$-$        & 633 & 6\\
01 10 22 & 0.5 (16:40)$^{d}$ & 0.4 (17:00)$^{d}$ & 0.92 (01:13) & 0.0076& M1.0 (00:22)& 57 & 772  & 20  & $-$8   & 53/$u$/186/196 & 553 & 14 \\
02 04 21 & 81.5 (01:30) & 81 (01:30)             & 84 (01:40)   & 3.5  & X1.5 (00:43) & 84 & 2393 & 120 & +17.3  & 15/11/3/2      & 485 & 35\\
02 08 03 &  $-$         & 0.007$^{w}$ (00:00)$^{nd}$& 0.9 (20:05)& 0.01 & X1.0 (18:59) & 76 & 1150 & 30 & +2.4 & $-$/$u$        & 497 & 29\\
02 08 20 & 0.07 (09:00) & 0.37 (09:00)            & 40 (08:52)   & 1.3  & M3.4 (08:22) & 38 & 1099 & 40  & $-$20.1& 54/37/14/13   & 466 & $-$13 \\
02 12 22 & $-$          & 0.02$^w$ (14:00)$^{d}$  & $-$          & $-$  & M1.1 (02:14) & 42 & 1071 & 80  & +5     & $-$           & 454 & $-$10 \\
03 05 31 & 0.73 (03:00)  & 0.57 (03:00)            & 14 (02:55)  & 0.65 & M9.3 (02:13) & 65 &1835 & 150 & $-$4.4 & 20/15/19/9    & 702 & 31 \\
03 08 19 & $-$          & 0.007 (10:00)       & 2.9$^{w}$ (08:30)& 0.045& M2.0 (07:38)& 63 & 412  & 40  & +5.9   & $-$/$-$/16/15 & 467 & 12 \\
03 10 29 & 57$^m$ (21:30) & 94$^m$ (21:30)        & 150 (22:05)  & 12   & X10 (20:37)  & 2  & 2029 & 360 & +5.8   & $u$   & 812$^{s}$ & $-$27\\
04 11 10 & 1.5$^m$ (03:00)& 5.5$^m$ (06:00)       & $-$          & $-$  & X2.5 (01:59) & 49 & 3387 & 120 & $-$6.4 & 14/37/$-$/$-$ & 758 & 18\\
\hline
\end{tabular}
\footnotetext{Column (7) lists the heliographic west longitude of the flare according to: the preliminary listings of the GOES solar X-ray flares in the {\it Solar Geophysical Data} (SGD), the corresponding H$\alpha$ flare longitude in the comprehensive reports in the SGD (in the parentheses), or the daily flare active region longitude reported in \url{SolarMonitor.org} (in square brackets). In column (9), the angular width \cite{2010JGRA..11508101C} of the corresponding CME is given. Column (10) lists the temporal offset in hours between the SEP start at GOES data (or {\it Wind}/EPACT when no event in GOES is observed) and the nearest ICME boundary, {\it i.e.}, a positive value denotes the time from the SEP onset to the end of the ICME and a negative value $-$ to the start of the ICME. \\
{\it C}: proton intensity from \inlinecite{2010JGRA..11508101C}; {\it d}: delayed SEP onset; {\it m}: multiple SEP intensity peaks; {\it nd}: next day; {\it s}: strong increase in the solar wind speed (1-h average data) during the 6-h period before the SEP onset; {\it w}: weak SEP intensity; {\it u}: uncertain.}
\end{sidewaystable}

\begin{sidewaystable}[h]
\caption[]{SoWi solar energetic particle events.\\}
\label{T-SoWi_events}
\tiny
\vspace{0.5cm}
\begin{tabular}{crrrr|rrrr|rrrr}
\hline
Event &\multicolumn{4}{c|}{Particle intensity$\quad$ (cm$^2$ s sr MeV)$^{-1}$} &\multicolumn{2}{c}{$\quad$Flare} & \multicolumn{2}{c|}{$\quad$CME} & \multicolumn{2}{c}{$\quad$SEP} &  & \\
date  &  GOES   & {\it Wind}/EPACT &  \multicolumn{2}{c|}{ACE/EPAM  ($\times 10^4$)} & Peak SXR      & Long. &       &      &       & Rise time & SoWi & Conn.\\
yymmdd& 15$-$40 & 19$-$28    &  38$-$53     & 175$-$135                        & flux         & W    & speed & AW  & offset &  (2)/(3)/(4)/(5)  & speed  & dist.\\
      &   [MeV]   &  [MeV]       &    [keV]       & [keV]                              & [W$\,$m$^{-2}$] & [deg]   & [km/s]  & [deg] & [days] &  [min] & [km/s]  & [deg] \\
(1)   & (2)     & (3)        &  (4)         & (5)                              & (6)           & (7)   &  (8)  &  (9)  & (10) & (11) & (12) & (13) \\
\hline
97 05 21 &  $-$         & 0.006$^w$ (21:00) & $-$           & $-$    & M1.3 (20:08) & 12 & 296 & 30 & +4.8 & $-$            & 311 & $-$64 \\
97 11 03 &  $-$         & 0.005$^u$ (12:00) & 0.23 (10:50)  & 0.0059 & M4.2 (10:18) &(22)& 352 & 100& +3.7 & $-$            & 317 & ($-$52) \\
97 11 04 & 2$^m$ (06:30)& 1 (07:00)         & 12 (06:24)    & 0.46   & X2.1 (05:52) & 33 & 785 & 110& +2.9 & 22/38/$-$/$-$  & 307 & $-$44 \\
98 11 05 & \multicolumn{2}{c}{{\it 0.0006 ($<$22:00)}$^C$}& $-$& $-$& M8.4 (19:00) & 18 & 1118 &60& +2 & $-$ & 433 & $-$36 \\
98 12 17 & \multicolumn{2}{c}{{\it 0.0003 (11:00)}$^C$}  & 0.4 (08:20) & 0.0086 & M3.2 (07:40) & 46 & 302 & $-$& +12.3 & $-$/$-$/13/60 & 382 & $-$16 \\
99 08 28 & \multicolumn{2}{c}{{\it 0.0005 (20:00)}$^C$} & $-$ & $-$ & X1.1 (17:52) & 14 & 462 & 110 & $-$5.4 & $-$ & 642 & $-$23 \\
00 03 03 & $-$          & 0.01$^w$ (03:00) & 1 (02:30)   & 0.029  & M3.8 (02:08) & 60 & 841  & 80 & $-$1  & $-$/$u$/24/$<$5    & 423 & 4 \\
00 03 22 & 0.03 (19:00) & 0.02 (19:00)     & $-$           & $-$    & X1.1 (18:34) & 57 & 478  & 80 & $-$3.25 & $u$/$-$ & 473 & 7\\
00 04 04 & 0.62 (16:30) & 0.57 (18:00)     & 17.5 (15:33)  & 0.0946 & C9.7 (15:12) & 66 & 1188 & 60& +2.7  & 29/53/4/$<$5 & 380 & 4  \\
00 05 01 & 0.04 (10:30) & 0.007$^{u}$ (11:00) & 74 (10:23) & 1.3    & M1.1 (10:16) & 54$^c$&1360 & 20& +1.4 & $u$/$u$/$<$5/$<$5 & 436 & 0 \\
00 06 15 &  \multicolumn{2}{c}{{\it 0.001 (21:00)}$^C$} & 7.3 (19:52) & 0.025 & M1.8 (19:38) & 65 & 1081 & 70 & $-$1.6 & $-$/$-$/$<$5/$<$5 & 606 & 26 \\
00 06 17 & 0.03 (05:00) & 0.01 (05:00)     & 16 (03:20)   & 0.086 & M3.5 (02:25) & 72 & 857  & 60 & +1.2   & $u$/$u$/12/20    & 472 & 22 \\
00 07 22 & 0.52 (12:00) & 0.34$^m$ (12:00) & $-$          & $-$   & M3.7 (11:17) & 56 & 1230 & 80 & +1.1   & 22/32/$-$/$-$    & 449 & 3 \\
00 09 12 & 5.5 (14:30)  & 3.4 (14:00)      & 4.5 (12:47)  & 0.26  & M1.0 (11:31) & 9  & 1550 & 100& $-$2.1 & 36/53/13/14      & 468 & $-$41 \\
00 11 24 & 0.23 (06:30) & 0.15 (06:00)     & 2.3 (05:50)  & 0.078 & X2.0 (04:55) & (5)& 1289 & 360& +3.1   & 45/80/63/59      & 318 & ($-$69)\\
00 11 24 & 2.2$^m$ (15:30) & 1.8$^m$ (15:30) & 14 (15:43) & 0.51  & X2.3 (14:51) & 7  & 1245 & 360& +2.7   & 64/79/28/29      & 410 & $-$50\\
01 01 28 & 0.73 (16:30) & 0.63$^m$ (16:30) & 5.1 (16:35)  & 0.2   & M1.5 (15:40) & 59 & 916  & 120& $-$2.4 & 24/68/30/15      & 326 & $-$13 \\
01 03 10 & \multicolumn{2}{c}{{\it 0.002 (08:00)}$^C$}  & 0.73 (05:40) & 0.028 & M6.7 (04:00) & 42 & 819 & 20& $-$5.3 & $-$/$-$/33/57 & 419 & $-$14 \\
01 04 10 & 2.7 (08:00)  & 2.6 (08:00)      & 5.8 (05:55)  & 0.21  & X2.3 (05:06) & 9  & 2411 & 360& $-$1.1 & 76/123/162/130   & 535& $-$35 \\
01 04 26 & \multicolumn{2}{c}{{\it 0.0003 ($<$22:00)}$^C$} & 0.7 (13:30)  & 0.0051 & M7.8 (11:26)  & 31 & 1006 & 360& +1.7 & $-$/$-$/19/$w$ & 433 & $-$25 \\
01 07 19 & \multicolumn{2}{c}{{\it 0.0003 (11:00)}$^C$} & 0.88 (10:20) & 0.018 & M1.8 (09:52) & 62 & 1668 & 40& $-$5.4 & $-$/$-$/70$^u$/77  & 601 & 23 \\
01 10 19 & 0.18 (02:00) & 0.14 (02:30)     & 1.1 (02:20) & 0.036  & X1.6 (00:47) & 18 & 558  & 180 & +2.8 & 34/81/32/51        & 309& $-$58 \\
01 10 19 & 0.18 (17:30) & 0.22$^m$ (17:30) & 1.9 (17:10) & 0.039  & X1.6 (16:13) & 29 & 901  & 160 & +2.1 & 81/139/27/35       & 326 & $-$43 \\
01 11 04 & 39$^m$ (16:30)& 284 (16:30)     & 88 (16:45)  & 5.7    & X1.0 (16:03) & 18 & 1810 & 130 & +1.1 & 101/19/8/4         & 310 & $-$58 \\
01 11 22 & 177 (21:00)  & 103 (21:00)      & 64 (21:00)  & 3.1    & M3.8 (20:18) & 67 & 1443 & 120 & $-$1.3 & 16/35/19/7       & 433 & 13 \\
01 12 26 & 23 (05:30)   & 22 (06:00)       & 80 (05:40)  & 2.5    & M7.1 (04:32) & 54 & 1446 & 90  & +1.8 & 11/16/6/6          & 384 & $-$7 \\
\hline
\end{tabular}
\footnotetext{Here, column (10) lists the temporal offset in days between the SEP start from GOES data (or {\it Wind}/EPACT when no event in GOES is observed) and the nearest ICME boundary: Positive (negative) values denote the time from the SEP onset to the start (end) of the ICME following (preceeding) the SEP event, respectively.}
\footnotetext{{\it c}: flare longitude as reported by \inlinecite{2010JGRA..11508101C}; {\it m}: multiple SEP intensity peaks; {\it w}: weak SEP intensity; {\it u}: uncertain.}
\end{sidewaystable}

\begin{sidewaystable}[h]
\addtocounter{table}{-1}
\caption[]{SoWi solar energetic particle events (cont'd).\\}
\label{T-SoWi_events1}
\tiny
\vspace{0.5cm}
\begin{tabular}{crrrr|rrrr|rrrr}
\hline
Event &\multicolumn{4}{c|}{Particle intensity$\quad$ (cm$^2$ s sr MeV)$^{-1}$} &\multicolumn{2}{c}{$\quad$Flare} & \multicolumn{2}{c|}{$\quad$CME} & \multicolumn{2}{c}{$\quad$SEP} &  & \\
date  &  GOES   & {\it Wind}/EPACT &  \multicolumn{2}{c|}{ACE/EPAM  ($\times 10^4$)} & Peak SXR      & Long. &       &      &       & Rise time & SoWi & Conn.\\
yymmdd& 15$-$40 & 19$-$28    &  38$-$53     & 175$-$135                        & flux         & W    & speed & AW  & offset &  (2)/(3)/(4)/(5)  & speed  & dist.\\
      &   [MeV]   &  [MeV]       &    [keV]       & [keV]                              & [W$\,$m$^{-2}$] & [deg]   & [km/s]  & [deg] & [days] &  [min] & [km/s]  & [deg] \\
(1)   & (2)     & (3)        &  (4)         & (5)                              & (6)           & (7)   &  (8)  &  (9)  & (10) & (11) & (12) & (13) \\
\hline
02 02 20 & 0.55 (06:30) & 0.24 (06:30)  & 43 (06:05)  & 0.9   & M5.1 (05:52) & 72 & 952 & 50 & +8.5  & 13/6/3/5    & 404 & 14 \\
02 03 15 & 0.05 (02:00)$^{nd}$& 0.01$^{u}$ (01:00)$^{nd}$& 0.9 (00:30)$^{nd}$& 0.027& M2.2 (22:09) & 3  & 957 & 360 & +3.2 & 350/485/39/35 & 345 & $-$65 \\
02 04 15 & \multicolumn{2}{c}{{\it... (03:00)}$^C$}& 4.8 (03:00) & 0.029 & C9.8 (02:46) & 79 & 674 & 45 & $-$1.6 & $-$/$-$/$<$5/$<$5 & 373 & 16 \\
02 07 15 & 1.3 (11:00)$^{nd}$& 1$^m$ (09:00)$^{nd}$& $-$& $-$ & X3.0 (19:59) & 1  & 1151 & 100& +2.1  & 91/120/$-$/$-$ & 344 & $-$68\\
02 08 14 & 0.29$^m$ (02:30)  & 0.32$^m$ (02:30) & 78 (02:00) & 0.48  & M2.3 (01:47) & 54 & 1309 & 60 & +5.45 & $u$/$u$/3/6  & 444 & 1\\
02 08 16 & 0.02 (07:00) & 0.01$^w$ (08:00) & 19 (06:30) & 0.054 & M2.4 (05:46) & 83 & 1378 & 70 & +3.3  & $u$/$u$/3/5 & 597 & 44 \\
02 08 24 & 10$^m$ (01:30) & 10$^m$ (01:30) & 19 (01:30) & 1.1 & X3.1 (00:49) & 81 & 1913 & 150& $-$2.4& 13/7/5/5 & 384 & 20\\
02 11 09 & 9.2 (15:00)  & 8.3 (15:30)   & 1.9 (14:00) & 0.055 & M4.6 (13:08) & 29 & 1838 & 90 & +7.8  & 31/31/23/26 & 365 & $-$36 \\
03 03 17 & 0.02 (20:00) & 0.01 (19:30)  & 2.9 (19:14) & 0.052 & X1.5 (18:50) & 39 & 1020 & 50 & $-$1.9& $u$/$u$/8/$<$5   & 722 & 6\\
03 03 18 & 0.03 (14:00) & 0.01 (14:00)  & 49 (12:30)  & 0.32  & X1.5 (11:51) & 46 & 1042 & 80 & $-$1.1& $u$/$u$/5/8   & 762 & 15 \\
03 04 23 & $w$          & 0.01 (02:00)  & 0.15 (01:15)& 0.0076& M5.1 (00:39) & 25 & 916  & 70 & +16.2 & $u$/$u$/$u$/15   & 518 & $-$21 \\
03 04 24 & 0.04 (13:30) & 0.02 (13:00)  & 0.91 (13:10)& 0.0093& M3.3 (12:45) & 39 & 609  & 45 & +14.75& 60/$u$/8/$<$5 & 459 & $-$12 \\
03 05 27 & \multicolumn{2}{c}{{\it 0.0003 (22:00)}$^C$} & $-$ & $-$ & X1.3 (22:56) & 17 & 964 & 360 & +1.6 & $-$  & 468 & $-$30 \\
03 11 04 & 11.7 (21:30) & 10.5 (21:25)  & 6 (20:25)  & 0.38  &X28 (19:29)   & 83 & 2657 & 130& $-$2.9& 123/100/$u$/187 & 637 & 46\\
04 02 04 & $-$          & 0.006$^w$ (12:00) & 0.5 (11:26) & 0.0097 & C9.9 (11:12)& 48 & 764 & 20  & $-$10.3& $-$/$-$/243/42 & 568 & 6 \\
04 04 11 & 0.64 (06:00) & 0.5 (06:00)   & 2.9 (04:36) & 0.14  & C9.6 (03:54) & 47 & 1645 & 90 & $-$5.5& 39/37/$<$5/10 & 441 & $-$6 \\
04 07 13 & 0.04 (01:00) & 0.03 (01:30)  & 0.17 (00:40)& 0.0063& M6.7 (00:09)&[60] & 607 & 60  & +9.7  & 71/327/122/74 & 506 & [13]\\
04 10 30 & 0.06 (07:00) & 0.03$^m$ (07:30) & 5.4$^u$ (06:25) & 0.062$^u$ & M4.2 (06:08)& 21 & 422 & 90& +8.65 & 110/251/14/21 & 387 & $-$40 \\
05 05 06 & 0.05 (07:30) & 0.03 (05:00)  & 8.4 (03:47) & 0.078 & C9.3 (03:05)& (74) & 1120 & 20& +9    & 380/202/52/62 & 338$^{s}$ & (4)\\
05 05 06 & 0.03 (16:30) & 0.03 (15:00)  & 30 (12:00)    & 0.046 & M1.3 (11:11)&(80)& 1144 & 30  & +8.5  & $u$/$u$/86/82 & 357 &(14)\\
05 05 11 & 0.03$^{nd}$ (21:30) & 0.01 (21:00) & 0.8 (20:00) & 0.0074& M1.1 (19:22)&(47)& 550 & 70 & +3.4  & $u$/$u$/38/79  & 461 &($-$4) \\
05 07 13 & \multicolumn{2}{c}{{\it 0.0003 (05:00)}$^C$} & 4.7 (04:10) & 0.013 & M1.1 (02:35) &[79]& 759 & 40 & $-$1.04 & $-$/$-$/16/30 & 525 & [34] \\
05 07 13 & 0.34 (16:30) & 0.22 (16:00)  & 23 (14:40)  & 0.22  & M5.0 (14:01) & (80) & 1423 & 70  & $-$1.5 & 133/168/26/16  & 580 & (39) \\
05 08 22 & 0.22 (02:00) & 0.2 (02:00)   & 11 (01:17)  & 0.15  & M2.6 (00:44) & (48) & 1194 & 160 & +1.95  & 47/42/5/13     & 537 & (4) \\
05 08 22 & 10.7 (19:00) & 8.5 (19:00)   & 58 (17:40)  & 1.2   & M5.6 (16:46) & [62] & 2378 & 100 & +1.25  & 47/66/16/15    & 545 & [19] \\
06 07 06 & 0.08 (09:00) & 0.07 (10:00)  & 0.24 (09:05)& 0.0057& M2.5 (08:13) & (32) & 911  & 160 & +4.5   & 65/22823$^u$/69$^u$  & 576 & ($-$9) \\
06 12 13 & 25 (02:30)   & 20.4 (03:00)  & 120 (02:41)   & 5.4   & X3.4 (02:14) & (24) & 1774 & 180 & +1.8   & 17/23/$<$5/$<$5& 665$^W$ & ($-$13)\\
\hline
\end{tabular}
\footnotetext{{\it d}: delayed SEP onset; {\it m}: multiple SEP intensity peaks; {\it nd}: next day; {\it s}: strong increase in the solar wind speed (1-h average data) during the 6-h period before the SEP onset; {\it w}: weak SEP intensity; {\it W}: solar wind data from {\it Wind}/SWE; {\it u}: uncertain.}
\end{sidewaystable}

\begin{sidewaystable}[h]
\caption[]{Solar energetic particle events in the vicinity of an ICME.\\}
\label{T-Other_events}
\tiny
\vspace{0.5cm}
\begin{tabular}{crrrr|rrrr|rrr}
\hline
Event &\multicolumn{4}{c|}{Particle intensity$\quad$ (cm$^2$ s sr MeV)$^{-1}$} &\multicolumn{2}{c}{$\quad$Flare} & \multicolumn{2}{c|}{$\quad$CME} &  &  & \\
date  &  GOES   & {\it Wind}/EPACT &  \multicolumn{2}{c|}{ACE/EPAM  ($\times 10^4$)} & Peak SXR      & Long. &        &      &  SEP    & SoWi & Conn.\\
yymmdd& 15$-$40 & 19$-$28    &  38$-$53     & 175$-$135                        & flux         & W    & speed & AW  & offset &  speed  & dist.\\
      &   [MeV]   &  [MeV]       &    [keV]       & [keV]                              & [W$\,$m$^{-2}$] & [deg]   & [km/s]   & [deg] & [hrs] &  [km/s]  & [deg] \\
(1)   & (2)     & (3)        &  (4)         & (5)                              & (6)           & (7)   &  (8)   &  (9)  & (10) & (11) & (12) \\
\hline
97 11 06 & 14.6 (12:30) & 13.2 (12:30) & 47 (12:35)   & 3.4     & X9.4 (11:49) & 63 & 1556 & 115 & +16.5   & 359 & $-$3 \\
99 06 04 & 1.4 (08:15)  & 1$^m$ (08:00)& 16 (07:22)   & 0.53    & M3.9 (06:52) & 69 & 2230 & 80  & $-$9.3  &  428 & 14  \\
99 06 27 & $-$          & 0.01 (10:00) & 1.1 (09:00)  & 0.0164  & M1.0 (08:34) & 25 & 903  & 40  & +10     &  479 & $-$24 \\
00 02 12 & 0.05 (05:20) & 0.05 (06:00) & 1.1 (04:50)  & 0.013   & M1.7 (03:51) & 23 & 1107 & 110 & sheath  &  573 & $-$18 \\
00 03 02 & 0.03 (09:00) & 0.02 (10:00) & 0.51 (08:46) & 0.025   & X1.1 (08:20) &(52)& 776  & 60  & $-$5.15 &  437 & ($-$2) \\
00 05 23 & \multicolumn{2}{c}{{\it 0.002 (22:00)} Cane {\it et al.} (2010)} & 4.7 ($-$) & 0.031  & C9.5 (20:48) & 43 & 475 & 50 & sheath &  591 & 3 \\
00 06 10 & 1.4 (17:30)  & 1.7 (17:30)                   & 60 (17:09)  & 0.53   & M5.2 (16:40) & 38 & 1108 & 120 & $-$0.01 &  512 & $-$8 \\
00 06 18 & 0.07 (02:30) & 0.05$^m$ (03:00)              & 3.2 (02:22) & 0.067  & X1.0 (01:52) & 85 & 629  & 70  & +7.6 & 432 & $-$30 \\
00 06 23 & 0.05 (15:45) & 0.02 (15:30)                  & 23 (14:45)  & 0.16   & M3.0 (14:18) & 72 & 847  & 60  & sheath  & 494 & 24 \\
01 04 09 & 0.16 (16:35) & 0.1 (17:00)                   & 1.7 (16:34) & 0.081  & M7.9 (15:20) & 4 & 1194  & 360 & $-$12.6 & 521 & $-$41 \\
01 04 14 & $-$          & 0.02$^{u}$ (18:00) & 51 (17:35) & 0.43& M1.0 (17:15) & 71 & 830  & 50  & $-$6 & 655 & 35 \\
01 04 15 & 30.4 (14:00) & 30.5 (14:00)                  & 62 (14:05)  & 5.5    & X14.4 (13:19)& 85 &1129  & 110 & +2.8    & 499 & 38 \\
01 09 12 & $-$          & 0.01$^{u}$ (22:30)            & $-$         & $-$    & C9.6 (21:05) & 62$^c$& 668 & 30& +19.5   & 356 & $-$4  \\
01 09 15 & 0.33 (12:10) & 0.19 (12:30)                  & 2.3 (12:07) & 0.028  & M1.5 (11:04) & 49 & 478  & 80  & $-$13.2 &  526 & 4 \\
02 04 11 & 0.03 (17:30) & 0.08 (17:00)                  & 6.1 (16:40) & 0.0155 & C9.2 (16:16) & 33 & 540  & 50  & +8      &  475 & $-$17  \\
02 04 14 & $-$          & 0.01 (13:00)$^d$              & 0.8 (09:00) & $-$    & C9.6 (07:28) & 57 & 757  & 50  & $-$23.9 &  386 & $-$4 \\
02 04 17 & 0.47 (11:20)$^d$ & 0.46 (11:00)$^d$          & 84 ($-$)    & 0.67   & M2.6 (07:46) & 34 & 1240 & 70  & +5.5   & 333 & $-$37 \\
02 08 18 & 0.05 (22:35) & 0.06 (23:00)                  & 12 (21:41)  & 0.11   & M2.2 (21:12) & 19 & 682  & 100 & sheath &  472$^s$& $-$31 \\
02 08 19 & \multicolumn{2}{c}{{\it 0.017 (10:00)}$^C$} & 55 (10:55)  & 0.47  & M2.0 (10:28) & 25 & 549 & 80 & sheath & 532 & $-$19 \\
02 08 22 & 1.1 (02:50)  & 0.5$^m$ (03:00)               & 10 (02:21)& 0.15     & M5.4 (01:47) & 62 & 998 & 80 & $-$12.1  &  416 & 5 \\
02 12 19 & 0.06 (00:55)$^{nd}$ & 0.11 (22:30)           & 6.6 (22:00) & 0.16   & M2.7 (21:34) &(9) & 1092 & 120 & $-$12 & 478 &($-$40) \\
03 10 26 & 14.5$^m$ (18:00) & 9.4$^m$ (18:00)           & 34 (17:53)  & 0.16   & X1.2 (17:21) & 38 & 1537 & 130 & +4.9   & 468 & $-$10 \\
03 11 02 & 60 (17:30)   & 52.6 (18:00)                  & 40 (17:42)  & 4.1    & X8.3 (17:03) & 56  & 2598 & 130 & $-$17  &  533 & $-$12 \\
03 11 20 & 0.14 (08:40) & 0.08 (08:00)                  & 14.2 (08:18)& 0.266  & M9.6 (07:35) & 8   & 669  & 90  & sheath &  503$^{s}$ & $-$39 \\
04 07 25 & 1.6 (16:30)  & 1.4$^m$ (16:00)               & 13 (15:27)  & 0.35   & M1.1 (14:19) & 33  & 1333 & 130 & sheath &  590 & $-$7 \\
04 11 07 & 14.2$^m$ (17:30) & 11.8$^m$ (18:00)          & 17 (17:00)  & 0.08   & X2.0 (15:42) & 17$^c$& 1759 & 150& +3.15 & 436 & $-$37 \\
04 11 09 & 1.7 (19:40)  & 1.1$^m$ (19:30)               & 46 (18:05)  & 0.46   & M8.9 (16:59) & 51  & 2000 & 130 & sheath &  690 & 17 \\
05 01 15 & 11.2 (00:00)$^{nd}$ & 0.2 (07:00)$^{nd}$     & 92 (23:15)  & 1.3    & X2.6 (22:25) &(3)  & 2861 & 130 & +14.8  & 567$^W$ &($-$37)\\
05 01 17 & 181$^m$ (13:30)$^d$ & 205$^m$ (13:00)$^d$    & 280 (10:00) & 19     & X3.8 (06:59) &(24) & 2094 & 110 & $-$5.7 & 577 &($-$16) \\
05 01 20 & 53 (07:00)   & 66.3 (07:00)                  & 110 (06:46) & 12     & X7.1 (06:36) & 58  & 882  & 80  & $-$3.5 & 822$^s$& 32 \\
05 07 09 & 0.08 (02:30)$^{nd}$&0.1$^m$ (01:00)$^{nd}$   & 2.3 (23:15) & 0.038  & M2.8 (21:47) &(27) & 1540 & 65  & +8.5   & 345 &($-$41) \\
05 07 12 & $-$          & 0.007$^{u}$ (18:00)           & 1.2 (17:13) & 0.012  & M1.5 (15:47) &(64) & 523  & 80  & $-$14  & 502 &(17) \\
06 12 14 & 8.1 (22:30)  & 0.6$^m$ (00:40)$^{nd}$        & 6 (22:50)   & 0.23   & X1.5 (21:07) &(46) & 1042 & 70  & $-$0.05& 936 & (21)\\
\hline
\end{tabular}
\footnotetext{In column (10) positive (negative) values denote the time from the SEP (GOES) onset to the start (end) of the following (preceding) ICME, respectively. \\
{\it c}: flare longitude as reported by \inlinecite{2010JGRA..11508101C}; {\it d}: delayed SEP onset; {\it m}: multiple SEP intensity peaks; {\it nd}: next day; {\it s}: strong increase in the solar wind speed (1-h average data) during the 6-h period before the SEP onset; {\it w}: weak SEP intensity; {\it W}: solar wind data from {\it Wind}/SWE; {\it u}: uncertain.}
\end{sidewaystable}

\begin{table}[t!]
\caption{Linear correlation coefficients (with standard deviations) between $J_{\rm max}$ and $\log I_{\rm SXR}$ or $\log V_{\rm CME}$ for GOES proton and ACE/EPAM low energy electron data, for the entire event sample ({\it i.e.} no event restriction) and for different sub-samples. The number of events in each group is given in brackets.}
\label{T-J-SXR-CME}
\begin{tabular}{lccc}
\hline
SEP event          & \multicolumn{3}{c}{IMF categories of SEP events}\\
sub-samples        & ICME & SoWi & All SEPs \\
\hline
\hline
Protons & \multicolumn{3}{c}{ GOES 15$-$40 MeV} \\
$\log J_{\rm p}$$-$$\log V_{\rm CME}$    & & & \\
No event restriction                     & 0.57$\pm 0.15$ (17) &  0.66$\pm 0.07$ (38) & 0.63$\pm 0.05$ (81)\\
$\,$ Flares $ > {\rm W} 30^{\circ}$      & 0.59$\pm 0.16$ (13) &  0.61$\pm 0.11$ (26) & 0.59$\pm 0.08$ (56) \\
$\,$ Flares $ > {\rm W} 50^{\circ}$      & 0.63$\pm 0.20$  (9) &  0.60$\pm 0.12$ (17) & 0.58$\pm 0.11$ (36) \\
$\,$ Flares $ > {\rm W} 60^{\circ}$      & 0.76$\pm 0.22$  (7) &  0.65$\pm 0.14$ (11) & 0.64$\pm 0.09$ (24) \\
%\hline
$\log J_{\rm p}$$-$$\log I_{\rm SXR}$    & & & \\
No event restriction                     & 0.67$\pm 0.13$ (17) &  0.36$\pm 0.13$ (38) & 0.59$\pm 0.07$ (81)\\
$\,$ Flares $ > {\rm W} 30^{\circ}$      & 0.58$\pm 0.16$ (13) &  0.36$\pm 0.18$ (26) & 0.56$\pm 0.09$ (56) \\
$\,$ Flares $ > {\rm W} 50^{\circ}$      & 0.61$\pm 0.17$  (9) &  0.52$\pm 0.14$ (17) & 0.62$\pm 0.09$ (36) \\
$\,$ Flares $ > {\rm W} 60^{\circ}$      & 0.51$\pm 0.34$  (7) &  0.54$\pm 0.17$ (11) & 0.60$\pm 0.11$ (24) \\
\hline
\hline
Electrons & \multicolumn{3}{c}{ ACE/EPAM 38$-$53 keV} \\
$\log J_{\rm e}$$-$$\log V_{\rm CME}$    & & &  \\
No event restriction                     & 0.64$\pm 0.14$ (18) &  0.55$\pm 0.09$ (46) & 0.53$\pm 0.07$ (96) \\
$\,$ Flares $ > {\rm W} 30^{\circ}$      & 0.60$\pm 0.16$ (14) &  0.57$\pm 0.11$ (33) & 0.56$\pm 0.07$ (68) \\
$\,$ Flares $ > {\rm W} 50^{\circ}$      & 0.62$\pm 0.17$ (11) &  0.45$\pm 0.22$ (21) & 0.53$\pm 0.10$ (45) \\
$\,$ Flares $ > {\rm W} 60^{\circ}$      & 0.72$\pm 0.16$ (9)  &  0.14$\pm 0.28$ (15) & 0.52$\pm 0.13$ (32) \\
%\hline
$\log J_{\rm e}$$-$$\log I_{\rm SXR}$    & & &  \\
No event restriction                     & 0.73$\pm 0.10$ (18) &  0.12$\pm 0.11$ (46) & 0.40$\pm 0.08$ (96) \\
$\,$ Flares $ > {\rm W} 30^{\circ}$      & 0.67$\pm 0.12$ (14) &  0.08$\pm 0.12$ (33) & 0.35$\pm 0.08$ (68) \\
$\,$ Flares $ > {\rm W} 50^{\circ}$      & 0.72$\pm 0.12$ (11) &$-$0.02$\pm 0.17$ (21)& 0.35$\pm 0.11$ (45) \\
$\,$ Flares $ > {\rm W} 60^{\circ}$      & 0.67$\pm 0.18$  (9) &  0.11$\pm 0.28$ (15) & 0.38$\pm 0.13$ (32)\\
\hline
\end{tabular}
\end{table}

%%% BIBLIOGRAPHY %%%%%%%%%%%%%%%%%%%%%%%%%%%%%%%%%%%%%%%%%%%%%%%%%%%%%%%%%%%
\mbox{}~\\
     % format of references provided by the journal (.bst)
\bibliographystyle{spr-mp-sola}
%\bibliographystyle{spr-mp-sola-cnd} %% Alternative style: no title,
                                      % no concluding page.

     % name your Bibtex file containing your references (.bib)
\bibliography{SOLA_bibliography_example}

     % Checking: look if the file containing the ``\bibitem'' exits
     %           so check if the .bbl file exist (bibTeX compilation)
\IfFileExists{\jobname.bbl}{} {\typeout{}
\typeout{****************************************************}
\typeout{****************************************************}
\typeout{** Please run "bibtex \jobname" to obtain} \typeout{**
the bibliography and then re-run LaTeX} \typeout{** twice to fix
the references !}
\typeout{****************************************************}
\typeout{****************************************************}
\typeout{}}

\end{article}

\end{document}